\documentclass[apjpt4,apj,numberedappendix,appendixfloats]{emulateapj}

\usepackage{color}

\usepackage{rotating}
\usepackage{longtable}
\usepackage{paralist}
\usepackage{graphics,graphicx}
\newcommand{\nar}{New Astronomy Reviews}
\newcommand{\pasa}{Publications of the Astronomical Society of Australia}

\begin{document}

\shortauthors{Mineo, Rappaport, Levine,  Pooley, Steinhorn \& Homan}
\shorttitle{A comprehensive X-ray and multiwavelength study of the Colliding Galaxy Pair NGC~2207/IC~2163}
\title{A comprehensive X-ray and multiwavelength study of the Colliding Galaxy Pair NGC~2207/IC~2163}

\author{S. Mineo\altaffilmark{1,2}, S. Rappaport\altaffilmark{3,6}, A. Levine\altaffilmark{4}, D. Pooley\altaffilmark{5,6}, B. Steinhorn\altaffilmark{7}, \& J. Homan\altaffilmark{4}}

\altaffiltext{1}{Harvard-Smithsonian Center for Astrophysics, 60 Garden Street Cambridge, MA 02138 USA; smineo@cfa.harvard.edu} 
\altaffiltext{2}{Max Planck Institut f\"ur Astrophysik, Karl-Schwarzschild-Str. 1 85741 Garching, Germany}
\altaffiltext{3}{37-602B, M.I.T. Department of Physics and Kavli
 Institute for Astrophysics and Space Research, 70 Vassar St.,
 Cambridge, MA, 02139; sar@mit.edu} 
\altaffiltext{4}{M.I.T. Kavli
Institute for Astrophysics and Space Research, Room 37-575, 70 Vassar St.,
Cambridge, MA, 02139; aml@space.mit.edu; jeroen@space.mit.edu} 
\altaffiltext{5}{Sam Houston State University, Department of Physics, Farrington Building, Suite 204, Huntsville, Texas 77341, dave@shsu.edu}
\altaffiltext{6}{Eureka Scientific, Inc., 2452 Delmer Street, Suite 100, Oakland, CA 94602}
\altaffiltext{7}{Harvard-MIT Division of Health Sciences and Technology, Harvard Medical School, 260 Longwood Avenue, Boston, MA 02115; bsteinho@mit.edu}

\begin{abstract}
We present a comprehensive study of the total X-ray emission from the colliding galaxy pair NGC~2207/IC~2163, based on {\em Chandra}, {\em Spitzer}, and {\em GALEX} data. We detect 28 ultra-luminous X-ray sources (`ULXs'), 7 of which were not detected previously due to X-ray variability. Twelve sources show significant long-term variability, with no correlated spectral changes. Seven sources are transient candidates. One ULX coincides with an extremely blue star cluster (B-V = -0.7). We confirm that the global relation between the number and luminosity of ULXs and the integrated star formation rate (`SFR') of the host galaxy also holds on {\em local} scales. We investigate the effects of dust extinction and/or age on the X-ray binary (`XRB') population on sub-galactic scales. The distributions of $N_{\rm{X}}$ and $L_{\rm{X}}$ are peaked at $L_{\rm{IR}}/L_{\rm{NUV}}\sim 1$, which may be associated with an age of $\sim$$10$ Myr for the underlying stellar population. We find that $\sim$$1/3$ of the XRBs are located in close proximity to young star complexes. The luminosity function of the X-ray binaries (`XRBs') is consistent with that typical for high-mass X-ray binaries, and appears unaffected by variability. We disentangle and compare the X-ray diffuse spectrum with that of the bright XRBs. The hot interstellar medium dominates the diffuse X-ray emission at $E\lesssim 1$ keV, has a temperature $kT=0.28^{+0.05}_{-0.04}$ keV and intrinsic 0.5-2 keV luminosity of $7.9\times 10^{40}\,\rm{erg}\,\rm{s}^{-1}$, a factor of $\sim$2.3 higher than the average thermal luminosity produced per unit SFR in local star-forming galaxies. The total X-ray output of NGC~2207/IC~2163 is $1.5\times10^{41}\,\rm{ergs}\,\rm{s}^{-1}$, and the corresponding total integrated SFR is $23.7\,M_{\odot}\,\rm{yr}^{-1}$.\end{abstract}


\section{Introduction}
\label{sec:intro}

\begin{figure*}
\begin{center}
\includegraphics[width=1.0\linewidth]{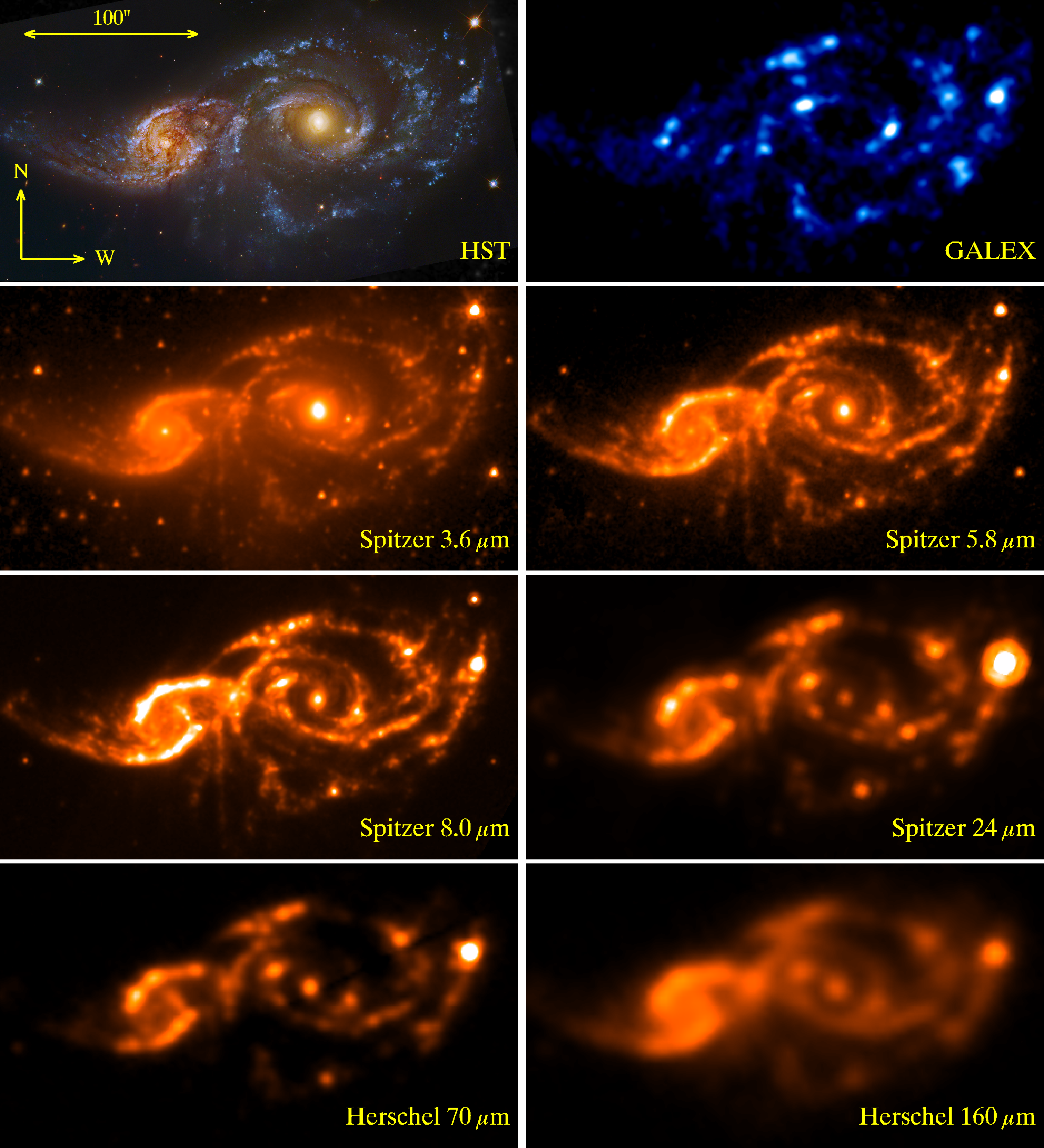} 
\caption{Montage of images of NGC 2207/IC 2163 taken with {\em HST} \citep[WFPC2 camera using filters F336W, F439W, F555W and F814W][]{2000AJ....120..630E}, {\em GALEX} (FUV), {\em Spitzer} (3.6, 5.8, 8.0 and 24 $\mu$m), and {\em Herschel} (70 and 160 $\mu$m).  NGC 2207 is the larger galaxy on the right.}
\label{fig:multipanel}
\end{center}
\end{figure*}

Galaxies in collision are known to host intense star formation activity. Presumably this is due to dynamical shocks that are induced by the supersonic relative speeds of the galaxies in comparison to the thermal speeds of the stars and gas clouds within the galaxies. These shocks, in turn, trigger the collapse of molecular clouds, leading to the formation of star clusters with a wide spectrum of stellar masses, including O and B stars \citep{1997ApJS..113..269S, 2005MNRAS.364...69S, 2006MNRAS.365...37B}. Furthermore, as is well known empirically, many of these O and B stars will naturally be found in binary systems.

Accompanying these star formation events in galaxy collisions are the production of numerous different classes of high-energy astrophysical objects such as core-collapse supernovae \citep{2000AJ....120.1479H, 2011PASJ...63S.363K}; high-mass X-ray binaries \citep[HMXBs;][]{1992ApJ...388...82D, 2003MNRAS.339..793G, 2004ApJS..154..519S, 2006A&A...455.1165L}; and gamma-ray burst sources \citep{2002ApJ...572L..45B}.  In the case of HMXBs  and gamma-ray burst sources, binary stars are an intrinsic part of the evolution of these objects \citep[see, e.g.,][]{1991PhR...203....1B, 2012ApJ...752...32W}, and they may also be relevant to the evolution of many supernovae \citep[see, e.g.,][]{1993Natur.364..509P}, both core collapse and thermonuclear events.  In general, the more massive star in the binary evolves first, may lose its envelope via mass transfer and/or ejection from the system, and this is followed by the collapse of the core which produces either a black hole or neutron star.  When that collapsed star accretes matter from the original secondary in the system, either via stellar wind accretion or Roche-lobe overflow, a massive X-ray binary is formed.

Among the many HMXBs that are found in collisional galaxies, a small fraction \citep[$\sim$10\%, according to][]{2012MNRAS.419.2095M} are so-called ``ultraluminous X-ray sources'' (ULXs).  These consist of off-nuclear sources with $L_{\rm{X}} > 10^{39}$ ergs s$^{-1}$, a luminosity which corresponds to the Eddington limit for an accreting 10 $M_\odot$ black hole and is taken as a fiducial reference point.  It is not known exactly what mass of black holes may power ULXs, but both super-Eddington accretion onto stellar-mass black holes \citep[see, e.g.,][]{2003MNRAS.342.1041D, 2008ApJ...688.1235M, 2009MNRAS.397.1836G} and sub-Eddington accretion onto intermediate mass black holes \citep[`IMBHs', with masses in the range of $10^3-10^4$ $M_\odot$;][]{1999ApJ...519...89C,2009Natur.460...73F} have been discussed and are plausible\footnote{However, recent evidence has been reported (Harrison et al.~2014, in prep.) that at least one ULX (in M~82) may in fact be an accreting neutron star.} \citep[see][for a review]{2011NewAR..55..166F}. 

On average $\sim$30\% of the ULXs hosted by star-forming galaxies have $L_{\rm{X}} > 4-5\times 10^{39}$ erg s$^{-1}$ and $\sim$10\% are very luminous ($L_{\rm{X}} >  10^{40}$ erg s$^{-1}$). Many of the high-luminosity ULXs are hosted by colliding galaxies and become X-ray bright $\approx$10-20 Myr after the end of star formation \citep{2004ApJS..154..519S, 2009ApJ...703..159S, 2011MNRAS.416.1844W}. 

The donor stars for most ULXs, i.e., for ULXs hosted by star-forming galaxies, are massive young stars and may be either blue supergiants \citep{2001MNRAS.325L...7R, 2002ApJ...580L..31L, 2012AJ....143..144S}, Wolf-Rayet stars \citep{2013Natur.503..500L} or red supergiants \citep{2005MNRAS.362...79C, 2008MNRAS.386..543P, 2014MNRAS.442.1054H}.

There are now numerous collisional galaxies that are known to host a substantial number of ULXs \citep[see, e.g.][]{2012AJ....143..144S}.  These include the famous Antennae galaxies \citep{1995AJ....109..960W, 2002ApJ...577..726Z}, the spectacular Cartwheel galaxy with its prominent ``spoke''-like features \citep{1995ApJ...455..524H, 2003ApJ...596L.171G, 2006MNRAS.373.1627W},  the ``cigar'' galaxy M82 \citep{2004MNRAS.348L..28K}, and the more recently studied NGC~2207/IC~2163 \citep{2013ApJ...771..133M}.

There is a well-established correlation between the total star formation rate (`SFR') in a galaxy and the total number of luminous X-rays sources harbored by that galaxy \citep[see, e.g.,][]{2003MNRAS.339..793G, 2005AJ....129.1350S, 2010MNRAS.408..234M, 2011ApJ...741...49S, 2012MNRAS.419.2095M}. There are numerous different techniques for determining the SFRs which include using separately, or in combination, UV continuum, H$\alpha$ recombination lines, forbidden lines ([\ion{O}{2}]), FIR continuum, and thermal radio emission \citep[see, e.g.,][for detailed discussions]{1998ARA&A..36..189K, 2012ARA&A..50..531K}. It has been suggested by a number of researchers that an appropriate linear combination of the {\em GALEX} UV and {\em Spitzer} FIR bands may be an especially robust indicator of SFR in nearby and starburst galaxies \citep[e.g.,][]{2003ApJ...586..794B,2003A&A...410...83H,2004A&A...419..109I,2006ApJS..164...38I, 2011ApJ...741..124H, 2012ARA&A..50..531K}.  In particular, \citet{2008AJ....136.2782L} proposed a formulation for computing the SFR based on a specific linear combination of {\em GALEX} FUV (centered at 1575 \AA) and {\em Spitzer} 24 $\mu$m fluxes to enable the creation of spatially-resolved (at the few arc-sec level) images of SFR per unit area (see specifically their Eqns.~(D10) and (D11)). 

In our previous work on this subject \citep{2013ApJ...771..133M} we developed a new approach toward investigating the correlation between the number and luminosity densities of luminous X-ray binaries and the local SFR in the regions immediately surrounding the X-ray sources, using the \citet{2008AJ....136.2782L} prescription. This novel technique enables us to probe these correlations on a galaxy-by-galaxy basis.  In this approach we quantitatively compare the location of the luminous X-ray sources imaged with {\em Chandra} with the spatial structures in the SFR images. Furthermore, as has been suggested by \citet[][]{2005ApJ...633..871C, 2007ApJ...666..870C}, \citet{2012ARA&A..50..531K}, M. Krumholz 2012, private communication; S. Rappaport et al.~2014, in preparation, the UV fluxes detected by {\em GALEX} tend to indicate older regions of star formation ($\sim$10-50 Myr), after the embedded dust has  been mostly cleared, in contrast with the 24-$\mu$m {\em Spitzer} images that highlight more recent star formation (i.e., $\sim$5-10 Myr) where the regions are still enshrouded by dust.  These latter regions may harbor more ULXs at the upper end of the luminosity function (i.e., with $L_X \gtrsim 10^{40}$ ergs s$^{-1}$). This kind of correlation analysis, done on the local level, enables some of the theoretical ideas concerning the formation and evolution of very massive binaries to be constrained.

Recent results by \citet{2014arXiv1410.1569L} show that in a sample of 17 nearby ($< 60$ Mpc) luminous infrared galaxies (LIRGs) with SFRs $> 7\,M_{\odot}\,\rm{yr}^{-1}$ and low foreground Galactic column densities ($N_{\rm{H}} \lesssim 5\times10^{20}\,\rm{cm}^{-2}$) there is a large deficit (a factor of $\sim$10) in the number of ULXs detected per unit SFR when compared to the detection rate in nearby normal star-forming galaxies. The study is based on {\em Chandra} observations with sufficiently sensitive imaging to permit the detection of all ULXs present in the galaxy. The authors suggest that it is likely that the high column of gas and dust in these galaxies, which fuels the high SFR, also acts to obscure many ULXs from our view. 

Based on a sample of Arp interacting galaxies, \citet{2012AJ....143..144S}, found a deficiency of ULXs in the most infrared-luminous galaxies, in agreement with the results mentioned above. They conclude that, although the active galactic nuclei may contribute to powering the far-infrared, ULXs in these galaxies may be highly obscured and therefore not detected by {\em Chandra}.

On the other hand, \citet{2013ApJ...762...45B, 2013ApJ...774..152B} show that the total X-ray luminosity output per unit SFR in distant star-forming galaxies weakly evolves with redshift. They suggest that the $L_{\rm{X}}/\rm{SFR}$ evolution is driven by metallicity \citep[see also][for similar conclusions on nearby galaxies]{2010MNRAS.408..234M, 2013ApJ...769...92P, 2014arXiv1404.3132B}, and show that the dust extinction ($L_{\rm{IR}}/L_{\rm{UV}}$) has insignificant effects on the observed values of $L_{\rm{X}}/\rm{SFR}$. These results seem consistent with that found by \citet{2012MNRAS.419.2095M}, who investigated the correlation of $L_{\rm{IR}}/L_{\rm{NUV}}$ with $L_{\rm{X}}-\rm{SFR}$ for HMXBs. They found no correlation, although the average $L_{\rm{X}}/\rm{SFR}$ seems to be decreasing with increasing values of $L_{\rm{IR}}/L_{\rm{NUV}}$ (see their Fig.~11d).

Our prior analysis of a 13 ks {\em Chandra} observation of NGC~2207/IC~2163 \citep{2013ApJ...771..133M}, an impressive pair of spiral galaxies in the initial stages of collision, revealed a total of 21 ULXs within the D25 ellipse \citep{1991trcb.book.....D}.  Such a production efficiency of luminous X-ray sources per unit stellar mass is comparable with that of the Antennae pair of colliding galaxies \citep{2002ApJ...577..726Z}.  Because NGC 2207/IC 2163 turned out be so rich in ULXs, based on an initially short 13 ks observation (Mineo et al.~2013; see also Kaufman et al.~2012), we were granted three further {\em Chandra} observations (for a collective additional exposure time of 50 ks).  

The colliding galaxies NGC~2207/IC~2163 are estimated to be at a distance of $39.6\pm 5.5$ Mpc \citep{1982ApJ...254....1A}, which is based on measurements of type Ia SNe (see NASA/IPAC Extragalactic Database `NED').  Figure \ref{fig:multipanel} shows a collection of images of NGC~2207/IC~2163 recorded with the Hubble Space Telescope ({\em HST}), {\em GALEX}, {\em Spitzer}, and {\em Herschel}, all to the same scale and orientation.  The larger (smaller) galaxy on the right (left) is NGC~2207 (IC 2163).  The collisional dynamics of such galaxy pairs in general, and of NGC~2207/IC~2163 in particular,  have been well modeled with N-body codes \citep[see, e.g.,][and references therein]{2005MNRAS.364...69S}.  Such simulations can, for example, indicate which of the currently observed features in these galaxies have been created or substantially modified by the collision. The consensus is that the major spiral arms of NGC~2207 existed prior to the collision, and have not been substantially perturbed by the interaction.  By contrast, the noteworthy ``ocular'' feature in IC~2163 (see, in particular, the {\em Spitzer} 8-$\mu$m image) was apparently produced in the encounter. These facts indicate that the collision between NGC~2207 and IC~2163 is likely in its initial phases, e.g., for perhaps only a single orbit of the galaxy pair. Given the timescale for this grazing encounter, the star formation that has been induced has likely been underway only for the past dynamical timescale, i.e., a few times $10^8$ yr. Mass estimates for NGC~2207 and IC~2163 (including dark matter), used in the \citet{2005MNRAS.364...69S} simulations as well as those measured by \citet{2013ApJ...771..133M}, are are quite comparable at $1.5 \times 10^{11}~{\rm and}~1.1 \times 10^{11}$ $M_\odot$, respectively, for the two galaxies.

In the present paper we reinvestigate the X-ray emission from this same galaxy pair with about five times the net {\em Chandra} exposure (63 ks vs.~13 ks) used for our previous study.  NGC 2207/IC 2163 was observed with {\em Chandra} on three subsequent occasions for a total additional exposure of 50 ks.  This deeper exposure enables more sensitive studies of the X-ray population ($3.4 \times 10^{38}$ erg s$^{-1}$ vs $10^{39}$ erg s$^{-1}$) and diffuse emission. The four different epochs also allow us to study the long-term variability of individual sources and that of the XLF. We also utilize multiwavelength data from the {\em GALEX}, {\em Spitzer}, {\em Herschel}, and Two Micron All Sky Survey (2MASS) archives. In Sect.~2 we describe the data products and the basic steps in the analysis.  We discuss the 56 point sources (excluding the central AGN) detected in the combined exposures, including 28 ULXs, in Sect.~3.  In this same section we present cumulative luminosity functions for the individual exposures as well as for the sum; we discuss the X-ray source variability; and we search for optical and infrared counterparts to the X-ray sources. In Sect.~4 we describe the construction of spatially-resolved maps of SFR density and $L_{\rm IR}/L_{\rm NUV}$, along with the related multiwavelength data acquisition. In Sect.~5 we repeat our correlation study between the {\em local} star formation rate in NGC~2207/IC~2163 and the number and X-ray luminosity of the ULXs -- this time with improved significance.  As a new feature of the analysis, in Sect.~6 we also compute the correlation between a {\em local} $L_{\rm IR}/L_{\rm NUV}$ and the number and luminosity of the ULXs, with quantitative considerations about the age of the stellar population associated with bright X-ray binaries. The spectrum of the diffuse emission is presented in Sect.~7.  Summary and Conclusions follow in Sect.~8.

\section{X-ray analysis}
\label{sec:Xray_analysis}

\subsection{Data preparation}
We analyzed four \textit{Chandra} ACIS-S observations of the galaxy pair NGC~2207/IC~2163 (see Table~\ref{tab:obs_log}). The data reduction was done following the standard CIAO\footnote{http://cxc.harvard.edu/ciao4.6/index.html} threads (CIAO version $4.6$, CALDB version $4.5.9$) for soft (0.5--2 keV), hard (2--8 keV) and broad (0.5--8.0 keV) energy bands. All \textit{Chandra} datasets were reprocessed using \texttt{chandra\_repro}, a script that automates the recommended data processing steps presented in the CIAO analysis threads. Using the script \texttt{fluximage} we computed a monochromatic exposure map for the mean photon energy of each band: $1.25$ keV, $5.0$ keV and $4.25$ keV for the soft, hard and broad band respectively. \texttt{fluximage} outputs both the instrument map for the center of each energy band using the tool \texttt{mkinstmap}, and the exposure maps in sky coordinates for each energy band using \texttt{mkexpmap}.

The four observations were also combined in order to improve the sensitivity. Prior to merging, we first 
corrected the individual aspect solution, using \texttt{wcs\_update}. In particular, we modified the right 
ascension ($\alpha$), the declination ($\delta$) and roll angle to match those of the observation with the 
longest exposure time (Obs. ID $14914$, see Table~\ref{tab:obs_log}). To do so, we ran CIAO 
\texttt{wavdetect} on each observation and used the output coordinates of one point source which is persistent 
and present in all 4 observations, $\alpha=94.071833$, $\delta=-21.380642$, as a common fiducial point. We 
did not choose the central AGN of NGC~2207 for that purpose, because \texttt{wcs\_update} uses the centroid 
of the chosen point source and the central AGN has a non-symmetric shape. The new aspect solution was then 
used to reproject all the original event files into the sky coordinates of Obs. ID $14914$, using 
\texttt{reproject\_events}. The new event files and aspect solutions were then merged using 
\texttt{reproject\_obs} and the individual and combined images, exposure maps, exposure-corrected images 
were created using \texttt{flux\_obs}.

The detection of point sources was carried out in the 0.5--8 keV band on all individual reprojected 
observations as well as on the combined observation, using CIAO \texttt{wavdetect}. To account  for the 
variation of the {\it Chandra} point spread function (PSF) effective width from the inner to the outer parts 
of the CCD chip, we used the $\sqrt{2}$-series from $1.0$ to  $8.0$ as the scale parameter. The value of 
the \texttt{sighthresh} parameter was set as the inverse of the total number of pixels in the image ($\sim 
10^{-6}$, $D25$ area only) in order to avoid false detections.
We used \texttt{maxiter} $= 10$, \texttt{iterstop} $= 0.00001$ and 
\texttt{bkgsigthresh} $= 0.0001$. The \texttt{wavdetect} parameter \texttt{psffile} was set differently for individual 
and combined observations. For single observations we used the tool \texttt{mkpsfmap} to compute a PSF map 
that carries information about the PSF-size for each pixel in the input image at $1.5$ keV for an $80\%$ 
enclosed counts fraction. However, \texttt{mkpsfmap} cannot be used on the combined image. In this case we 
created an exposure-map-weighted PSF map using \texttt{dmimgcalc}. This is the best approach when the roll 
angles of individual observations differ\footnote{http://cxc.harvard.edu/ciao/threads/wavdetect\_merged/}, 
which is the case for our data.

\begin{deluxetable}{ccccc}
\tablewidth{0pt}
\tabletypesize{\scriptsize}
\tablecaption{\label{tab:obs_log} NGC2207/IC2163 \textit{Chandra}
  observation log.}
\tablehead{
	\colhead{Obs.ID} &
	\colhead{Obs.Start} &
	\colhead{Exp.Time}  & 
	\colhead{Obs.Mode} &
	\colhead{Camera} \\
	\colhead{} & 
	\colhead{(UT)} &
	\colhead{(ks)} & 
      	\colhead{} &
	\colhead{} \\
	(1)  & (2) & (3) & (4) & (5)
}
\startdata 
11228 & 2010-07-18 11:04:33 & 12.88 & VF & ACIS-S\\
14914 & 2012-12-30 04:22:44 & 19.85 & F & ACIS-S\\
14799 & 2013-04-07 21:49:16 & 9.84 & VF & ACIS-S\\
14915 & 2013-08-24 04:20:31 & 19.84 & F & ACIS-S
\enddata
\tablecomments{(1) {\em Chandra} identification numbers, (2) Start date and time of the observation, (3)
  Exposure Time in kilo-seconds, (4) Observing mode (F: faint mode;
  VF: very faint mode), (5) Observing instrument.}
\end{deluxetable}

\subsection{Source photometry}
\label{sec:photometry}
The aperture-corrected X-ray photometry of compact sources was performed using the same approach and scripts as in our first paper on NGC~2207/IC~2163 \citep[see][and references therein]{2013ApJ...771..133M}.
Briefly, the count rate for each detected point source was calculated inside a circular region centered on the source coordinates given by the \texttt{wavdetect} output. In order to determine the radius of the circle, for each observation we extracted the PSF using CIAO 4.4 \texttt{mkpsf} task.
For the merged observation, the PSFs in single images were combined using the values of the exposure maps as weights. The PSFs were mapped into the World Coordinate System (WCS) reference frame of the relative point source image using \texttt{reproject\_image} task. The radius of the circle was determined individually for each source so that the encircled PSF energy was 90\%. The background region was defined as an annulus with inner radius equal to the radius of the source region and outer radius 3 times larger. For a detailed description of the procedure and its caveats we refer to Section 3.2 in \citet{2012MNRAS.419.2095M}.

A number of compact sources were found to have background regions overlapping their neighboring sources. In these cases we excluded the 90\% PSF circular regions of the overlapping point sources from the image in order to subtract the source contribution from the background counts; these regions were also subtracted from the exposure map to correct the source area. The procedure for the aperture-corrected photometry described above was then repeated using the corrected image and exposure map.

\subsection{Luminosities and hardness ratios}
\label{sec:lums}
The net count rates measured in the 0.5--8.0 keV band for each X-ray point source (Sect.~\ref{sec:photometry}) were converted into fluxes, i.e., units of erg\,cm$^{-2}$\,s$^{-1}$. The count-rate-to-flux conversion factor was obtained as follows. For each observation, we first extracted the {\em combined} spectrum of all point sources detected within the $D25$ ellipse that have a number of net counts $\geq 10$, in order to avoid undue contamination from diffuse emission and faint unresolved X-ray sources\footnote{At this stage we only have the count rates for each X-ray point source, not yet their luminosities. We note that after having obtained the source luminosities (Sect.~\ref{sec:lums}) and after having performed the completeness analysis (Sect.~\ref{sec:completeness}), the selection of sources with net counts $\geq 10$ corresponds to an average of $\sim$90\% of the sources detected above the completeness luminosity in each observation (Table~\ref{tab:compl}).}. The task \texttt{dmextract} was used for this purpose and the central source was excluded from the multi-source region, as it may be an active galactic nucleus (AGN) \citep{2006ApJ...642..158E, 2012AJ....144..156K}. Using the \texttt{sky2tdet} tool we obtained the weights maps that are needed  to create the weighted Ancillary Response Files (ARF) with \texttt{mkwarf}. The weighted Response Matrix Files (RMF) were created using \texttt{mkrmf}. The background spectrum was similarly extracted from multiple large regions between, and far enough from, the point sources, within the $D25$ ellipse, on the same detector chip. Finally, we used the script \texttt{combine\_spectra} to sum the four composite source spectra, associated background spectra and source and background ARF and RMF instrument responses. 

The final spectrum was binned in order to have a minimum of 20 counts per channel to apply $\chi^{2}$ statistics. This background-subtracted co-added spectrum (see Sect.~\ref{sec:diffuse}) was modeled as an absorbed power law using XSPEC v. 12.7.1b. The best-fit parameters for this model are $N_{\rm{H}}=(3.0 \pm 0.3)\times 10^{21}$ cm$^{-2}$ and $\Gamma = 1.95 \pm0.08$, with $\chi^{2}=78.4$ for 76 degrees of freedom (i.e., reduced $\chi^{2} = 1.03$). This is in full agreement with the results of the population study by \citet{2004ApJS..154..519S}, who showed that the distribution of the power-law photon indexes for luminous compact X-ray sources in star-forming galaxies is centered on $\Gamma=1.97\pm 0.11$. 

Our best-fit absorbed power law was adopted to convert the net count rate of each detected point source into a flux (erg\,cm$^{-2}$\,s$^{-1}$). The fluxes were then converted into luminosities (erg\,s$^{-1}$) assuming a distance of 39.6 Mpc \citep{1982ApJ...254....1A}.

Since we do not have sufficient statistics for X-ray spectral fitting of individual sources, to investigate the ULX spectral properties we used hardness ratios. The procedure described in Sect. \ref{sec:photometry} was applied to the reference source list to obtain count rates in both soft ($S$: 0.5--2 keV) and hard ($H$: 2--8 keV) bands. The respective source counts were used to calculate the X-ray hardness ratio as:
\begin{equation}
\label{eq:hr}
\rm{HR} = \frac{\rm{H}-\rm{S}}{\rm{H}+\rm{S}}
\end{equation}

\begin{figure*}
\begin{center}
\includegraphics[width=0.85\linewidth]{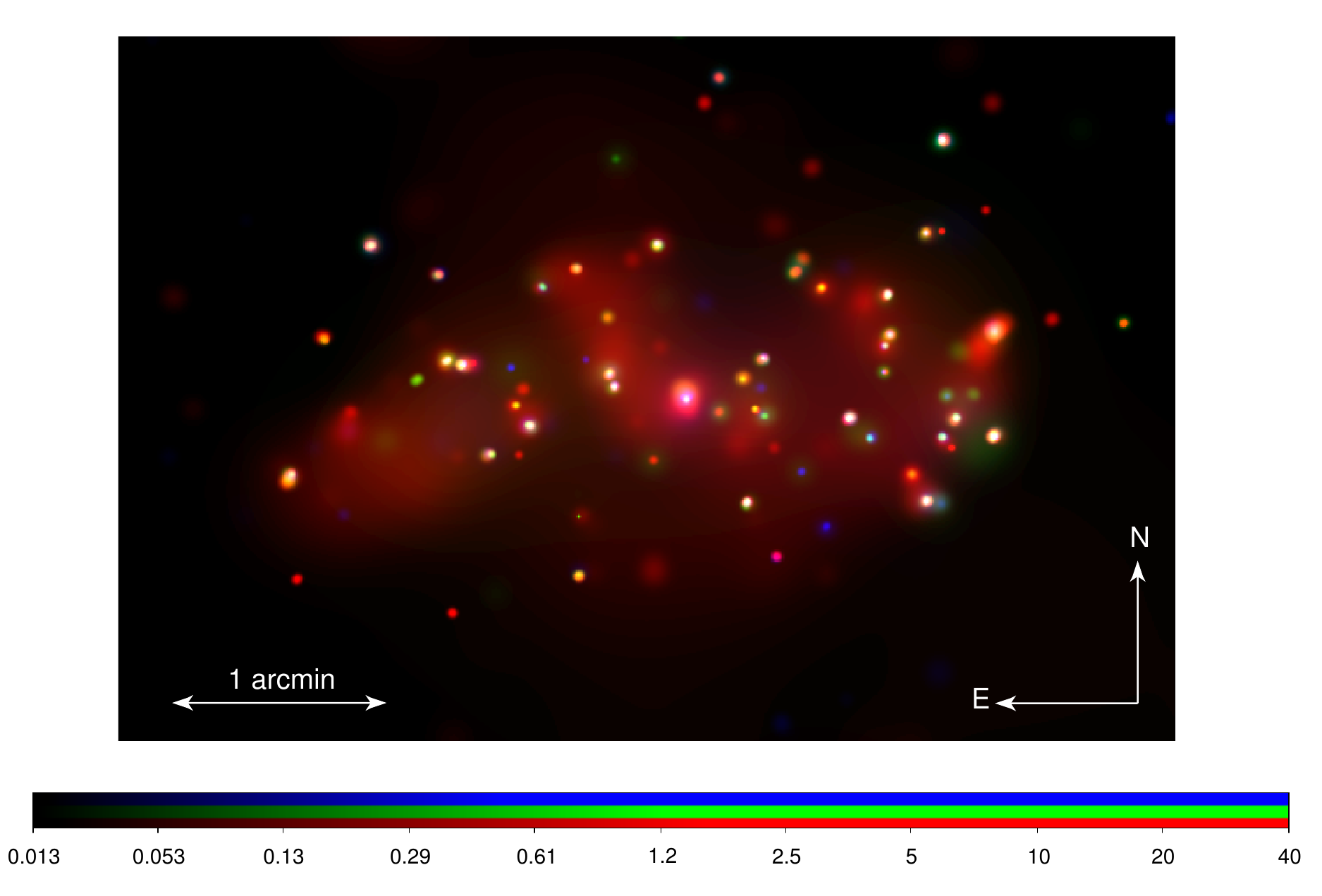} 
\caption{\textit{Chandra} X-ray image of the colliding galaxies NGC~2207 and IC~2163.  Red corresponds to soft ($0.2-1.5$ keV), green to medium ($1.5 - 2.5$ keV), and blue to hard ($2.5 - 8.0$ keV) X-ray photons.  We note that the definition of ``soft" and ``hard" photons used to generate this figure is different than that utilized in the data analysis and calculation of hardness ratio. A soft diffuse X-ray component is quite prominent in the image.  This image was adaptively smoothed using the CIAO task {\tt csmooth} (with the minimum signal-to-noise set to 2).}
\label{fig:xray}
\end{center}
\end{figure*}

\begin{figure*}
\begin{center}
\includegraphics[width=0.9\linewidth]{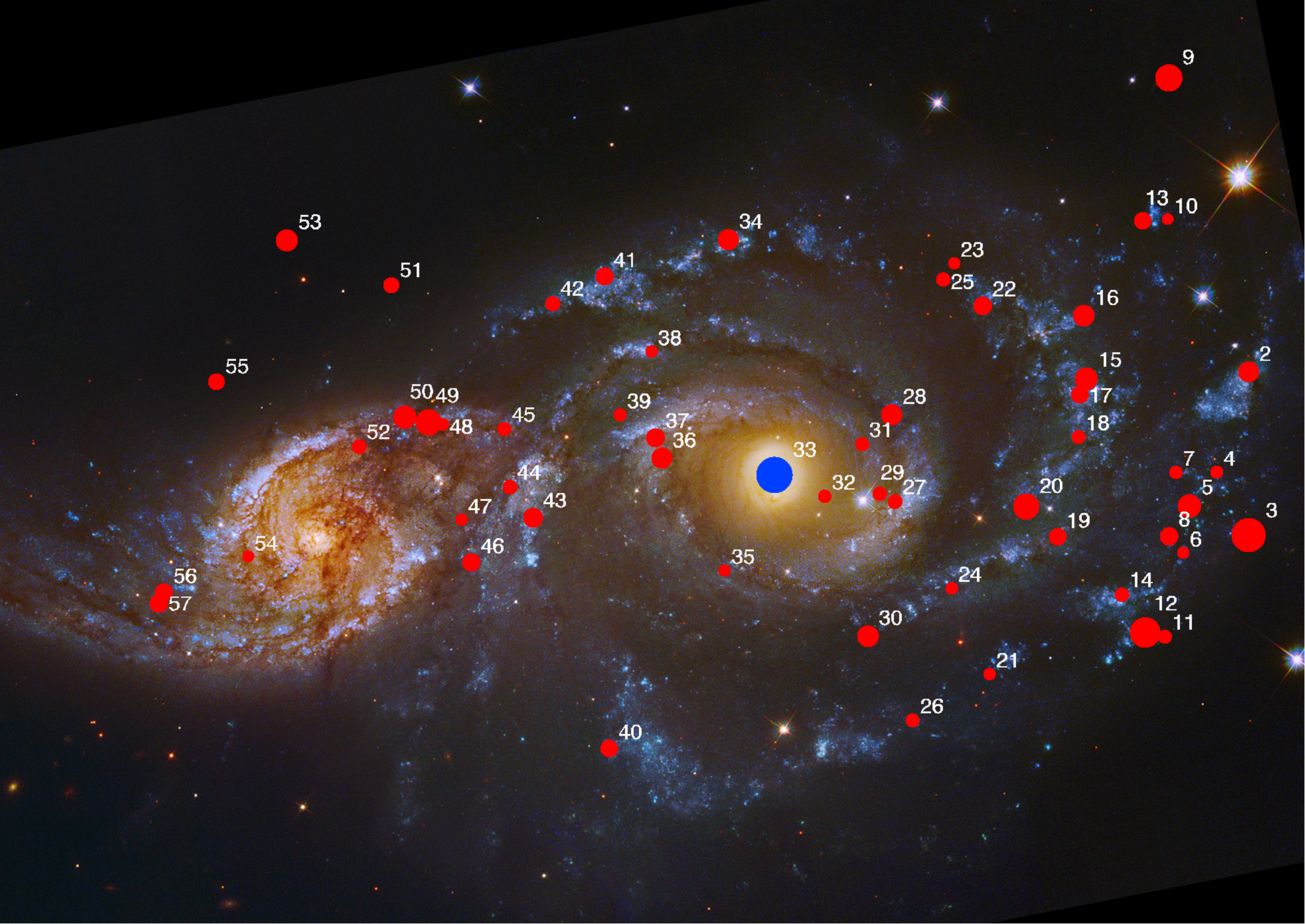} 
\caption{HST image of the galaxy pair NGC~2207/IC~2163 \citep[WFPC2 camera using filters F336W, F439W, F555W and F814W][]{2000AJ....120..630E}. The superposed filled red circles mark the locations of the 56 point X-ray sources detected in the four combined {\it Chandra} observations above the completeness limit ($3.4\times10^{38}\,\rm{erg}\,\rm{s}^{-1}$; Table~\ref{tab:compl}). The size of each circle is proportional to the cube root of the luminosity of the X-ray source.}
\label{fig:hst}
\end{center}
\end{figure*}

\subsection{Completeness analysis}
\label{sec:completeness}
Following the method and tools of \citet{2006A&A...447...71V}, we computed the completeness function $K(L_{\rm{X}})$ for all individual \textit{Chandra} observations, as well as for the combined one, within the $D25$ region. We define the ``completeness luminosity'' $L_{\rm{C}}$, as the luminosity at which $> 90\%$ of point sources are detected in the 0.5--8 keV band, corresponding to $K(L_{\rm{C}}) = 0.9$ (i.e., no more than 10\% of the sources within the $D25$ are missing). The values of $L_{\rm{C}}$ at the assumed distance of $39.6$ Mpc for each observation are listed in Table~\ref{tab:compl}.

\section{Luminous X-ray sources}
\label{sec:srcs}

\subsection{Discrete source content}
\label{sec:src_content}
In Table~\ref{tab:compl} we list the number of compact sources detected with luminosities above $L_{\rm{C}}$ within the $D25$ region, with significance of at least $3\sigma$, along with their integrated luminosity. The same table also includes the expected number and luminosity of background AGN computed above an equivalent $L_{\rm{C}}$ within the $D25$ region. The contribution of background AGNs was estimated using the $\log N-\log S$ function of \citet{2008MNRAS.388.1205G}. We converted their $\log N-\log S$ function defined over the 0.5--10 keV band, to apply in our somewhat more narrow band (0.5--8 keV). After obtaining the predicted total flux of background AGNs, $S_{\rm{AGN}}$, above the completeness threshold flux, $S_{\rm{C}}$, we computed the equivalent AGN luminosities as $4\pi\,D^{2}\,S_{\rm{AGN}}(>S_{\rm{C}})$, where $D$ is the distance to the galaxy pair. From Table~\ref{tab:compl} it is evident that the contribution of background AGNs to the bright X-ray compact source population of NGC~2207/IC~2163 is within the 5--7\% range.

The properties of the X-ray point sources detected within the $D25$ ellipse are listed in Tables \ref{tab:11228_properties}, \ref{tab:14914_properties}, \ref{tab:14799_properties}, \ref{tab:14915_properties} and \ref{tab:merged_properties}. For all sources in each observation, we provide the {\it Chandra} positions, the net counts after background subtraction in several bands, the hardness ratio, the X-ray luminosity, and the X-ray flux. There were 74 sources detected in the composite image, 57 of which ($\sim$77\%) were above the completeness luminosity. One source in each observation is associated with the low-luminosity AGN near the center of NGC~2207 \citep{2012AJ....144..156K} and is indicated with a $\dagger$ symbol in the Tables in the Appendix. 

\begin{figure}
\begin{center}
\includegraphics[trim=5mm 50mm 2mm 10mm, width=1.0\linewidth]{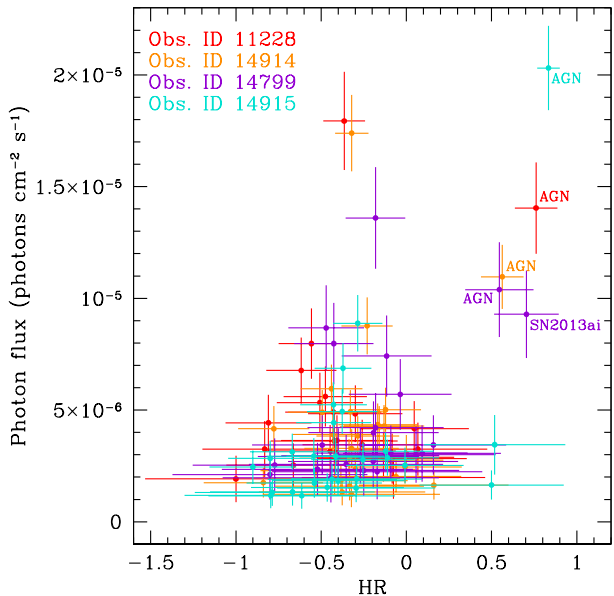} 
\caption{Net photon fluxes in 0.5--8 keV band versus hardness ratios (HR, Eq.~(\ref{eq:hr})), for all point sources detected above the completeness luminosity of each individual observation (see Table~\ref{tab:compl} for details). The sources are indicated with four different colors, depending on the observation where they were detected. The central AGN is well separated from the ULX population, having a harder ($0.5<\rm{HR}< 1$) spectrum and being bright. We also note that the spectrum of the supernova SN~2013ai, detected in Obs.ID~14799, is harder than that of the bulk of ULXs.}
\label{fig:f_hr}
\end{center}
\end{figure}

We indicate, with a $\ddagger$ symbol, the X-ray source associated with the elongated soft X-ray feature, called {\em feature i} by \citet{2000AJ....120..630E}. This source has a spectrum compatible with the rest of ULXs (see Fig.~\ref{fig:f_hr}) and could actually be a ULX embedded in the diffuse emission. \citet{2014AJ....147...60S} also found that this source is quite extended, but with a low surface brightness, and that it has a soft X-ray spectrum as well as a similar X-ray flux to that measured in the present work.

The feature is located in the outer spiral arm of NGC~2207, $\sim$$1.5\arcmin$ N-W from its center in the middle of a dusty starburst region. Other symbols ($\star$, $\diamond$ and $\ast$) are used to indicate X-ray sources that match (within $\sim$3$\arcsec$) the position of the supernovae (SNe) hosted by NGC~2207: SN~2003H, SN~2013ai, SN~1999ec.

A {\em Chandra} X-ray image of NGC~2207/IC~2163 is shown in Fig.\,\ref{fig:xray}. It was obtained by combining the four available observations (see Table~\ref{tab:obs_log}) and it unveils the presence of soft diffuse emission (discussed in Sect.~\ref{sec:diffuse}) in addition to the bright X-ray compact source population. The signal from diffuse emission is weak and was apparently not visible in the single pointing (Obs.~ID~11228) analyzed in our previous paper \citep{2013ApJ...771..133M}. The detection of diffuse emission in the current work is also aided by the fact that we now use the CIAO task {\tt csmooth}, while in the previous paper only a Gaussian smoothing kernel was applied.

In Fig.~\ref{fig:hst} we show the locations of the 57 point X-ray sources above the completeness threshold superposed on the {\em HST} image of NCC 2207/IC 2163.  Not surprisingly, many of these sources lie along the prominent spiral arms of NGC 2207 (larger galaxy on the right), though the same does not seem to follow for the smaller galaxy IC 2163.

In Fig.~\ref{fig:f_hr} the net photon fluxes (photons cm$^{-2}$ s$^{-1}$) in the 0.5--8 keV band are plotted versus the hardness ratios defined in Eq.~(\ref{eq:hr}), for all point sources detected above the completeness luminosity of each individual observation (see Table~\ref{tab:compl} for details). The sources are indicated with four different colors, depending on the observation in which they were detected. The figure shows that the bulk of ULXs in NGC~2207/IC~2163 have a soft spectrum ($HR < 0$). The hardness ratios measured for the central AGN are well separated from the ULX population, having a harder ($0.5<HR < 1$) spectrum and being bright. 

We also note that the spectrum of the X-ray source matching the position of the supernova SN~2013ai, detected in Obs.ID~14799, is harder than the spectrum of the bulk of the ULXs. Most type II SNe have X-ray spectra well described by a $kT=1~\mathrm{keV}$ to $kT=10~\mathrm{keV}$ mekal model, but a couple type IIn SNe have been observed with rather hard X-ray spectra, e.g., a power law with photon index of $-0.2$ in SN~2005kd \citep{2007ATel.1023....1P} and a power law with photon index of 1.1 in SN~2001em \citep{2004IAUC.8323....2P}.

\begin{deluxetable*}{cccccccc}
\tablewidth{0pt}
\tabletypesize{\scriptsize}
\tablecaption{\label{tab:compl} Completeness luminosities and global source
  properties.}
\tablehead{
	\colhead{Obs. ID} &
	\colhead{$L_{\rm{C}}$} &
	\colhead{$N_{\rm{X}}(>L_{\rm{C}}$)}  & 
	\colhead{$L_{\rm{X}}(>L_{\rm{C}}$)}  & 
	\colhead{$N_{\rm{AGN}}(>L_{\rm{C}}$)} &
	\colhead{$L_{\rm{AGN}}(>L_{\rm{C}}$)} &
	\colhead{XLF slope $\gamma$} \\
	\colhead{} & 
	\colhead{(erg/s)} &
	\colhead{} & 
	\colhead{(erg/s)} &
      	\colhead{} &
	\colhead{(erg/s)} &
      	\colhead{} \\
	(1)  & (2) & (3) & (4) & (5) & (6) & (7) 
}
\startdata 
11228 & $1.2\times 10^{39}$ & $20$ & $6.4\times10^{40}$ & 1.4 & $7.0\times10^{39}$ & $1.32^{+0.33}_{-0.29}$ \\
14914 & $8.2\times 10^{38}$ & $27$ & $7.1\times10^{40}$ & 1.9 & $7.5\times10^{39}$ & $1.04^{+0.23}_{-0.21}$ \\
14799 & $1.3\times 10^{39}$ & $23$ & $6.9\times10^{40}$ & 1.3 & $6.9\times10^{39}$ & $1.56^{+0.36}_{-0.31}$ \\
14915 & $7.7\times 10^{38}$ & $26$ & $5.2\times10^{40}$ & 2.0 & $7.6\times10^{39}$ & $1.27^{+0.27}_{-0.24}$ \\
Combined & $3.4\times 10^{38}$& $56$ & $7.8\times10^{40}$ & 3.6 & $8.4\times10^{39}$ & $^{\dagger}$$0.92^{+0.14}_{-0.13}$ 

\enddata
\tablecomments{All values are referred to the $D25$
  ellipse of NGC2207/IC2163. (1) {\em Chandra} identification numbers; (2) completeness
  luminosity at which $> 90\%$ of point sources are detected in the
  0.5--8 keV band; (3) number of point sources detected above
  $L_{\rm{C}}$ (the central AGN was excluded from this number); (4) cumulative 0.5--8 keV luminosity
  of compact sources above $L_{\rm{C}}$; (5) expected number of
  background AGNs above the completeness luminosity; (6) expected equivalent
  luminosity of the background AGNs. See Sect.~\ref{sec:completeness}
  for details; (7) best-fitting XLF slope of the simple powerl-law
  model (eq.~\ref{eq:xlf_pl}), in cumulative form. $\dagger$ We fitted the combined XLF
  with a power law model with an exponential cut-off
  (eq~\ref{eq:xlf_plexp}), obtaining a best-fitting slope $\alpha =
  0.66\pm0.04$ ($1.66$ in differential form) and the exponential
  cut-off at $L_{o} = (3.39 \pm 0.28)\times 10^{39}$ erg\,s$^{-1}$
  (see Section~\ref{sec:xlf} for details).}
\end{deluxetable*}

\subsection{X-ray luminosity function}
\label{sec:xlf}

\begin{figure*}
\begin{center}
\hbox
{
\includegraphics[trim=5mm 50mm 2mm 10mm, width=0.7\columnwidth]{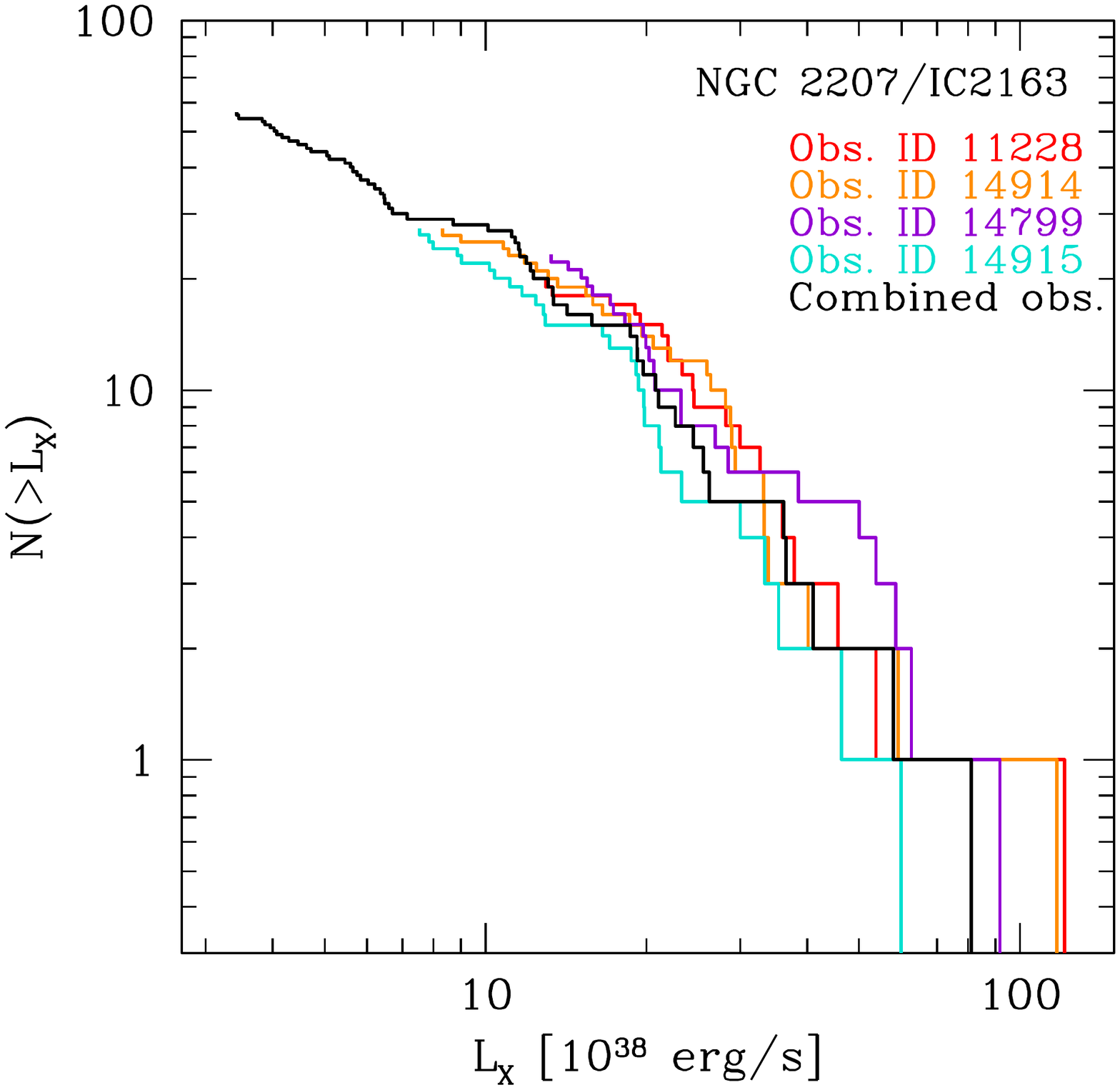}
\includegraphics[trim=5mm 50mm 2mm 10mm, width=0.7\columnwidth]{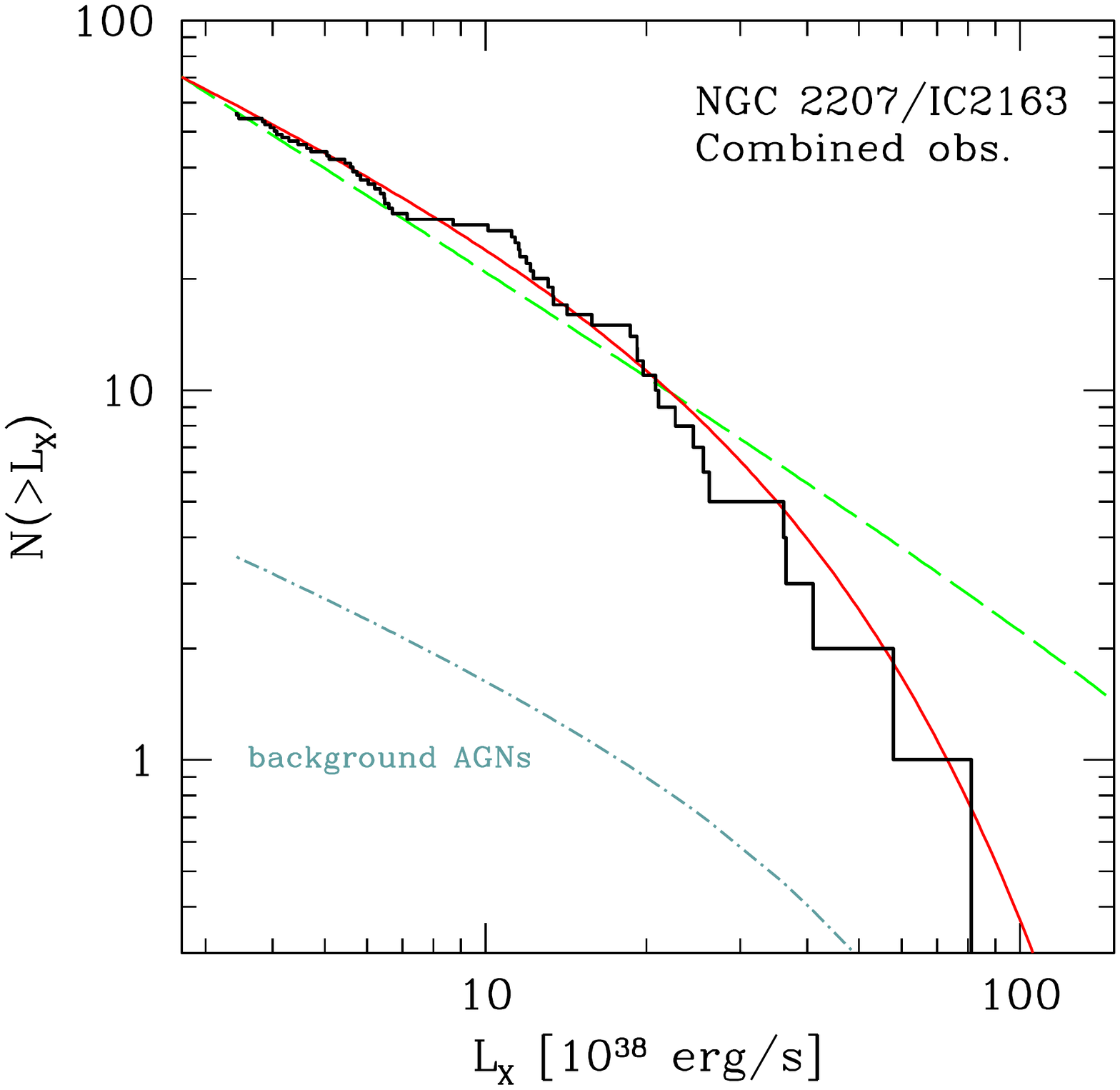}
\includegraphics[trim=5mm 50mm 2mm 10mm, width=0.7\columnwidth]{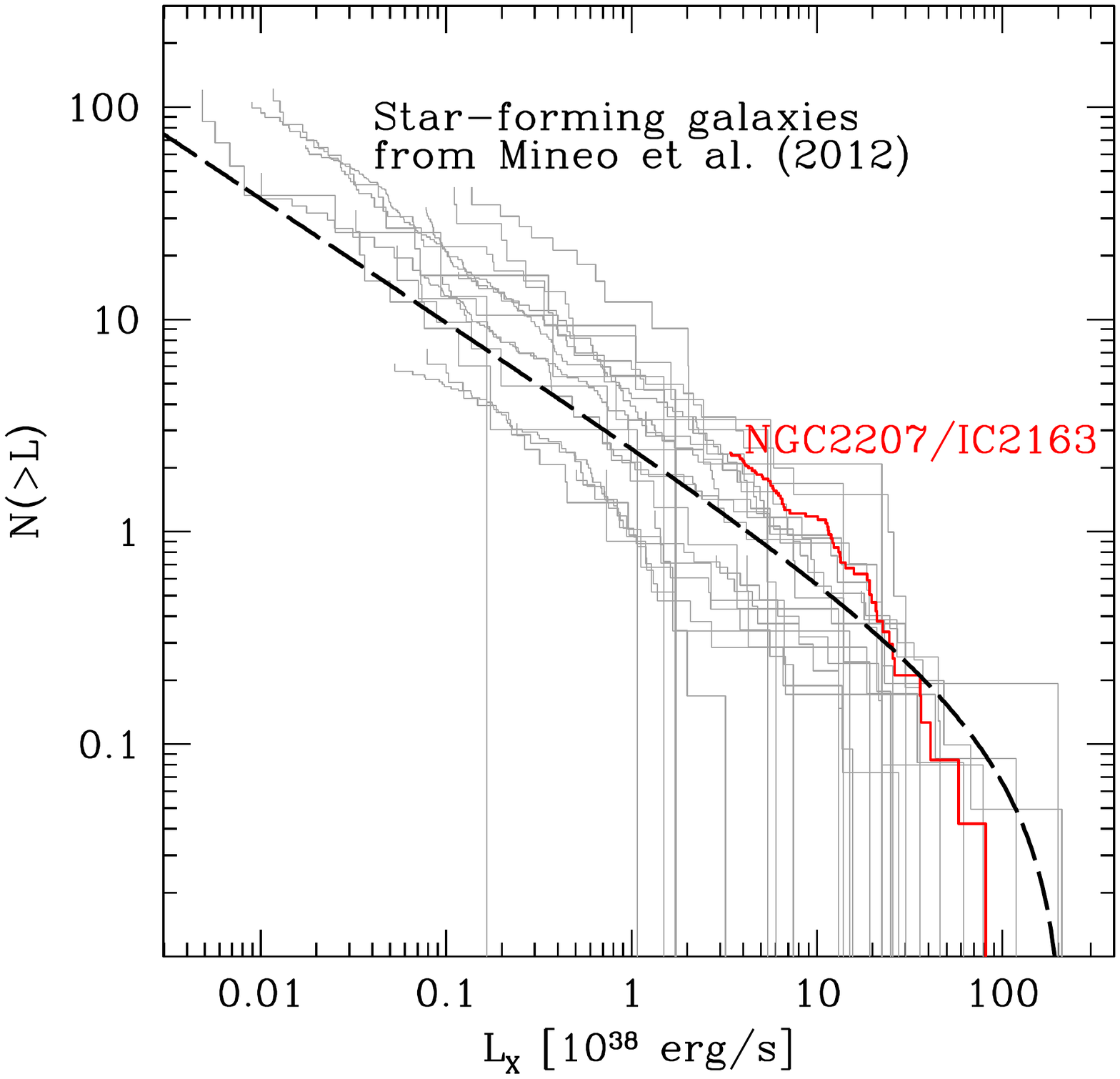}
}
\caption{({\em Left}) Cumulative X-ray luminosity distributions of the compact sources detected within the $D25$ ellipse, above the completeness luminosity of the galaxy pair NGC~2207/IC~2163: Obs.~ID~11228 (red), Obs.~ID~14914 (orange), Obs.~ID~14799 (purple), Obs.~ID~14915 (cyan), combined observations (black). The central AGN is not included. ({\em Center}) Cumulative X-ray luminosity function of the combined observations (black), along with its best-fitting models: simple power-law (Eq.~\ref{eq:xlf_pl}) with $\gamma = 0.92^{+0.14}_{-0.13}$ (dashed green) and power-law model with $\alpha = 0.66\pm0.04$ and exponential cut-off  at $L_{o} = (3.39 \pm 0.28)\times 10^{39}$ erg\,s$^{-1}$ (solid red, Eq.~\ref{eq:xlf_plexp}). The grey, dot-dashed curve shows the predicted level of resolved background AGNs, relative to the combined XLF and based on results from \citet{2008MNRAS.388.1205G}. ({\em Right}) The cumulative XLF of NGC~2207/IC~2163 (red) from the combined observations, normalized by its SFR ($23.7\,M_{\odot}\,\rm{yr}^{-1}$, \citet[][Table~2]{2013ApJ...771..133M}), plotted along with the cumulative XLFs of individual star-forming galaxies from \citet{2012MNRAS.419.2095M} (grey), normalized by their respective SFRs. The dashed black line is the average cumulative XLF per unit SFR, given by integration of their equation (18).}
\label{fig:xlf}
\end{center}
\end{figure*}

After excluding the central AGN, we constructed the cumulative X-ray luminosity function (XLF) of the X-ray point sources in NGC~2207/IC~2163, for each individual observation and for the combined observations (left and middle panels of Fig.~\ref{fig:xlf}). The cumulative luminosity distribution of background AGNs (see Sect.~\ref{sec:src_content} for details) is marked in the middle panel of Fig.~\ref{fig:xlf} by a dot-dashed (grey) curve. Fig.~\ref{fig:xlf} shows that combining the four {\em Chandra} observations allowed for a significant improvement in sensitivity: $3.4 \times 10^{38}$ erg s$^{-1}$ vs $\sim$$10^{39}$ erg s$^{-1}$ for the individual observations (see also column~2 in Table~\ref{tab:compl}).

The XLF was modeled with a simple power law. We fixed the cut-off at $L_{\rm{cut}} = 10^{41}\,\rm{erg}\,\rm{s}^{-1}$, i.e., a luminosity exceeding that of the brightest detected compact source ($L_{0.5-8\,\rm{keV}}=1.2\times 10^{40}\,\rm{erg}\,\rm{s}^{-1}$, detected in Obs.ID 11228):
\begin{equation}
N(>L_{38})=\xi \times L_{38}^{-\gamma}, ~~~~~L\le 10^{41}\,\rm{erg}\,\rm{s}^{-1}.
\label{eq:xlf_pl}
\end{equation} 
We fitted the cumulative XLFs using a Maximum Likelihood (ML) method. The best-fitting power law slopes for individual observations are listed in column (7) of Table~\ref{tab:compl}. They range between $1.04$ and $1.56$ (corresponding to $2.04$ and $2.56$ in differential form). The ML fit of the combined observations yielded a slope of $\gamma = 0.92^{+0.14}_{-0.13}$, ($1.92$ in differential form) which is indicated with a green line on the middle panel of Fig.~\ref{fig:xlf}. This slope is steeper than the typical slope of $\approx$0.6 ($1.58\pm 0.02$ in differential form) found for the high-mass X-ray binary (HMXB) luminosity distributions below $\sim$$10^{40}\,\rm{erg}\,\rm{s}^{-1}$ \citep{2003MNRAS.339..793G,2011ApJ...741...49S, 2012MNRAS.419.2095M}. In our previous work, \citep{2013ApJ...771..133M}, we found a similar result and speculated that this this might have been related to the limited sensitivity of the single {\it Chandra} observation (Obs. ID 11228) that we analyzed. In fact, with a sensitivity of $\sim$$10^{39}$ erg s$^{-1}$, we could be observing only the high-luminosity roll-off of the power-law distribution that extends to lower luminosities with a slope similar to $1.6$.
In the present work we improved the sensitivity down to $3.4\times 10^{38}$ erg s$^{-1}$ by combining four observations. We also fitted the combined XLF using a power law model with an exponential cut-off:
\begin{equation}
N(>L_{38}) = \xi\times L_{38}^{-\alpha}\,\exp(-L_{38}/L_{o}),
\label{eq:xlf_plexp}
\end{equation}
which yielded a best-fitting slope $\alpha = 0.66\pm0.04$ ($1.66$ in differential form) and the exponential cut-off at $L_{o} = (3.39 \pm 0.28)\times 10^{39}$ erg\,s$^{-1}$. The slope is now in full agreement with that of the average XLF for HMXBs, and the exponential cut-off somewhat confirms the speculation that we may be observing only the bright end roll-off of a more extended power law distribution with slope $1.6$. 

We performed a Kolmogorov-Smirnov (KS) test to determine the goodness of fit. The test statistic $D_{KS}$, i.e., maximum absolute value of the differences between the two distributions, is $0.31$ and $0.27$ for the power-law and exponentially cut-off power-law models respectively. The $p$-values for the two-sided hypothesis are $5.5\times 10^{-4}$ and  $3.8\times 10^{-3}$ respectively, i.e. smaller than the canonically assumed significance level of 0.05 for rejection. Formally the null hypothesis is rejected, but it still appears to be a good qualitative fit.

The observed power law roll-off is due to the lack of sources brighter than $\sim$$10^{40}\,\rm{erg}\,\rm{s}^{-1}$, also observed in our previous work \citep{2013ApJ...771..133M}. However, this could well be expected based on the star formation rate (SFR) of NGC~2207/IC~2163: the number predicted by the XLF from   \citet{2012MNRAS.419.2095M} for bright HMXBs is marginally consistent ($2 \pm 1.4$) with what we observe.

\subsection{ULX variability}
\label{sec:variability}

\begin{figure*}
\begin{center}
\includegraphics[trim=5mm 10mm 2mm 10mm, width=0.8\linewidth]{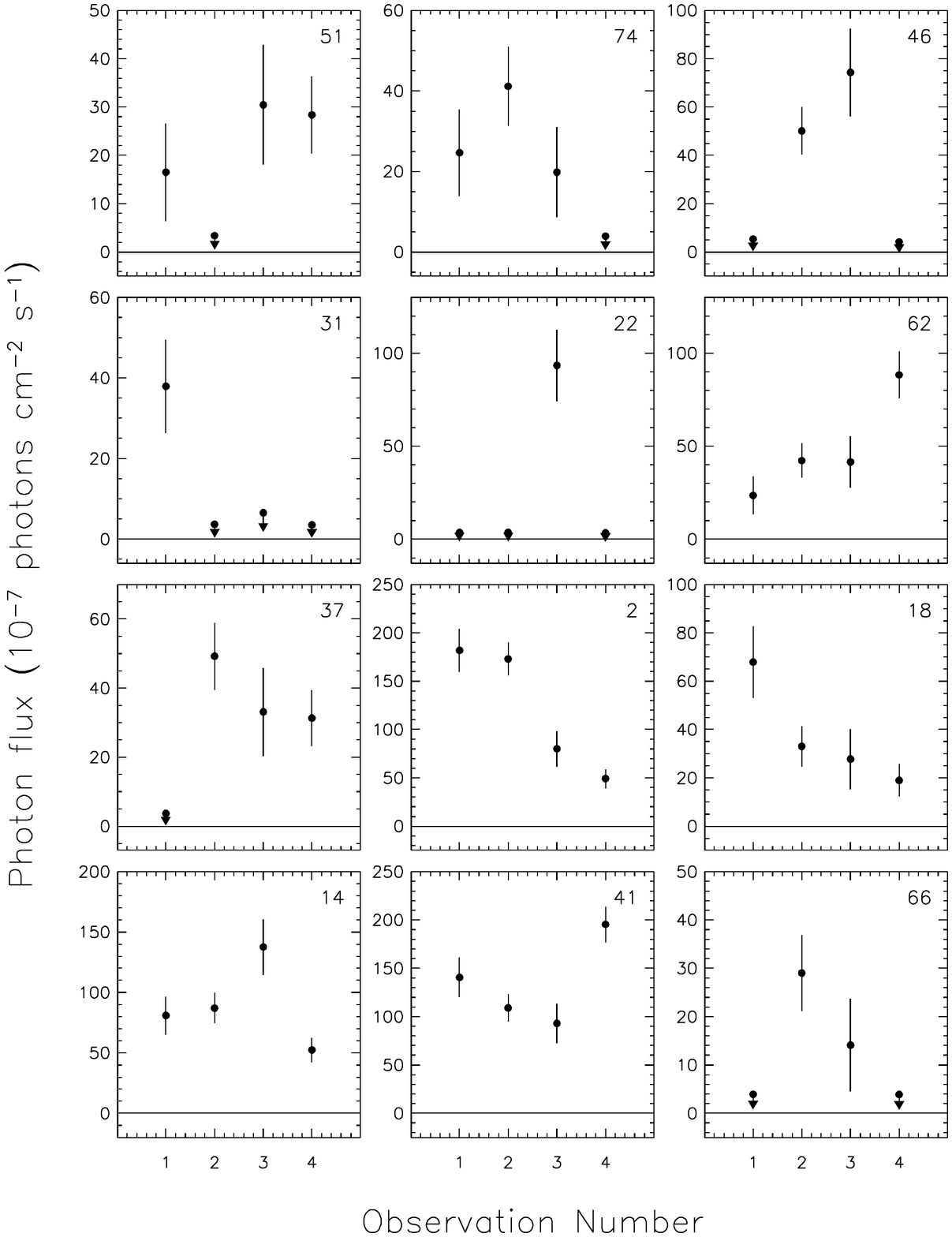} 
\caption{Long-term \textit{Chandra} light curves of the 12 sources with significant variability listed in Table~\ref{tab:ulx_variability}. The observation numbers on the x-axis correspond to the four individual observations: 1: Obs.~ID 11228 (2010-07-18), 2: Obs.~ID 14914 (2012-12-30), 3: Obs.~ID 14799 (2013-04-07), 4: Obs.~ID 14915 (2013-08-31). The flux on the y-axis is in units of $10^{-7}$ photons cm$^{-2}$ s$^{-1}$ and it was measured as described in Sections~\ref{sec:photometry} and \ref{sec:variability}. Sources \#51, 31, 37, 74, 22, 46 and 66 are ULX transient candidates. Source 22 is associated ($0.11\arcsec$ separation) with the supernova SN~2013ai. Source 41 is the central AGN in NGC~2207. Source numbers are based on Table~\ref{tab:merged_properties}.}
\label{fig:var_plot}
\end{center}
\end{figure*}

To investigate the variability of the ULX population of the colliding galaxy pair NGC~2207/IC~2163, we started from the list of sources detected in the combined observation. At the position of each detection, we used the same tools and technique as described in Sect.~\ref{sec:photometry} to measure the net number of counts and the net photon flux (photons cm$^{-2}$ s$^{-1}$) in each individual observation, in the soft (0.5--2 keV), hard (2--8 keV) and full (0.5--8 keV) bands. The soft and hard source counts were used to calculate the X-ray hardness ratios as in Eq.~(\ref{eq:hr}). The full band count rates were used to investigate the source variability, following \citet{2008ApJS..177..465F}, by means of the significance $S_{\rm{flux}}$ for long-term flux variability:
\begin{equation}
\label{eq:sign}
S_{\rm{flux}} = \rm{max}_{i,j}\frac{|F_{i}-F{j}|}{\sqrt{\sigma^{2}_{F_{i}}+\sigma^{2}_{F_{j}}}},
\end{equation}
where: $F_{i}$, $F_{j}$ are the net photon fluxes in the 0.5--8 keV band of the given source in the $i^{\rm{th}}$ and $j^{\rm{th}}$ {\em Chandra} observations, and $\sigma_{F_{i}}$, $\sigma_{F_{j}}$ are the respective uncertainties. A source is {\em variable} if $S_{\rm{flux}} > 3$, and it is a {\em transient} candidate if it is variable ($S_{\rm{flux}} > 3$) and its measured flux is consistent with zero during at least one observation.
For sources with zero counts (therefore null count rate $F_{i,j} = 0$) we used the $1\sigma$ upper limit for zero counts assuming Poisson statistics, based on the tables in \citet{1986ApJ...303..336G}, which is 1.84 counts. We converted to photon flux uncertainty $\sigma_{F_{i,j}}$ by dividing the $1\sigma$ upper limit for zero counts by the value of the exposure map (cm$^{2}$ s counts photon$^{-1}$) at the position of the source.

In total, 12 sources out of 57 ($\sim$20\% including the AGN at the center of NGC~2207) show significant long-term variability ($S_{\rm{flux}} > 3$). Of these, 7 are transient candidates (sources No. 51, 31, 37, 74, 22, 46, 66). We list all the variable sources and transient candidates in Table~\ref{tab:ulx_variability}, along with their count rates in the 0.5-8 keV band, the hardness ratios (computed with Eq.~\ref{eq:hr}) and the significance, $S_{\rm{flux}}$, of the variability. The long term light curves of these sources are shown in Fig.~\ref{fig:var_plot}. 
The ratio between maximum and minimum count rate ranges between 2 and 3.8 for variable sources, and between $\sim$34 and $\sim$72 for transient candidates. One of the variable sources (\#41 in Table~\ref{tab:ulx_variability}) is the central AGN in NGC~2207. Another one (\#22 in Table~\ref{tab:ulx_variability}), is associated ($0.11\arcsec$ separation) with the supernova SN~2013ai. The source was detected only during Obs. ID~14799, where it happens to have a rather hard X-ray spectrum (HR$\sim$$0.7\pm0.2$) and 0.5--8 keV luminosity $L_{\rm{X}}=(6.3\pm1.3)\times 10^{39}\,\rm{erg}\,\rm{s}^{-1}$ (see Table~\ref{tab:14799_properties}). 

Correlated X-ray spectral and luminosity changes have been observed in a number of ULXs \citep[e.g.,][]{2001ApJ...547L.119K, 2014MNRAS.439.3461P}. Fig.~\ref{fig:f_hr} shows no significant long term variability in the hardness ratio of the ULXs in NGC~2207/IC~2163. It also shows that there are no other sources located close to the AGN in the $HR>0.5$ area of the diagram, suggesting there are probably not many background AGN among the sources shown here, in line with our other estimates (see Sect.~\ref{sec:src_content}). 

We looked more carefully for possible spectral changes in the 12 variable sources mentioned above, and found none. The lack of such changes is common when the flux variability is only a factor of a few \citep{2008ApJS..177..465F, 2009ApJ...695.1614S, 2014PASA...31....9W}.

\subsection{HST and Spitzer counterparts}
\label{sec:crossmatch}

Using {\em HST} images of NGC~2207/IC~2163, \citet{2001AJ....121..182E} identified 17 optical ``super-star clusters'' (SSC, mass 1--20$\times 10^4\,M_{\odot}$) within the galaxy pair, with $M_{V} \leq -10.3$. We cross checked the coordinates of the SSCs (D. Elmegreen \& M. Kaufman, private communication) with the coordinates of the 74 X-ray sources detected within the $D25$ ellipse in the co-added {\em Chandra} image (Table~\ref{tab:merged_properties}). We found that only one source is coincident, within a $1.5\arcsec$ tolerance limit, with an SSC. Our source \#48, $L_{\rm{X}} = (4.1\pm1.5)\times 10^{38}\,\rm{erg}\,\rm{s}^{-1}$, is a very close match ($0.13\arcsec$) to the SSC \#16 in \citet{2001AJ....121..182E}, which has the following colors and magnitude: $M_{V} = -11.5$, $U-B=-1.6$, $B-V=-0.7$, $V-I=0.5$. The $V-I$ color is compatible with a young or an intermediate age SSC \citep[][and references therein]{2001AJ....121..182E}. However, the B-V color is much bluer than is typical for either a young or intermediate SSC \citep[][and references therein]{2001AJ....121..182E}, and is therefore indicative of a younger population. Similarly, \citet{2011MNRAS.418L.124V} found two ULXs coincident with young massive stellar clusters in M~82 and NGC~7479 (M82 X-1 and CXOU J230453.0+121959 respectively). They concluded that rarity of observing ULXs inside massive clusters makes it unlikely that most ULXs are formed inside clusters, unless they are kicked out of the clusters at birth.

The 74 X-ray sources in Table~\ref{tab:merged_properties} were also cross-correlated with the positions of the 225 {\em Spitzer} 8-micron clumps identified by \citet{2006ApJ...642..158E}. We found a statistically significant set of $\sim$$1/3$ of our X-ray sources which align with {\em Spitzer} 8-micron clumps, and half of the matching sources are ULXs. Among the matches there are two SNe (SN~1999ec and SN~2013ai), as well as the extended X-ray source at the location of the dusty starburst region called  ``{\em feature i}'' \citep{2000AJ....120..630E}. Since the young star-forming regions represented by the IR clumps are unresolved below $\sim$370 pc (due to the {\em Spitzer} angular resolution at 8 $\mu$m), and this is much larger than individual OB associations, it is possible that these star complexes may also include some older stars \citep{2012AJ....144..156K}.

Overall, we found almost no interesting correlation between the bright X-ray sources and the {\em HST}-detected SSCs, and we conclude that the SSCs in this galaxy pair don't typically host luminous X-ray sources. On the contrary, $\sim$$1/3$ of the X-ray sources detected in our co-added {\em Chandra} observation match up with the $8\,\mu$m-detected young star complexes.

As an added note, we report that our X-ray source \#18 corresponds to source ``X1'' in \citet{2012AJ....144..156K}, centered on a collection of blue star clusters, the most prominent of which lies close to a discrete radio source. Based only on one {\em XMM-Newton} observation, the latter authors interpreted their source X1 as a possible radio SN, a SNR, or a background quasar. We believe that our X-ray source \#18 is a ULX for two reasons.  First, the source shows significant variability (see Fig.~\ref{fig:var_plot}).  It has been detected in all four {\em Chandra} pointings, with a peak luminosity of $L_{\rm{X}} = (4.6\pm1.0)\times 10^{39}\,\rm{erg}\,\rm{s}^{-1}$ in Obs.~ID~11228, and decreasing down to $L_{\rm{X}} = (1.3\pm 0.5)\times 10^{39}\,\rm{erg}\,\rm{s}^{-1}$ by Obs.~ID~14915.  Second, the source is soft during all four observations, with a spectral hardness ranging between $-0.32 \pm 0.28$ and $-0.62 \pm0.21$ and, therefore, is incompatible with an AGN that typically have harder spectra \citep[see, e.g.][as well as Fig.~\ref{fig:f_hr} this work]{2008A&A...482..517S}.

\section{Spatially-resolved maps of SFR density and extinction}
\label{sec:maps}

Using only the data from {\it Chandra} Obs.~ID~11228, along with {\it Spitzer} $24\,\mu$m and {\it GALEX} FUV images, \citet{2013ApJ...771..133M} investigated, for the first time, the spatial and luminosity distributions of ULXs as a function of the {\em local} SFRs within a galaxy. They found that the relation between the total number of ULXs and the integrated SFR of the host galaxy \citep{2010MNRAS.408..234M, 2012MNRAS.419.2095M, 2012AJ....143..144S} is also valid on {\em sub-galactic} scales, i.e., a local $N_{\rm{X}}-\rm{SFR}$ relation. Due to the small number of X-ray sources (21 ULXs) detected in Obs.~ID~11228, \citet{2013ApJ...771..133M}  were not able to study the local $L_{\rm{X}}-\rm{SFR}$ relation in a statistically meaningful manner. Using the same SFR density image constructed by \citet{2013ApJ...771..133M} (see their Sect.~4.1 for details) and following exactly the same technique (described in their Sect.~5), we now revise the spatially-resolved $N_{\rm{X}}-\rm{SFR}$ and $L_{\rm{X}}-\rm{SFR}$ relations for ULXs in NGC~2207/IC~2163, using the combined data from all available {\it Chandra} observations (Table~\ref{tab:obs_log}). In addition, adopting the same technique, we explore the effects of age and dust extinction on the bright XRB population in NGC~2207/IC~2163 on sub-galactic scales. To characterize the dust extinction and/or age effects, we use the ratio of $8-1000\,\mu$m luminosity ($L_{\rm{IR}}$) to observed (i.e., uncorrected for attenuation effects) NUV ($2267\,\AA$) luminosity ($L_{\rm{NUV}}$). We note that the SFR density map was constructed using {\em GALEX} FUV ($1516\,\AA$) and Spitzer MIPS $24\,\mu$m, following the calibration of \citet{2008AJ....136.2782L}. However, the FUV image of NGC~2207/IC~2163 has poorer statistics compared with the NUV image, which makes it less suitable for a pixel-by-pixel analysis; this is the reason why we use the NUV image to construct the $L_{\rm{IR}}/L_{\rm{NUV}}$ map.

For the basic IR data, we used a MIPS $24\,\mu$m Large Field image from the ``post Basic Calibrated Data" products provided by the Spitzer Space Telescope Data archive\footnote{http://irsa.ipac.caltech.edu/applications/Spitzer/Spitzer/}. These are images calibrated in $\rm{MJy}/\rm{sr}$, suitable for photometric measurements. We measured the $24 \,\mu\rm{m}$ background in a region away from the galaxy, and subtracted it from the image. The total net counts were then converted from units of  $\rm{MJy}/\rm{sr}$ into Jy using a conversion factor $C_{24\,\mu\rm{m}}=1.41\times10^{-4}$. The monochromatic fluxes (Jy) at $24\,\mu$m were then converted into spectral luminosities ($\rm{erg}\,\rm{s}^{-1}\,\rm{Hz}^{-1}$). The total IR luminosity ($8-1000\,\mu$m) was estimated using the relations from \citet{2008A&A...479...83B}: $L_{\rm{IR}} (L_{\odot}) =6856\times (\nu L_{\nu}/L_{\odot})_{24\,\mu\rm{m, rest}}^{0.71}$.

We based the ultraviolet analysis on {\em GALEX} NUV and FUV background-subtracted intensity map data that are publicly available in the archive of GR6/GR7 Data Release\footnote{http://galex.stsci.edu/GR6/}. These are images calibrated in units of counts per pixel per second, corrected for the relative response, with the sky background subtracted. We converted the net counts to flux using the conversion factors\footnote{http://galexgi.gsfc.nasa.gov/docs/galex/FAQ/counts\_background.html} between {\em GALEX} count rate ($\rm{cts}/\rm{s}$) and flux ($\rm{erg}\,\rm{cm}^{-2}\,\rm{s}^{-1}\,\AA^{-1}$): $C_{\rm{NUV}}=2.06\times 10^{-16}$ and $C_{\rm{FUV}}=1.40\times 10^{-15}$. The fluxes were then converted into a broad band NUV luminosity by taking the product $\lambda F_\lambda$ and multiplying by $4 \pi D^2$.

Using the routine HASTROM, from the NASA IDL Astronomy User's Library\footnote{http://idlastro.gsfc.nasa.gov/}, we oversampled the $24\,\mu\rm{m}$ image ($2.45 \arcsec/\rm{pix}$) so as to match the better angular resolution ($1.5 \arcsec/\rm{pix}$) of the {\it GALEX} images. This routine interpolates without adding spatial frequency information beyond the intrinsic resolution of the original image. The resulting maps, displayed in Figs.~\ref{fig:sfr_map} and \ref{fig:dust_map}, have the same spatial resolution and pixel coordinates as the {\it GALEX} images, but are limited by the {\em Spitzer} resolution. Figs.~\ref{fig:multipanel}, \ref{fig:hst} and \ref{fig:dust_map} show that IC~2163 is dustier than NGC~2207.

\section{ULX and star formation activity}
\label{sec:sfr}

\begin{figure*}
\begin{center}
\includegraphics[width=0.8\linewidth]{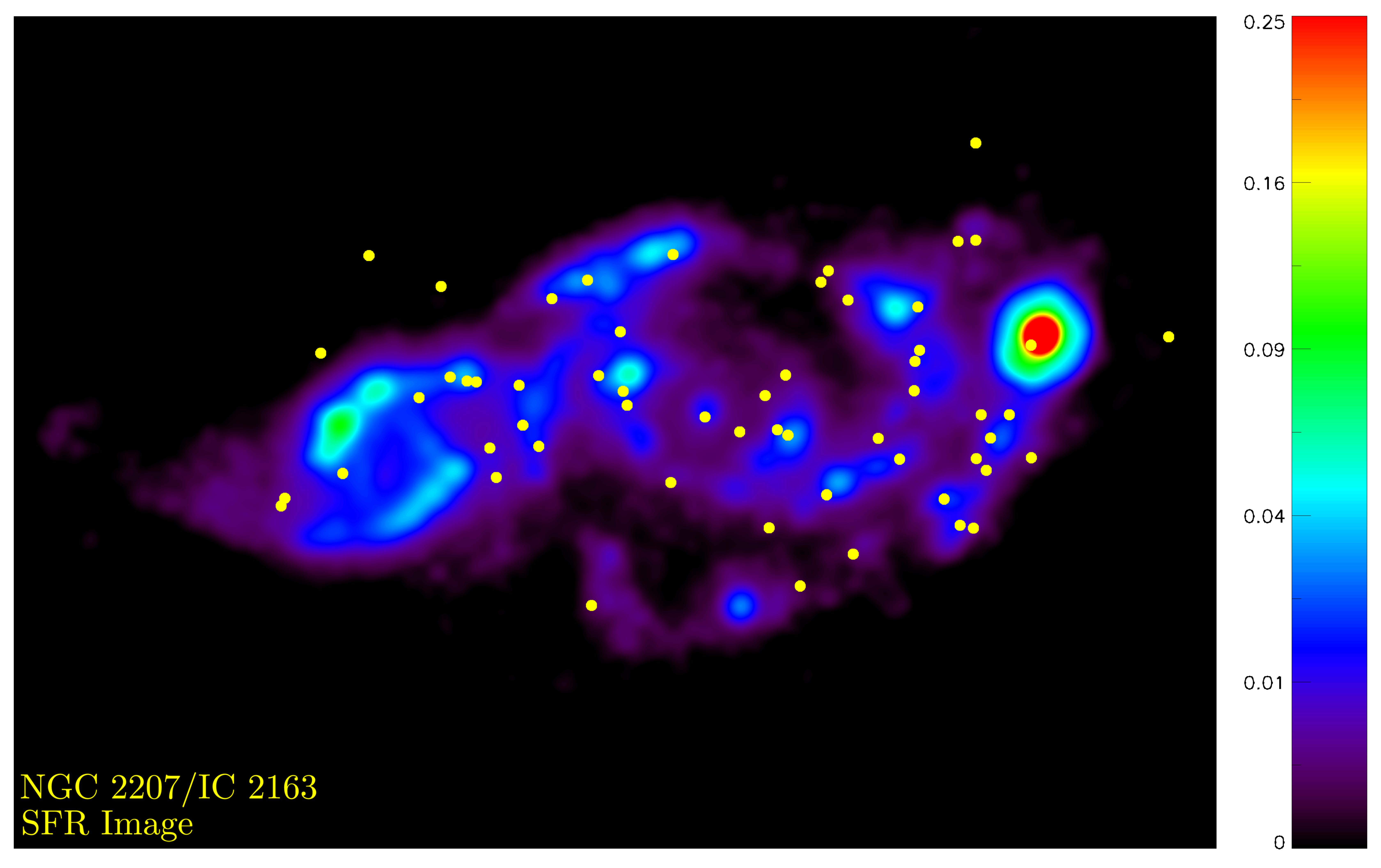} 
\caption{Star formation rate density map for the colliding galaxies NGC~2207/IC~2163.  This image was generated from a linear combination of {\it GALEX} FUV and Spitzer $24\,\mu\rm{m}$ images according the prescription of \citet{2008AJ....136.2782L}.  The units are $M_{\odot}\,\rm{yr}^{-1}\,\rm{kpc}^{-2}$. The filled yellow circles mark the locations of the 57 X-ray sources detected in the four combined {\it Chandra} observations, above the completeness limit  ($3.4\times10^{38}\,\rm{erg}\,\rm{s}^{-1}$; see Table~\ref{tab:compl}). For details see Sect.~4.1 in \citet{2013ApJ...771..133M}.}
\label{fig:sfr_map}
\end{center}
\end{figure*}

With the SFR density image \citep[Section~\ref{sec:maps} and Fig.~\ref{fig:sfr_map} this paper, and Section~4.1 of][]{2013ApJ...771..133M}, we now improve the spatially-resolved $N_{\rm{X}}-\rm{SFR}$ and $L_{\rm{X}}-\rm{SFR}$ relations for ULXs in NGC~2207/IC~2163 \citep{2013ApJ...771..133M}, using the combined data from all available {\it Chandra} observations (Table~\ref{tab:obs_log}).
We defined a grid of uniformly spaced logarithmic bins of SFR density based on the pixel values in the SFR image (see Fig.~\ref{fig:sfr_map}). Based on the SFR density value at the position of a given X-ray source, we counted the number of sources brighter than $10^{39}\,\rm{erg}\,\rm{s}^{-1}$ and their collective luminosity for each bin of SFR density. We counted the number of pixels in each bin of SFR density (over the $D25$ region) and thereby computed the corresponding integrated area. We use this area to normalize the background AGN $\log{N}-\log{S}$ function, which we take from \citet{2008MNRAS.388.1205G}, and thereby calculate the predicted number of background AGNs, $N_{\rm{AGN}}(L> 10^{39}\,\rm{erg}\,\rm{s}^{-1})$, and their equivalent luminosity, $L_{\rm{AGN}}(>10^{39}\,\rm{erg}\,\rm{s}^{-1})$, above the same luminosity threshold. The respective values for the bright X-ray sources and the AGN were then subtracted to yield $N_{\rm{X}}(L> 10^{39}\,\rm{erg}\,\rm{s}^{-1})$ and $L_{\rm{X}}(> 10^{39}\,\rm{erg}\,\rm{s}^{-1})$, respectively. Dividing these values by the area underlying each bin of SFR density, we obtained the surface density for X-ray point sources (source number kpc$^{-2}$) and luminosity density (erg s$^{-1}$ kpc$^{-2}$) as a function of SFR density. 

\begin{figure*}
\begin{center}
\hbox
{
\includegraphics[trim=5mm 50mm 2mm 10mm, width=1.0\columnwidth]{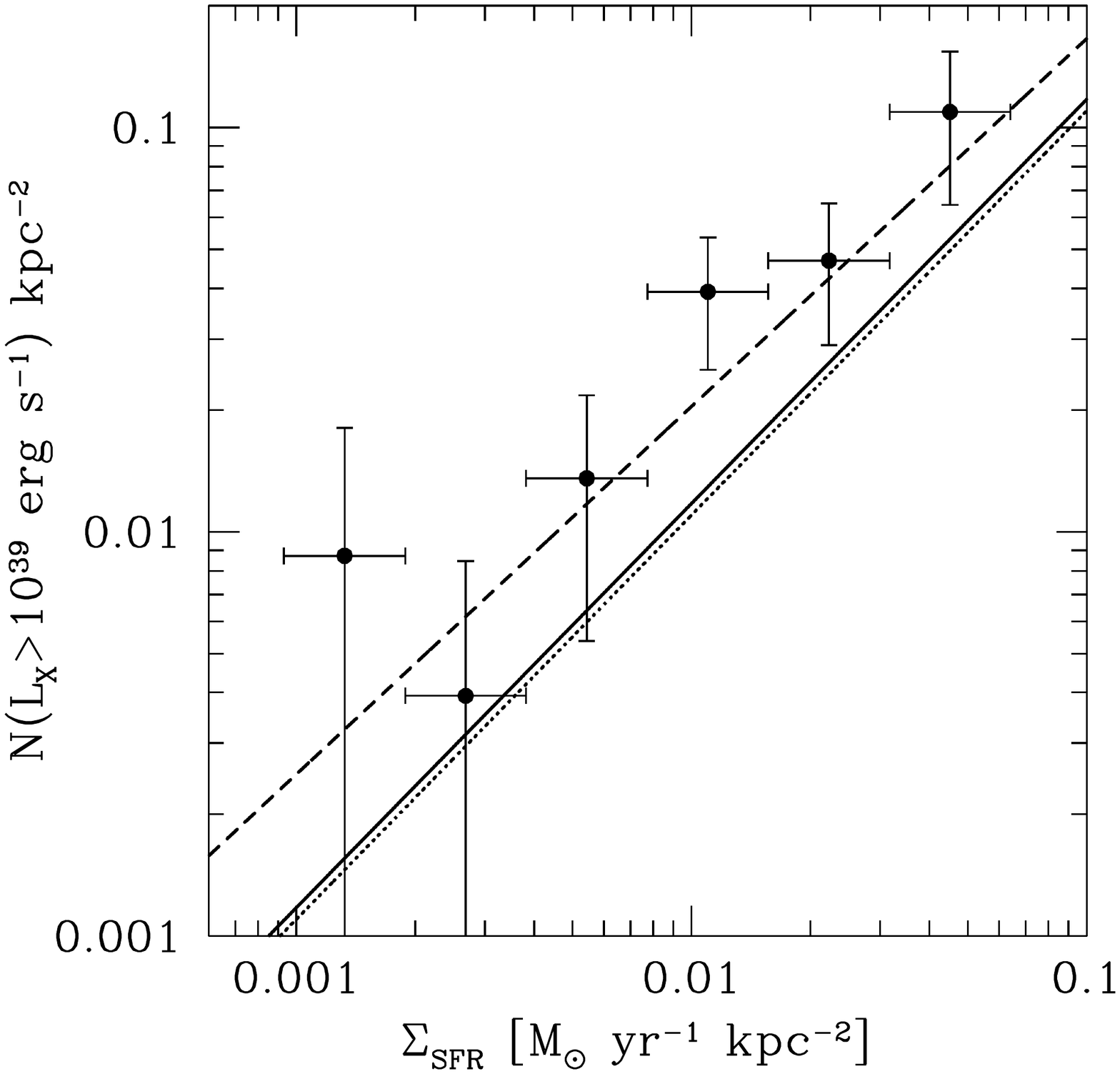}
\includegraphics[trim=5mm 50mm 2mm 10mm, width=1.0\columnwidth]{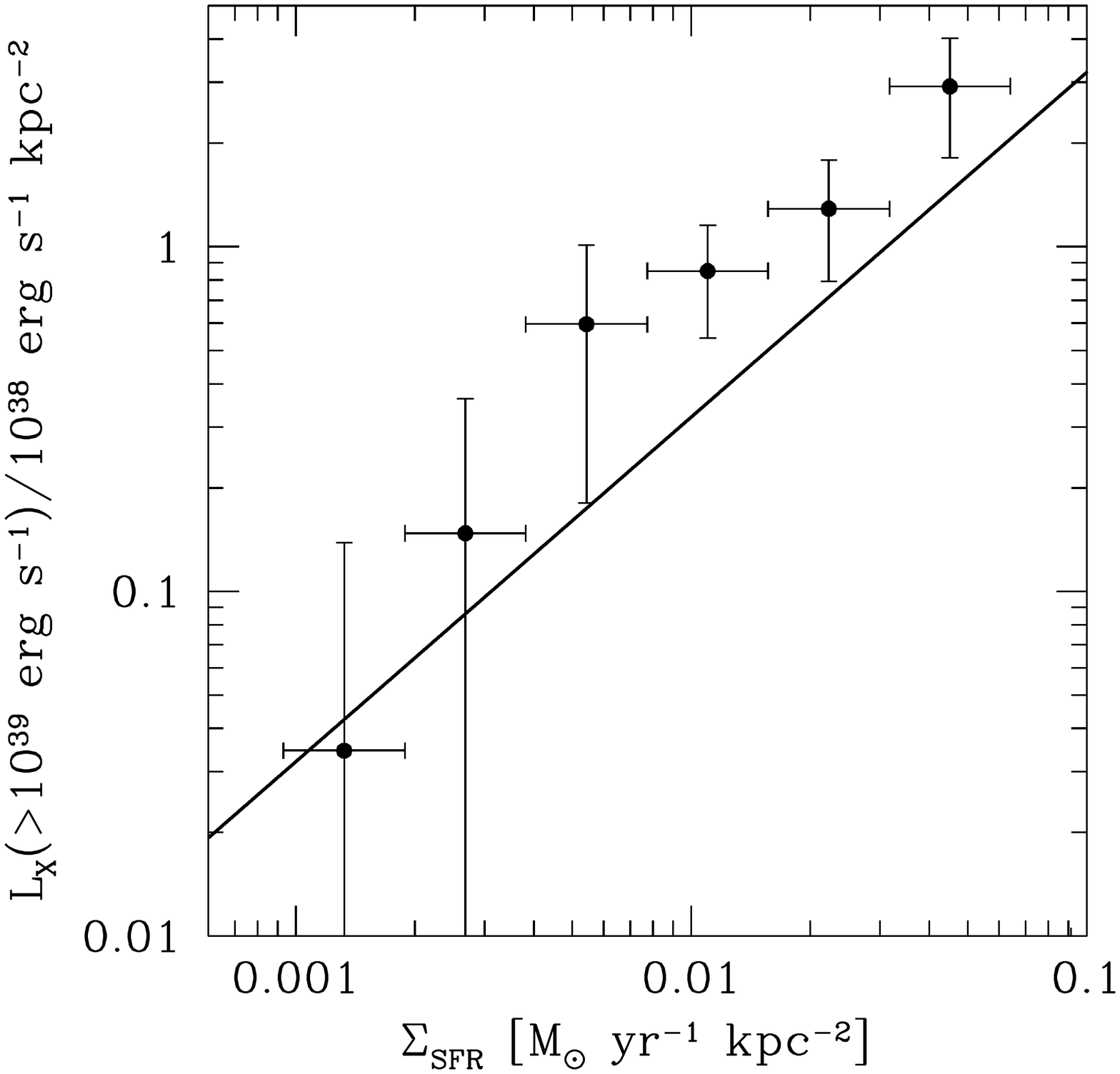}
}
\caption{Relation between the {\em local} SFR density in NGC~2207/IC~2163 ($M_{\odot}\,\rm{yr}^{-1}\,\rm{kpc}^{-2}$) and the number density ($N_{\rm{X}}/\rm{kpc}^{2}$, left panel) and luminosity density ($L_{\rm{X}}/\rm{kpc}^{2}$, right panel) of ULXs. Both of these quantities have been corrected for the contribution of background AGNs. The SFR density was computed from the \citet{2008AJ....136.2782L} algorithm as described in Sect.~\ref{sec:maps}. The solid curves are the multiple-galaxy-wide average $N_{\rm{X}}-\rm{SFR}$ (left panel) and $L_{\rm{X}}-\rm{SFR}$ (right panel) relations for ULXs obtained in \citet[][their eqs.(20) and (22) respectively]{2012MNRAS.419.2095M}. These curves are shown for comparison after having been rescaled to match the IMF assumption in the SFR recipe from \citet{2008AJ....136.2782L}. The corresponding ULX vs.~SFR relation from \citet[][their Eqn.~6]{2010MNRAS.408..234M} is shown as a dashed line in the left panel. The dotted line represents the $N_{\rm{ULX}}-\rm{SFR}$ relation from \citet{2012AJ....143..144S}.}
\label{fig:nx_lx_sfr}
\end{center}
\end{figure*}

These quantities are plotted against the SFR surface density in Fig.~\ref{fig:nx_lx_sfr}. Table~\ref{tab:fig8} describes the data used to prepare Fig.~\ref{fig:nx_lx_sfr}. Due to ULX variability (Sect.~\ref{sec:variability}) we now detect 28 ULXs, 7 of which were not visible in Obs.~ID~11228 alone. This $\sim$30\% increase in ULX number, and a better average value for $L_X$, allow for an improvement in both the $N_{\rm{X}}-\rm{SFR}$ and $L_{\rm{X}}-\rm{SFR}$ relations, which is more appreciable in the latter case. Pixels with SFR density less than $6\times 10^{-4}\,M_{\odot}\,\rm{yr}^{-1}\,\rm{kpc}^{-2} $ were not used as they are dominated by background noise \citep[see Section~5 in][]{2013ApJ...771..133M}. Two ULXs were located in these regions of low SFR density, therefore they are not included in our plot. In Fig.~\ref{fig:nx_lx_sfr} we also plot the average $N_{\rm{X}}-\rm{SFR}$ and $L_{\rm{X}}-\rm{SFR}$ relations for ULXs obtained for large samples of nearby star-forming galaxies by \citet[][their eqs.(20) and (22); solid lines in Fig.~\ref{fig:nx_lx_sfr}]{2012MNRAS.419.2095M} after having converted the SFR estimate for the different IMF used in the present work \citep[IMF as in \citet{2007ApJ...666..870C}, i.e., slope $-1.3$ for the $0.1-0.5\,M_{\odot}$ mass range and $-2.3$ for $0.5-120\,M_{\odot}$, see Sect.~4.2 and Table~2 in][]{2013ApJ...771..133M}. Similarly, we compare our results with the $N_{\rm{ULX}}-\rm{SFR}$ relation from \citet{2010MNRAS.408..234M}. The dashed line in the top panel of Fig.~\ref{fig:nx_lx_sfr} shows their Eq.~(6), which is slightly non-linear. We also plot the $N_{\rm{ULX}}-\rm{SFR}$ relation from \citet{2012AJ....143..144S} as a dotted line, after having applied all the necessary conversions to make it compatible with the units used in Fig.~\ref{fig:nx_lx_sfr}. The latter relation is almost identical to the scaling from \citet{2012MNRAS.419.2095M}.

Fig.~\ref{fig:nx_lx_sfr} confirms that the multiple-galaxy-averaged relation between the total number of X-ray point sources and the integrated SFR of the host galaxy also holds on {\em sub-galactic} scales. 

\begin{deluxetable}{cccccc}
\tablewidth{0pt}
\tabletypesize{\scriptsize}
\tablecaption{\label{tab:fig8} Table of the data used to prepare Figure~8}
\tablehead{
	\colhead{$\Sigma_{\rm{SFR}}$} &
	\colhead{$N_{\rm{X}}$} &
	\colhead{$L_{\rm{X}}$} &
	\colhead{$N_{\rm{AGN}}$}  & 
	\colhead{$L_{\rm{AGN}}$}  & 
	\colhead{Area} \\
	\colhead{($M_{\odot}/\rm{yr}/\rm{kpc}^{2}$)} & 
	\colhead{} &
	\colhead{($10^{38}\,\rm{erg}/\rm{s}$)} &
	\colhead{} & 
	\colhead{($10^{38}\,\rm{erg}/\rm{s}$)} &
      	\colhead{(kpc$^{2}$)} \\
	(1)  & (2) & (3) & (4) & (5) & (6) 
}
\startdata 
0.0013 & 1 & 10 & 0.13 & 6.7 & 100\\
0.0027 & 1 & 41 & 0.25 & 13 & 190\\
0.0054 & 3 & 130 & 0.26 & 13 & 200\\
0.011 & 8 & 180 & 0.26 & 13 & 195\\
0.022 & 7 & 200 & 0.19 & 9.7 & 145\\
0.045 & 6 & 160 & 0.07 & 3.6 & 55\\
\enddata
\tablecomments{(1) Value of $\Sigma_{\rm{SFR}}$ at the bin
  center, (2) Number of ULXs per bin of $\Sigma_{\rm{SFR}}$,
  (3) Luminosity of ULXs per bin of $\Sigma_{\rm{SFR}}$, (4)
   Expected number of background AGNs per bin of
   $\Sigma_{\rm{SFR}}$, (5) Expected luminosity of background AGNs per
   bin of $\Sigma_{\rm{SFR}}$, (6) Area underlying each bin of
   $\Sigma_{\rm{SFR}}$ in kpc$^{2}$.}
\end{deluxetable}

\section{Dust extinction and age effects on bright XRB}
\label{sec:dust_maps_results}

\citet{2012MNRAS.419.2095M} investigated the correlation of $L_{\rm{IR}}/L_{\rm{NUV}}$ with $L_{\rm{X}}/\rm{SFR}$ for HMXBs and found virtually no correlation (Spearman's rank correlation coefficient is $r_{S} = -0.35$, corresponding to a probability of $P = 7\%$ for the null hypothesis. On the other hand, although within large error bars, the multiple-galaxy average $L_{\rm{X}}$ to SFR ratio suggests a decreasing trend with increasing values of $L_{\rm{IR}}/L_{\rm{NUV}}$ (see their Fig.~11d). The ratio $L_{\rm{IR}}/L_{\rm{NUV}}$ is affected by both dust extinction and/or age. We investigated the possible effects of dust extinction and age on the bright XRB population in NGC~2207/IC~2163 on sub-galactic scales.

\begin{figure*}
\begin{center}
\includegraphics[width=0.8\linewidth]{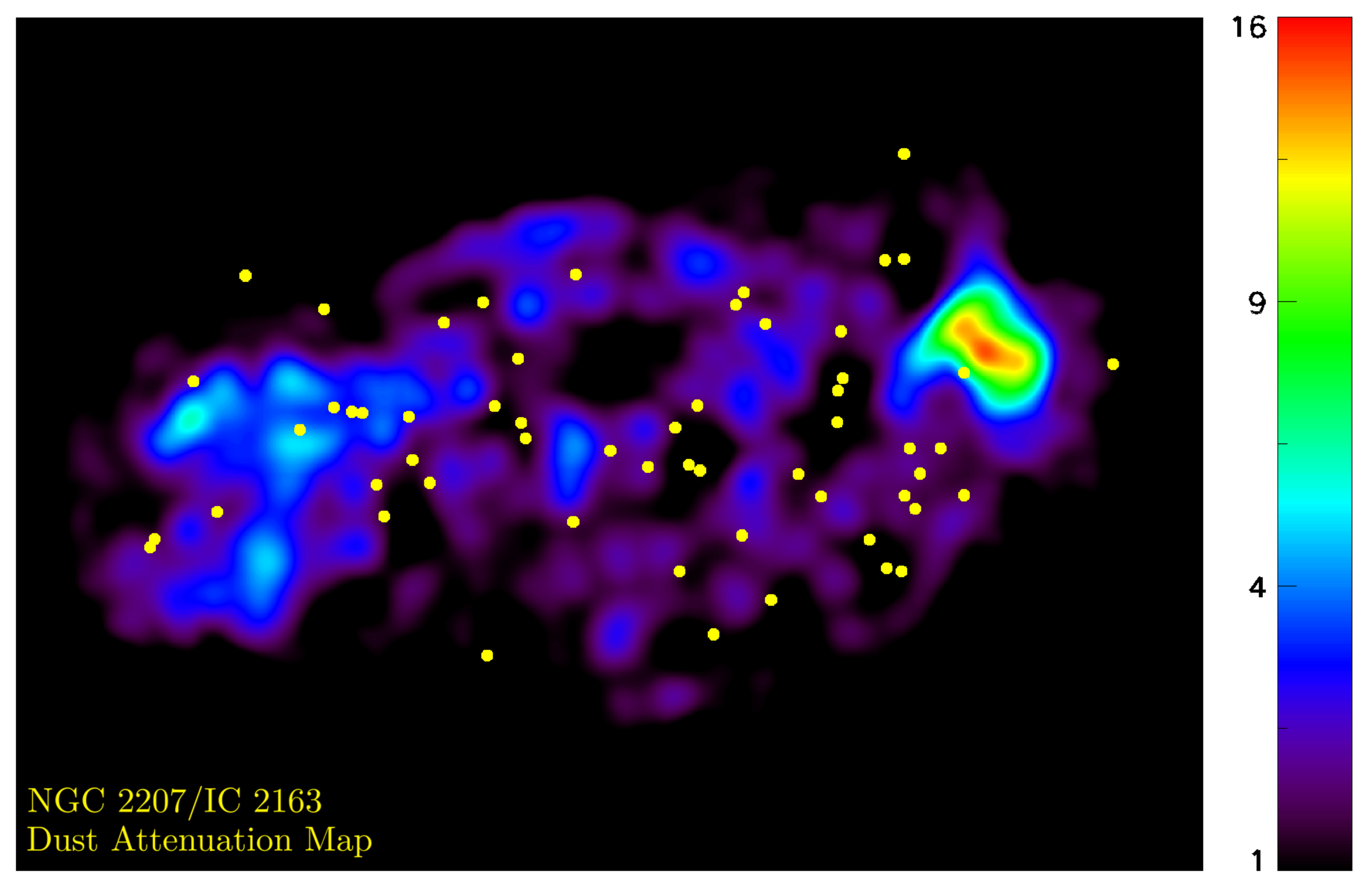}
\caption{Map of $L_{\rm{IR}}/L_{\rm{NUV}}$ for NGC~2207/IC~2163 obtained by combining {\it GALEX} FUV and Spitzer $24\,\mu\rm{m}$ images according the prescription described in Sect.~\ref{sec:maps}. The small yellow circles mark the locations of the 57 X-ray sources detected in the four combined {\it Chandra} observations, above the completeness limit of $3.4\times10^{38}\,\rm{erg}\,\rm{s}^{-1}$ (Table~\ref{tab:compl}).}
\label{fig:dust_map}
\end{center}
\end{figure*}

\begin{figure}
\begin{center}
\includegraphics[trim=5mm 60mm 2mm 10mm, width=1.0\linewidth]{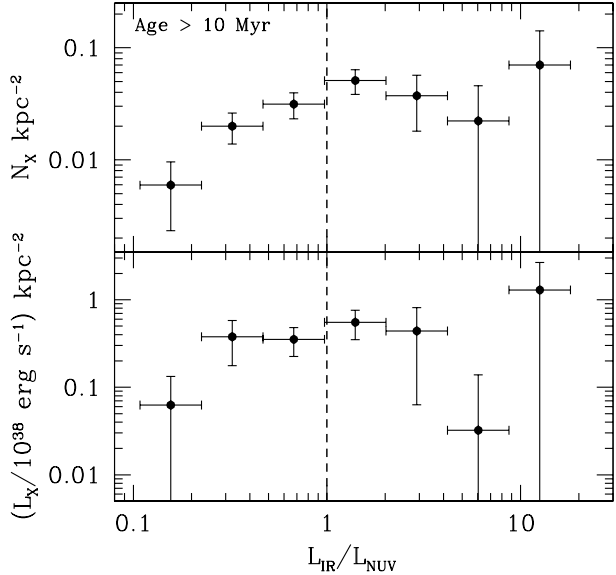}
\caption{Relation between the {\em local} $L_{\rm{IR}}/L_{\rm{NUV}}$ in NGC~2207/IC~2163, as indicated by $L_{\rm{IR}}/L_{\rm{NUV}}$, and the number density $N_{\rm{X}}/\rm{kpc}^{2}$ (upper panel) and luminosity density $L_{\rm{X}}/\rm{kpc}^{2}$ (lower panel) of X-ray sources detected above our completeness luminosity threshold in the combined image ($3.4\times 10^{38}\,\rm{erg}\,\rm{s}^{-1}$). Both of these quantities have been corrected for the contribution of background AGNs, see Sect.~\ref{sec:dust_maps_results} for details.}
\label{fig:nx_dust}
\end{center}
\end{figure}

With the spatially-resolved image of $L_{\rm{IR}}/L_{\rm{NUV}}$ (Section~\ref{sec:maps}, Fig.~\ref{fig:dust_map}), we studied the number of X-ray point sources above the completeness luminosity and their luminosities as a function of the {\em local} $L_{\rm{IR}}/L_{\rm{NUV}}$. We applied the same pixel-by-pixel analysis to the $L_{\rm{IR}}/L_{\rm{NUV}}$ image that we utilized to study the local correlation between numbers and luminosities of X-ray sources and the local SFR density, which is described in Sections~\ref{sec:maps} and \ref{sec:sfr}. We thereby obtained the densities $N_{\rm{X}}/\rm{kpc}^{2}$ and $L_{\rm{X}}/\rm{kpc}^{2}$ for the luminous X-ray sources and plotted these against the value of $L_{\rm{IR}}/L_{\rm{NUV}}$ in Fig.~\ref{fig:nx_dust} (upper and lower panel respectively). Table~\ref{tab:fig10} describes the data used to prepare Fig.~\ref{fig:nx_dust}. The figure shows a modest increase of $N_{\rm{X}}/\rm{kpc}^{2}$ and $L_{\rm{X}}/\rm{kpc}^{2}$ with $L_{\rm{IR}}/L_{\rm{NUV}}$ at small $L_{\rm{IR}}/L_{\rm{NUV}}$ values, up to $L_{\rm{IR}}/L_{\rm{NUV}}\sim1$. This is followed by a decrease of the $N_{\rm{X}}/\rm{kpc}^{2}$ and $L_{\rm{X}}/\rm{kpc}^{2}$ values when $L_{\rm{IR}}/L_{\rm{NUV}}$ increases. Fig.~\ref{fig:nx_dust} shows these quantities plotted against 7 discrete bins in $L_{\rm{IR}}/L_{\rm{NUV}}$. In order to take into account possible binning effects on the statistical significance of the observed number and luminosity density trends with $L_{\rm{IR}}/L_{\rm{NUV}}$, we also analyzed plots with 5 bins in $L_{\rm{IR}}/L_{\rm{NUV}}$. The observed trend for $N_{\rm{X}}/\rm{kpc}^{2}$ has a higher statistical significance ($\sim$3.3$\sigma$ and $\sim$2.8$\sigma$ for the central bins in the case of 5 bins of $L_{\rm{IR}}/L_{\rm{NUV}}$ vs 7 bins, respectively) than that for $L_{\rm{X}}/\rm{kpc}^{2}$ ($\sim$1.3$\sigma$ and $\sim$2$\sigma$ for the central bins in the case of 5-bins and 7-bins, respectively). The significance was estimated by comparing the data points and their uncertainties, with the best-fit average values: $N_{\rm{X}}/\rm{kpc}^{2}=(1.5\pm 0.3)\times 10^{-2}$, $L_{\rm{X}}/(10^{38}\,\rm{erg}\,\rm{s}^{-1})/\rm{kpc}^{2}=(1.5\pm 0.3)\times 10^{-1}$ (7-bin plot). We view all this as tentative evidence for a smoothly peaked dependence of $N_{\rm{X}}/\rm{kpc}^{2}$ and $L_{\rm{X}}/\rm{kpc}^{2}$ on $L_{\rm{IR}}/L_{\rm{NUV}}$.

The $L_{\rm{IR}}/L_{\rm{NUV}}$ ratio may also be seen as a rough indicator of the timescale of the star formation event. The UV emission originating from the photospheres of O and B stars is absorbed and re-emitted in the IR band by dust grains heated by the embedded young ionizing stars. The IR should trace somewhat younger star formation \citep[$<10$ Myr,][]{2007ApJ...666..870C}, while UV traces older star formation \citep[$30-100$ Myr,][]{2005ApJ...633..871C}, see also \citet{2012ARA&A..50..531K}. This might suggest, as a rough timescale, that at $L_{\rm{IR}}/L_{\rm{NUV}} < 1$, the age may be $\gtrsim 10$ Myr, and therefore, the trend in Fig.~\ref{fig:nx_dust} might be tentatively ascribed to an age effect on the bright XRB formation and evolution. The UV dominates where the dust absorption is very low. In Milky Way type galaxies, this means that the molecular clouds have dissipated, or at least the stars have moved out of the clouds. 
\begin{inparaenum}[i\upshape)]
To summarize: \item high $L_{\rm{IR}}$ could mean such high dust obscuration that we fail to detect some X-ray sources. \item High $L_{\rm{NUV}}$ could mean an older stellar population which has become free of dust. \item  The $L_{\rm{IR}}/L_{\rm{NUV}}$ ratio could indicate age which might, importantly, affect how many ULXs have already evolved away. Additionally, there are further complications due to the chemical composition, which we do not address in the present work. 
\end{inparaenum}

Interpreting the trend observed in Fig.~\ref{fig:nx_dust} as an effect of age or star formation timescale, we see that the number of bright XRBs and their luminosity peaks at $\approx$10 Myr, i.e., around the epoch where young stars are escaping from the dust clouds that enshroud them. We note that these qualitative conclusions are in agreement with the results found by \citet{2009ApJ...703..159S} for a large sample of ULXs detected in 58 nearby galaxies. They found that the most luminous ULXs (or equivalently, the most common phases of very high mass transfer) are biased toward early B-type donors with an initial mass of $\approx 10-15\,M_{\odot}$ and an age $\sim 10-20$ Myr, perhaps at the stage where the B star expands to become a blue supergiant. Similarly, an age of $\approx 10-20$ Myr was inferred for the stars around NGC~4559~X-1 by \citet{2005MNRAS.356...12S} and $\approx$ 20 Myr for those around NGC~1313~X-2 by \citet{2008A&A...486..151G}.

However, this interpretation remains mostly tentative. Not only do the data points in Fig.~\ref{fig:nx_dust} have rather large error bars, but there is also some uncertainty introduced as a result of the pixel interpolation applied to obtain the $L_{\rm{IR}}/L_{\rm{NUV}}$ map (see Sect.~\ref{sec:maps}). Moreover, the $24\,\mu$m emission may depend on dust geometry and the conversion from $24\,\mu$m to total IR luminosity has a very large uncertainty. The $L_{\rm{IR}}/L_{\rm{NUV}}$ is not a perfect age indicator, but it should be at least statistically meaningful. 

The $L_{\rm{IR}}/L_{\rm{NUV}}$ ratio of galaxies also depends upon metallicity, with lower metallicity systems having less dust and therefore lower $L_{\rm{IR}}/L_{\rm{NUV}}$ \citep[see, e.g.,][]{2000A&ARv..10....1K, 2007ApJS..173..392J, 2013ApJ...774..152B}; therefore, the range $L_{\rm{IR}}/L_{\rm{NUV}} < 1$ may also show the low metallicity regions of NGC~2207/IC~2163. Based on the emission line analysis of \ion{H}{2} regions in interacting galaxies, \citet{2010ApJ...723.1255R} found that the metallicity in NGC~2207/IC~2163 is in the range $12+\log([O/H])\approx 8.8-9.2$ and they show that IC~2163 hosts regions with, on average, higher metallicity ($12+\log([O/H]) \gtrsim 9$) than those in NGC~2207 ($12+\log([O/H]) \lesssim 9$). On the other hand, a full range of $\approx0.4$ dex may not be enough in terms of the implied dust to gas ratio, suggesting that in the case of NGC~2207/IC~2163 the effects of metallicity may not be important. We may conclude that either bright XRBs have not yet formed in the dustier galaxy (IC~2163), or they are hidden by high extinction \citep[in agreement with the results of][]{2012AJ....143..144S, 2014arXiv1410.1569L}.

\begin{deluxetable}{cccccc}
\tablewidth{0pt}
\tabletypesize{\scriptsize}
\tablecaption{\label{tab:fig10} Table of the data used to prepare Figure~10}
\tablehead{
	\colhead{$L_{\rm{IR}}/L_{\rm{NUV}}$} &
	\colhead{$N_{\rm{X}}$} &
	\colhead{$L_{\rm{X}}$} &
	\colhead{$N_{\rm{AGN}}$}  & 
	\colhead{$L_{\rm{AGN}}$}  & 
	\colhead{Area} \\
	\colhead{} & 
	\colhead{} &
	\colhead{($10^{38}\,\rm{erg}/\rm{s}$)} &
	\colhead{} & 
	\colhead{($10^{38}\,\rm{erg}/\rm{s}$)} &
      	\colhead{(kpc$^{2}$)} \\
	(1)  & (2) & (3) & (4) & (5) & (6) 
}
\startdata 
0.16 & 4 & 59 & 1.3 & 30 & 455\\
0.32 & 12 & 230 & 1.5 & 35 & 525\\
0.67 & 16 & 200 & 1.3 & 31 & 465\\
1.4 & 17 & 200 & 0.9 & 21 & 315\\
2.9 & 4 & 50 & 0.28 & 6.7 & 100\\
6.1 & 1 & 4 & 0.11 & 2.7 & 40\\
13 & 1 & 19 & 0.039 & 0.91 & 15\\
\enddata
\tablecomments{(1) Value of $L_{\rm{IR}}/L_{\rm{NUV}}$ at the bin
  center, (2) Number of X-ray binaries above the
completeness luminosity per bin of
  $L_{\rm{IR}}/L_{\rm{NUV}}$, (3) Luminosity of X-ray binaries per bin of
  $L_{\rm{IR}}/L_{\rm{NUV}}$, 
   (4) Expected number of background AGNs per bin of
   $L_{\rm{IR}}/L_{\rm{NUV}}$, (5) Expected luminosity of background AGNs per bin of
   $L_{\rm{IR}}/L_{\rm{NUV}}$,  (6) Area underlying each bin of
   $L_{\rm{IR}}/L_{\rm{NUV}}$ in kpc$^{2}$.}
\end{deluxetable}

\section{Diffuse X-ray emission}
\label{sec:diffuse}

Star-forming galaxies are known to emit significant amounts of X-rays at $\sim$sub-keV temperatures (typically in the range of 0.2--0.3 keV, sometimes with evidence of a second thermal component at $\sim$0.7 keV), due to hot ionized gas. The luminosity of the diffuse thermal X-ray emission correlates with the SFR of the host galaxy. The gas is thought to be in a state of outflow, driven by the collective effects of supernovae and winds from massive stars \citep{1985Natur.317...44C, 2000AJ....120.2965S, 2004ApJS..151..193S, 2005ApJ...628..187G, 2006A&A...448...43T, 2009MNRAS.394.1741O, 2012ApJ...758..105Y, 2012MNRAS.426.1870M, 2013MNRAS.428.2085L}.  

\subsection{Isolating the hot ISM in NGC~2207/IC~2163}
\label{sec:isolating_ism}

Recently, \cite{2012MNRAS.426.1870M} isolated the contribution of hot, diffuse ISM in a sample of 21 local, star-forming galaxies. They took special care of various systematic effects and controlled the contamination by ``spill-over" counts from bright resolved compact sources to the diffuse emission. Here we use the same procedures to isolate the truly diffuse emission due to the hot ISM from contaminating components and to obtain its luminosity.

We first  searched for the optimal size of the regions to be used to remove the point source counts from the image and minimize the contamination by ``spill-over" counts. We used the same procedure described in Sect.\ref{sec:Xray_analysis} to search for point-like sources in the soft (0.5--2 keV), hard (2--8 keV), and total (0.5--8 keV) energy bands. For each source, we used the information about the shape of the point spread function (PSF) at the source position to determine the radius of the circular region containing 90\% of the source counts, i.e., $R_{90\%\,\rm{PSF}}$. From the source lists obtained in each energy band, we created a set of source regions having radii ranging from $0.5 \, R_{90\%\,\rm{PSF}}$ to $3.5\,R_{90\%\,\rm{PSF}}$ with a step of 0.1. A corresponding set of diffuse emission images was created for each observation of NGC~2207/IC~2163, adopting the following method. We removed the source regions from the image and, using the CIAO task {\tt dmfilth} (POISSON method), we filled in the holes left by the source removal with pixel values interpolated from surrounding background regions. The background region for this interpolation purpose was defined as a circle with radius 3 times the radius of the source region. We ensured that the chosen background annuli did not contain neighboring point sources. For each background region listed in the input file, we subtracted all the overlapping neighboring point source regions and merged them into a single source removal region. For each of the resulting images, we estimated the count rate within the $D25$ ellipse using the CIAO task {\tt dmextract} and plotted it against the radius of the removed-source region. This plot showed a sharp decrease of the count rate at small source radii $\leq R_{90\%\,\rm{PSF}}$, followed by a flattening of the curve at $R>$1.5--2$\,R_{90\%\,\rm{PSF}}$. We found that on average the difference between excluding source regions with $R=1.5\,R_{90\%\,\rm{PSF}}$ and $R=2\,R_{90\%\,\rm{PSF}}$ is only $\sim$4\% of the background-subtracted soft band count rate. Based on this analysis, we adopted a source region radius of  $R=1.5 \,R_{90\%\,\rm{PSF}}$ to minimize the contamination of diffuse emission by point-source counts without compromising the statistics for the diffuse emission itself. The diffuse emission spectrum was extracted, in each individual observation, from the $D25$ region, after having removed the circular regions with $R=1.5 \, R_{90\%\,\rm{PSF}}$ for all detected point sources.

The background spectrum was extracted, in each observation, from a region defined by the whole {\it Chandra} chip outside $1.3\times D25$, to avoid contamination from true diffuse emission in the outskirts of the $D25$ region and to retain good count statistics for the background spectrum itself. This takes into account both the instrumental and cosmic X-ray background. 

The co-added spectrum of diffuse X-ray emission from all four observations in the 0.5--8 keV band is shown in Fig.~\ref{fig:spectra}, along with the composite point source spectrum that was obtained in  Sect.~\ref{sec:lums}.
 
\subsection{Spectral analysis}
\label{sec:spectra}

\begin{deluxetable}{cccc}
\tablewidth{0pt}
\tablecaption{\label{tab:flux_ratio} Diffuse and point source count rate ratios}
\tablehead{
	\colhead{Energy band} &
	\colhead{$f_{\rm{diff}}$} &
	\colhead{$f_{\rm{XRBs}}$} &
	\colhead{$f_{\rm{diff}}/f_{\rm{XRBs}}$} \\
	\colhead{} & 
	\colhead{($10^{-3}$ cts/s)} & 
	\colhead{($10^{-3}$ cts/s)} & 
	\colhead{} \\
       	(1)  & (2) & (3) & (4) 
}
\startdata 
0.5--8 keV  & $38 \pm 1.3$ & $31 \pm 0.71$ & $1.2 \pm 0.051$ \\
0.5--2 keV  & $32 \pm 0.94$ & $20 \pm 0.57$ & $1.6 \pm 0.064$ \\
0.5--1 keV  & $18 \pm 0.65$ & $4.4 \pm 0.27$ & $4.1 \pm 0.29$ \\
1--2 keV  & $13 \pm 0.67$ & $16 \pm 0.5$ & $0.86 \pm 0.051$ \\
2--3 keV & $2 \pm 0.42$ & $4.7 \pm 0.27$ & $0.42 \pm 0.094$ \\
3--8 keV & $3.8 \pm 0.84$ & $5.6 \pm 0.3$ & $0.67 \pm 0.15$ \\
\enddata
\tablecomments{(1) Energy band, (2) net count rate for the spectrum of
  diffuse emission, (3) net count rate for the spectrum of resolved
  point sources, (4) ratio between diffuse emission and resolved point
  source count rates. See Sect.~\ref{sec:diffuse} for details.}
\end{deluxetable}

Using the {\tt combine\_spectra} script we co-added the four spectra extracted from the individual observations. We binned the composite spectrum so as to have a minimum of 20 counts per channel in order to apply $\chi^{2}$ statistics. The spectral analysis was performed with XSPEC v. 12.7.1b. 

\begin{figure}
\begin{center}
\includegraphics[trim=5mm 10mm 2mm 10mm, width=1.1\linewidth]{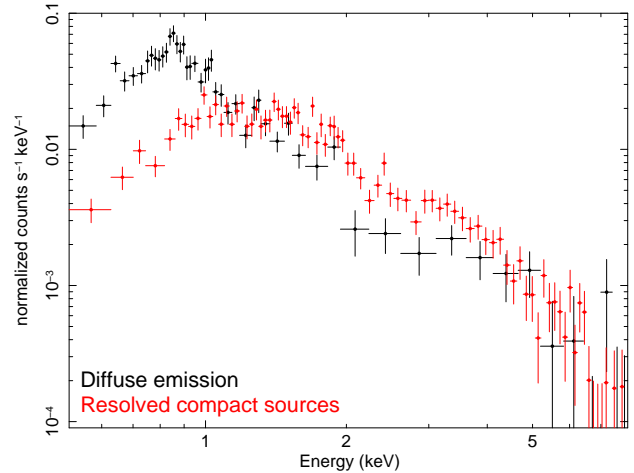} 
\caption{The co-added X-ray spectra of diffuse emission (black, Sect.~\ref{sec:diffuse}) and resolved point sources (red, Sect.~\ref{sec:lums}) over the four individual observations in the full (0.5--8 keV) band. Both spectra were extracted from the $D25$ ellipse of NGC~2207/IC~2163 and are binned in order to have a minimum of 20 counts per channel. The hot ISM dominates the emission in the 0.5--1 keV band.}
\label{fig:spectra}
\end{center}
\end{figure}

The co-added background-subtracted spectrum of diffuse X-ray emission was first modeled in the 0.5--8 keV band, with two components, thermal ({\tt mekal}) and power-law, to which we applied photo-electric absorption. We used two absorbing components ({\tt phabs(1)*phabs(2)}), one fixed to the Galactic value, $N_{\rm{H}}= 8.8 \times 10^{20}$  cm$^{-2}$ \citep{2005A&A...440..775K}. The other one was instead left free to constrain the absorption local to NGC 2207/IC 2163. Overall the model provides a good description of the full-band spectrum with $\chi^{2}=214$ for 208 degrees of freedom (reduced $\chi^{2} = 1.03$).
The power-law model accommodates the emission at $E > 1-1.5$ keV. This can be visually recognized in Fig.~\ref{fig:spectra} and can also be noted by looking at the ratios between diffuse and point source count rates in several bands listed in Table~\ref{tab:flux_ratio}. Since the contribution of resolved point sources was excluded from the diffuse spectrum, this power-law component may be due to unresolved accreting compact sources \citep[see, e.g.,][]{2012MNRAS.426.1870M}. The best fit photon index is $\Gamma = 1.83^{+0.39}_{-0.36}$, which makes it consistent with unresolved compact objects, but it is not very well constrained (note the uncertainties) due to the weak count statistics of the diffuse spectrum in the 2--8 keV band. Our analysis  suggests that the local absorption in NGC~2207/IC~2163 is $N_{\rm{H}}=(2.0 \pm 1.4)\times 10^{21}$ cm$^{-2}$, in agreement with the column density obtained for the co-added point source spectrum (see Sect.~\ref{sec:lums}). The thermal component fits the spectrum rather well at $E < 1$ keV and its best-fit plasma temperature is well constrained: $kT = 0.28^{+0.05}_{-0.04}$ keV, for solar metal abundances.

We repeated the fit over only the 0.5--1 keV band, where the thermal emission dominates. The soft spectrum is well described by an absorbed thermal model, with $\chi^{2}=35.5$ for 30 degrees of freedom (reduced $\chi^{2} = 1.18$). The local absorbing component was constrained to $N_{\rm{H}}=(3.3 \pm 1.1)\times 10^{21}$ cm$^{-2}$, in agreement with the values reported above as well with that obtained for the point source spectrum. The best-fit value for the plasma temperature is $kT = 0.25^{+0.04}_{-0.03}$ keV. Although the soft band emission, 0.5--2 keV, is not totally dominated by diffuse gas, these spectral characteristics are typical for normal star-forming and starburst galaxies.

\subsection{X-ray luminosity of the hot ISM}
\label{sec:lum_ism}
The 0.5--2 keV X-ray luminosity of the thermal plasma, calculated based on the full-band fitting, corrected for Galactic absorption, is $L_{\rm{0.5-2 keV}}^{\rm{mekal}}=2.8\times 10^{40}\,\rm{erg}\,\rm{s}^{-1}$, which is a factor of $\sim$2.3 larger than the average thermal luminosity produced per unit SFR in local star-forming galaxies \citep[][see their Eq.~(2)]{2012MNRAS.426.1870M}. 

This calculation was based on the integrated SFR of NGC~2207/IC~2163 \citep[$23.7\,M_{\odot}\,\rm{yr}^{-1}$, see also][]{2013ApJ...771..133M} estimated using observed (i.e.~uncorrected for dust absorption) NUV and total, 8-1000$\,\mu$m, IR, assuming a Salpeter initial mass function (IMF) from 0.1 to 100 $M_{\odot}$ \citep{2006ApJS..164...38I}. This is one of the best SFR estimators for nearby galaxies, and is the same proxy as used by \citet{2012MNRAS.419.2095M, 2012MNRAS.426.1870M}, providing the most robust value for the integrated SFR of NGC~2207/IC~2163, and full consistency when we compare our results with previous, more extensive work.

The measured value of $L_{\rm{0.5-2 keV}}^{\rm{mekal}}/\rm{SFR}$ is also a factor of $\sim$1.5 larger than the average 0.5--2 keV luminosity of the HMXBs per unit SFR \citep{2012MNRAS.419.2095M}, in full agreement with the values in Table~\ref{tab:flux_ratio}, and corresponds to $\sim$40\% of the integrated luminosity of all point sources detected above the completeness limit for the combined observations in the 0.5--8 keV band (listed in Table~\ref{tab:compl}). Such a result is about 10\% larger that what is typically in local star-forming galaxies, but within the dispersion. After having corrected for both Galactic and local absorption, we obtained the intrinsic luminosity of the hot ISM, $L_{\rm{0.5-2 keV}}^{\rm{mekal, int}}=7.9\times 10^{40}\,\rm{erg}\,\rm{s}^{-1}$. 

After having rescaled our SFR estimate to a Kroupa IMF, the intrinsic luminosity of the hot ISM obtained above is in the same range as the values found by \citet[][their Table 12, only including sources with extended X-ray emission]{2014AJ....147...60S} for a sample of hinge clumps in five interacting galaxies, and similar to that for the Antennae.

\section{Summary and conclusions}

We have presented a comprehensive study of the total X-ray emission from the colliding galaxy pair NGC~2207/IC~2163.

We repeated our correlation study between the {\em local} SFR in NGC~2207/IC~2163 and the number and X-ray luminosity of the ULXs with improved significance with respect to our pilot study \citep{2013ApJ...771..133M}, which was based on only one (Obs.~ID~11228) of the four observations utilized in the present work. Due to ULX variability (Sect.~\ref{sec:variability}) we now detect 28 ULXs, 7 of which were not visible previously. This $\sim$30\% increase in ULX number allows for an improvement in both the spatially-resolved $N_{\rm{X}}-\rm{SFR}$ and $L_{\rm{X}}-\rm{SFR}$ relations, which is most evident in the latter case. 
We confirm that the global relation between the {\em number} of X-ray point sources and the integrated SFR of the host galaxy also holds on {\em local} scales within a given galaxy. Thanks to the improved statistics utilized in the present analysis, we can now show that the relation between {\em X-ray luminosity} and {\em local} SFR is also in general agreement with the multiple-galaxy-wide average relation between the cumulative luminosity of ULXs and the integrated SFR.

We investigated the long-term flux and spectral variability of the ULX population and found that 12 sources out of 57 ($\sim$20\% including the AGN at the center of NGC~2207) show significant long-term variability ($S_{\rm{flux}} > 3$; see Sect.~\ref{sec:variability}). Of these, 7 are transient source candidates. The ratio between maximum and minimum count rate ranges between 2 and 3.8 for the variable sources, but between $\sim$34 and $\sim$72 for the  transient source candidates. One of the variable sources (\#41 in Tables~\ref{tab:ulx_variability} and \ref{tab:merged_properties}) is the central AGN in NGC~2207. One of the transient sources (\#22 in Tables~\ref{tab:ulx_variability} and \ref{tab:merged_properties}), is associated ($0.11\arcsec$ separation) with the supernova SN~2013ai. We find no evidence for X-ray spectral changes in connection with the ULX's flux variability. All the ULXs (except the one associated with SN~2013ai) have a soft spectrum during all four observations. Such behavior has been observed when the flux variability ranges over only a factor of a few.

We cross checked the coordinates of 17 optical ``super-star clusters'' (SSC, mass 1--20$\times 10^4\,M_{\odot}$) identified with {\em HST} as well as the coordinates of 225 {\em Spitzer} 8-micron clumps with the positions of the 74 X-ray sources detected within the $D25$ ellipse in the co-added {\em Chandra} image. Within a $1.5\arcsec$ tolerance limit, we found that only one young SSC is coincident with a bright XRB ($L_{\rm{X}} < 10^{39}\,\rm{erg}\,\rm{s}^{-1}$); by contrast, we found a statistically significant set of $\sim$$1/3$ of our X-ray sources which align with {\em Spitzer} $8\,\mu$m-detected young star complexes, and half of the matching sources are ULXs. Among the matches there are two SNe, SN~1999ec and SN~2013ai, as well as the extended X-ray source at the location of the dusty starburst region called  ``{\em feature i}''.

We report that our X-ray source \#18 corresponds to source ``X1'' in \citet{2012AJ....144..156K}, whose nature was still uncertain. The latter authors interpreted their source X1 as a possible radio SN, a SNR, or a background quasar. The significant variability and spectral hardness of our X-ray source \#18 are incompatible with an AGN and, according to its luminosity, $L_{\rm{X}} > 10^{39}\,\rm{erg}\,\rm{s}^{-1}$ in all {\em Chandra} pointings, we conclude that the source is a ULX.

We constructed the X-ray luminosity function of the bright X-ray point sources in each individual observation and found that the best-fitting slopes are in agreement with each other within the uncertainties (Table~\ref{tab:compl}). This suggests that the X-ray variability exhibited by about 20\% of the detected sources does not significantly influence their XLFs. We also obtained the XLF from the four combined {\em Chandra} observations. The best-fitting model for the average XLF is a power-law with slope $\alpha = 0.66\pm0.04$ ($1.66$ in differential form) and an exponential cut-off  at $L_{o} = (3.39 \pm 0.28)\times 10^{39}$ erg\,s$^{-1}$. The slope is in full agreement with that of the average XLF for HMXBs and the exponential cutoff in our model may correspond to the roll-off of the bright end of a more extended power-law distribution with slope $1.6$. 

We studied the possible effects of dust extinction and age on a bright XRB population in NGC~2207/IC~2163 on sub-galactic scales. We applied the same technique as that used to obtain the image of the SFR density. To characterize the dust extinction within the galaxy pair, we used the ratio of $8-1000\,\mu$m luminosity ($L_{\rm{IR}}$) to observed (i.e., uncorrected for attenuation effects) NUV luminosity ($L_{\rm{NUV}}$ at  2267\,\AA). We found that the number and luminosity of bright XRBs show a trend with the {\em local} $L_{\rm{IR}}/L_{\rm{NUV}}$ ratio. In particular, we observe a peak in the $N_{\rm{X}}$ and $L_{\rm{X}}$ distributions at $L_{\rm{IR}}/L_{\rm{NUV}}\sim 1$, which is more significant for $N_{\rm{X}}$. The peak may be tentatively interpreted as an effect of the different star formation timescales traced by the IR and NUV proxies, and we speculate that at $L_{\rm{IR}}/L_{\rm{NUV}} \approx 1$, the age of the underlying stellar population may be around 10 Myr. That these qualitative conclusions are in agreement with more quantitative previous results suggests that the most luminous ULXs are biased towards donor stars having an age $\sim$$10-20$ Myr \citep{2005MNRAS.356...12S, 2008A&A...486..151G, 2009ApJ...703..159S}. However, this interpretation still remains somewhat tentative. 

We disentangled and compared the X-ray spectra of the diffuse emission with the population of bright XRBs hosted by NGC~2207/IC~2163. The hot ISM has a temperature $kT = 0.28^{+0.05}_{-0.04}$ keV, assuming solar metal abundances, and dominates the overall X-ray output of NGC~2207/IC~2163 at $E \lesssim 1$ keV. Unresolved accreting compact objects most likely dominate the diffuse X-ray emission at $E \gtrsim 1$ keV. 

The co-added spectrum of resolved X-ray point sources is well described by an absorbed power-law with index $\Gamma = 1.95 \pm0.08$, consistent with the centroid of the power law photon index distribution for luminous X-ray compact sources in star-forming galaxies, $\Gamma=1.97\pm 0.11$ \citep{2004ApJS..154..519S}. 

The 0.5--2 keV X-ray luminosity of the thermal plasma, based on the full-band fitting, and corrected for both Galactic and local absorption, is $L_{\rm{0.5-2 keV}}^{\rm{mekal, int}}=7.9\times 10^{40}\,\rm{erg}\,\rm{s}^{-1}$, which is a factor of $\sim$2.3 larger than the average thermal luminosity produced per unit SFR in local star-forming galaxies and corresponds to $\sim$100\% of the collective luminosity of all point sources detected above the completeness limit for the combined observations ($7.8\times10^{40}\,\rm{erg} \,\rm{s}^{-1}$). Such a result is about 10\% larger that what is typical in local star-forming galaxies, but within the dispersion of values. After having subtracted the estimated contribution of background AGNs, the total X-ray output of NGC~2207/IC~2163 is $1.5\times10^{41}\,{\rm ergs}\,\rm{s}^{-1}$. The corresponding total, integrated SFR is $23.7\,M_{\odot}\,\rm{yr}^{-1}$.\\

\begin{sidewaystable*}
\caption{NGC2207/IC2163: ULX variability}
\label{tab:ulx_variability}
\resizebox{\textwidth}{!}{
\begin{tabular}{lcccccccccccccc}
\hline
\hline
& & & \multicolumn{2}{c}{Obs. ID 11228} & &\multicolumn{2}{c}{Obs. ID 14914} && \multicolumn{2}{c}{Obs. ID 14799} & &\multicolumn{2}{c}{Obs. ID 14915} \\
\cline{4-5}  \cline{7-8} \cline{10-11} \cline{13-14} \\ 
Source & $\alpha_{J2000}$ & $\delta_{J2000}$  & Photon flux & HR & &Photon flux & HR & & Photon flux & HR & &Photon flux & HR &$S_{\rm{flux}}$ \\
 & (deg) & (deg) & ($10^{-7}$ ph cm$^{-2}$ s$^{-1}$) & (cts) & & ($10^{-7}$ ph cm$^{-2}$ s$^{-1}$) & (cts) & & ($10^{-7}$ ph cm$^{-2}$ s$^{-1}$) & (cts) & & ($10^{-7}$ ph cm$^{-2}$ s$^{-1}$) & (cts) & \\
(1)  & (2) & (3) & (4) & (5) & & (6) & (7) & & (8) & (9) & & (10) & (11) & (12) \\
\hline
51 & 94.10071 & -21.3865 & $16 \pm 10$ & $-1.00 \pm 0.63$ & &  $0.91 \pm 3.4$ & $-1.00 \pm 4.28$ & & $30 \pm 12$ & $-0.08 \pm 0.46$ & & $28 \pm 7.9$ & $-0.80 \pm 0.25$ & $3.2$ \\
31 & 94.08431 & -21.3851 & $38 \pm 12$ & $-0.41 \pm 0.32$ & & $\leq3.7$ & \nodata & & $2.4 \pm 6.5$ & $1.00 \pm 3.75$ & & $1.3 \pm 3.5$ & $1.00 \pm 4.30$ & $3.1$ \\
37 & 94.08675 & -21.3808 & $\leq3.7$ & \nodata & &  $49 \pm 9.7$ & $-0.33 \pm 0.20$ & & $33 \pm 13$ & $0.16 \pm 0.43$ & & $31 \pm 8.1$ & $-0.67 \pm 0.25$ & $4.7$ \\
14 & 94.07177 & -21.3807 & $81 \pm 16$ & $-0.55 \pm 0.19$ & &  $87 \pm 13$ & $-0.23 \pm 0.15$ & & $140 \pm 23$ & $-0.18 \pm 0.17$ & & $52 \pm 10$ & $-0.42 \pm 0.20$ & $3.4$ \\
74 & 94.12505 & -21.3792 & $25 \pm 11$ & $-0.83 \pm 0.43$ & &  $41 \pm 9.8$ & $-0.77 \pm 0.21$ & & $20 \pm 11$ & $-0.74 \pm 0.62$ & & $\leq3.9$ & \nodata & $3.9$ \\
22$\star$ & 94.07649 & -21.3758 & $\leq3.6$ & \nodata & & $\leq3.6$  & \nodata & & $93 \pm 19$ & $0.71 \pm 0.18$ & & $0.91 \pm 3.4$ & $-1.00 \pm 4.95$ & $4.8$ \\
2 & 94.06618 & -21.3757 & $180 \pm 22$ & $-0.36 \pm 0.12$ & &  $170 \pm 17$ & $-0.32 \pm 0.10$ & & $80 \pm 18$ & $-0.43 \pm 0.23$ & & $49 \pm 10$ & $-0.38 \pm 0.21$ & $6.3$ \\
41$\dagger$ & 94.09177 & -21.3727 & $140 \pm 20$ & $0.76 \pm 0.12$ & &  $110 \pm 14$ & $0.56 \pm 0.13$ & & $93 \pm 21$ & $0.79 \pm 0.20$ & & $200 \pm 18$ & $0.81 \pm 0.07$ & $3.7$ \\
46 & 94.09788 & -21.3719 & $1 \pm 5.3$ & $1.00 \pm 4.37$ & &  $50 \pm 9.8$ & $-0.12 \pm 0.21$ & & $74 \pm 18$ & $-0.12 \pm 0.26$ & & $3.1 \pm 4.2$ & $0.83 \pm 2.00$ & $4.4$ \\
62 & 94.11047 & -21.3701 & $23 \pm 10$ & $-0.80 \pm 0.44$ & &  $42 \pm 9.2$ & $-0.15 \pm 0.23$ & & $41 \pm 14$ & $-0.19 \pm 0.37$ & & $88 \pm 13$ & $-0.28 \pm 0.15$ & $4$ \\
18 & 94.07493 & -21.3678 & $68 \pm 15$ & $-0.62 \pm 0.21$ & &  $33 \pm 8.4$ & $-0.26 \pm 0.27$ & & $28 \pm 12$ & $-0.37 \pm 0.49$ & & $19 \pm 6.7$ & $-0.44 \pm 0.38$ & $3$ \\
66 & 94.11249 & -21.3632 & $\leq3.9$ & \nodata & &  $29 \pm 7.9$ & $-0.34 \pm 0.29$ & & $14 \pm 9.6$ & $-0.20 \pm 0.79$ & & $2.7 \pm 3.9$ & $-1.00 \pm 1.87$ & $3.3$ \\
\hline
\end{tabular}}
\tiny Note: (1) Source number based on Table~\ref{tab:merged_properties}, (2) Right Ascension (RA), (3)
  Declination (Dec). (4), (6), (8), (10) Net photon fluxes in the 0.5--8 keV
  band, in units of $10^{-7}$ cts s$^{-1}$; (5), (7), (9), (11)
  Hardness ratios, computed with eq. (\ref{eq:hr}). (12) Significance
  of long-term flux variability, computed with eq. (\ref{eq:sign}). $\dagger$ Central Active Galactic Nucleus, $\star$ SN~2013ai ($0.11\arcsec$ match)
\end{sidewaystable*}


\acknowledgments
We thank the anonymous referee for helpful suggestions that greatly improved this paper. 
We acknowledge support from the NASA's Astrophysics Data Analysis Program (ADAP) grant NNH13CH56C and by NASA {\em Chandra} grant GO3-14092A. We acknowledge Steven Willner, Luca Cortese, Bret Lehmer and Antara Basu-Zych for insightful discussions on dust extinction, star formation and their relation with the starburst age. We are grateful to Michele Kaufman and Debra Elmegreen who kindly supplied us with the coordinates of the 17 super star clusters that they had identified in the HST image, and for further discussions. We made use of \textit{Chandra} archival data and software provided by the \textit{Chandra} X-ray Center (CXC) in the application package CIAO. We also utilized the software tool SAOImage DS9, developed by the Smithsonian Astrophysical Observatory. The FUV, 3.6 $\mu$m, and 24 $\mu$m images were taken from the \textit{GALEX} and {\em Spitzer} archives, respectively. The \textit{Spitzer Space Telescope} is operated by the Jet Propulsion Laboratory, California Institute of Technology, under contract with NASA. \textit{GALEX} is a NASA Small Explorer, launched in 2003 April. We also made use of data products from the Two Micron All Sky Survey (2MASS), which is a joint project of the University of Massachusetts and the Infrared Processing and Analysis Center/California Institute of Technology, funded by NASA and the National Science Foundation. Helpful information was found in the NASA/IPAC Extragalactic Database (NED) which is operated by the Jet Propulsion Laboratory, California Institute of Technology, under contract with the National Aeronautics and Space Administration.

\newpage
\clearpage
\begin{appendix}
\section{Source lists for all individual and combined observations}

\begin{deluxetable}{lccccccccc}
\tablewidth{0pt}
\tabletypesize{\tiny}
\tablecaption{\label{tab:11228_properties} NGC2207/IC2163: X-Ray Source Properties
for ObsID~11228}
\tablehead{
	\colhead{Source} &
	\colhead{$\alpha_{J2000}$} &
	\colhead{$\delta_{J2000}$}  & 
	\colhead{$0.5-8\,\rm{keV}$} &
	\colhead{Signif} &
	\colhead{$0.5-2\,\rm{keV}$} &
	\colhead{$2-8\,\rm{keV}$} &
	\colhead{HR} &
    	\colhead{$L_{\rm{X}}$} &
	\colhead{$F_{\rm{X}}$} \\
	\colhead{} & 
	\colhead{(deg)} &
	\colhead{(deg)} & 
      	\colhead{(cts)} &
      	\colhead{($\sigma$)} &
	\colhead{(cts)} &
	\colhead{(cts)} &
         \colhead{} &
	\colhead{($10^{38}\,\rm{erg}\,\rm{s}^{-1}$)} &
	\colhead{($10^{-14}\,\rm{erg}\,\rm{cm}^{-2}\,\rm{s}^{-1}$)}\\
	(1)  & (2) & (3) & (4) & (5) & (6) & (7) & (8)  & (9) & (10)
}
\startdata 
1$\ddagger$ & 94.06612 & -21.3673 & $5.8 \pm 4.5$ & 3.4 & $3.9 \pm 4.1$ & $1.8 \pm 3$ & $-0.37 \pm 0.84$ & $7.6 \pm 5.9$ & $0.40 \pm 0.32$ \\
2 & 94.06617 & -21.3757 & $92 \pm 11$ & 38 & $63 \pm 9.5$ & $29 \pm 6.9$ & $-0.36 \pm 0.12$ & $120 \pm 15$ & $6.45 \pm 0.79$ \\
3 & 94.06938 & -21.3743 & $22 \pm 6.2$ & 8.5 & $20 \pm 5.9$ & $2.1 \pm 3$ & $-0.81 \pm 0.25$ & $30 \pm 8.5$ & $1.59 \pm 0.45$ \\
4 & 94.07004 & -21.3726 & $6.5 \pm 4$ & 2.8 & $3.3 \pm 3.2$ & $3.2 \pm 3.3$ & $-0.02 \pm 0.71$ & $8.9 \pm 5.5$ & $0.48 \pm 0.29$ \\
5 & 94.07046 & -21.3527 & $28 \pm 6.8$ & 13 & $21 \pm 6$ & $7.4 \pm 4.2$ & $-0.48 \pm 0.24$ & $38 \pm 9.1$ & $2.02 \pm 0.48$ \\
6 & 94.07048 & -21.3758 & $20 \pm 6$ & 8.6 & $9.7 \pm 4.5$ & $11 \pm 4.7$ & $0.05 \pm 0.32$ & $28 \pm 8.3$ & $1.50 \pm 0.44$ \\
7 & 94.07176 & -21.3806 & $41 \pm 8$ & 16 & $32 \pm 7.2$ & $9 \pm 4.6$ & $-0.56 \pm 0.19$ & $54 \pm 11$ & $2.87 \pm 0.57$ \\
8 & 94.07191 & -21.3599 & $18 \pm 5.7$ & 7.8 & $11 \pm 4.8$ & $6.5 \pm 4$ & $-0.26 \pm 0.35$ & $23 \pm 7.5$ & $1.24 \pm 0.40$ \\
9 & 94.07487 & -21.3679 & $33 \pm 7.2$ & 14 & $27 \pm 6.6$ & $6.4 \pm 3.9$ & $-0.62 \pm 0.21$ & $46 \pm 9.9$ & $2.44 \pm 0.53$ \\
10 & 94.07500 & -21.3647 & $14 \pm 5.3$ & 8.3 & $8 \pm 4.4$ & $6.1 \pm 4$ & $-0.13 \pm 0.41$ & $20 \pm 7.4$ & $1.04 \pm 0.39$ \\
11 & 94.07521 & -21.3686 & $6.5 \pm 4.1$ & 3.3 & $5.4 \pm 3.9$ & $1.1 \pm 2.7$ & $-0.65 \pm 0.70$ & $9.1 \pm 5.7$ & $0.48 \pm 0.30$ \\
12 & 94.07536 & -21.3707 & $5.1 \pm 3.7$ & 2.3 & $4.2 \pm 3.5$ & $0.94 \pm 2.5$ & $-0.63 \pm 0.84$ & $7.7 \pm 5.5$ & $0.41 \pm 0.29$ \\
13 & 94.07818 & -21.3743 & $25 \pm 6.4$ & 11 & $16 \pm 5.4$ & $8.6 \pm 4.3$ & $-0.30 \pm 0.28$ & $33 \pm 8.5$ & $1.74 \pm 0.46$ \\
14 & 94.08054 & -21.3641 & $13 \pm 5$ & 5.9 & $11 \pm 4.7$ & $2.2 \pm 2.9$ & $-0.66 \pm 0.39$ & $17 \pm 6.6$ & $0.91 \pm 0.35$ \\
15 & 94.08430 & -21.3851 & $18 \pm 5.7$ & 7.5 & $13 \pm 5$ & $5.4 \pm 3.7$ & $-0.41 \pm 0.33$ & $25 \pm 7.6$ & $1.31 \pm 0.41$ \\
16 & 94.08525 & -21.3740 & $5.8 \pm 3.9$ & 2.5 & $1.7 \pm 2.9$ & $4.1 \pm 3.5$ & $0.40 \pm 0.79$ & $7.8 \pm 5.3$ & $0.42 \pm 0.28$ \\
17 & 94.08544 & -21.3697 & $27 \pm 6.7$ & 11 & $20 \pm 6$ & $6.6 \pm 4.1$ & $-0.51 \pm 0.25$ & $36 \pm 9$ & $1.92 \pm 0.48$ \\
18 & 94.08567 & -21.3720 & $7.3 \pm 4.1$ & 2.9 & $4.2 \pm 3.5$ & $3.1 \pm 3.2$ & $-0.15 \pm 0.64$ & $9.8 \pm 5.5$ & $0.52 \pm 0.30$ \\
19 & 94.08604 & -21.3736 & $5.1 \pm 3.8$ & 2.1 & $5.6 \pm 3.8$ & $0 \pm 2.1$ & $-1.00 \pm 0.74$ & $6.9 \pm 5.2$ & $0.37 \pm 0.28$ \\
20$\dagger$ & 94.09177 & -21.3727 & $70 \pm 10$ & 24 & $8.4 \pm 4.8$ & $61 \pm 9.4$ & $0.76 \pm 0.12$ & $95 \pm 14$ & $5.05 \pm 0.74$ \\
21 & 94.09430 & -21.3608 & $24 \pm 6.6$ & 11 & $17 \pm 5.7$ & $7 \pm 4.3$ & $-0.43 \pm 0.28$ & $33 \pm 8.9$ & $1.77 \pm 0.48$ \\
22 & 94.09828 & -21.3707 & $7.7 \pm 4.3$ & 3.3 & $5.7 \pm 3.9$ & $2 \pm 2.9$ & $-0.47 \pm 0.62$ & $11 \pm 5.9$ & $0.57 \pm 0.32$ \\
23 & 94.10083 & -21.3866 & $7.6 \pm 4.1$ & 3.4 & $7.6 \pm 4.1$ & $0 \pm 2$ & $-1.00 \pm 0.53$ & $13 \pm 7$ & $0.69 \pm 0.37$ \\
24 & 94.10100 & -21.3627 & $9.7 \pm 4.6$ & 3.9 & $5.2 \pm 3.8$ & $4.5 \pm 3.6$ & $-0.08 \pm 0.54$ & $13 \pm 6.4$ & $0.71 \pm 0.34$ \\
25 & 94.10377 & -21.3641 & $6.4 \pm 3.9$ & 2.6 & $4.2 \pm 3.5$ & $2.2 \pm 2.9$ & $-0.32 \pm 0.71$ & $8.9 \pm 5.5$ & $0.47 \pm 0.29$ \\
26 & 94.10472 & -21.3749 & $13 \pm 5.3$ & 5.9 & $7.3 \pm 4.3$ & $6.2 \pm 4.1$ & $-0.08 \pm 0.44$ & $19 \pm 7.5$ & $1.02 \pm 0.40$ \\
27 & 94.10630 & -21.3704 & $6.4 \pm 4$ & 2.6 & $0 \pm 2.1$ & $6.7 \pm 4$ & $1.00 \pm 0.63$ & $9.3 \pm 5.8$ & $0.49 \pm 0.31$ \\
28 & 94.10824 & -21.3771 & $14 \pm 5.1$ & 5.2 & $6.5 \pm 3.9$ & $7.4 \pm 4.1$ & $0.07 \pm 0.41$ & $22 \pm 8$ & $1.17 \pm 0.43$ \\
29 & 94.11017 & -21.3701 & $7.8 \pm 4.3$ & 5.2 & $6.8 \pm 4.2$ & $0.95 \pm 2.6$ & $-0.75 \pm 0.59$ & $11 \pm 6.3$ & $0.60 \pm 0.34$ \\
30 & 94.11180 & -21.3697 & $13 \pm 5$ & 5.2 & $12 \pm 4.8$ & $1.1 \pm 2.5$ & $-0.83 \pm 0.37$ & $22 \pm 8.5$ & $1.17 \pm 0.45$ \\
31 & 94.11419 & -21.3712 & $5.4 \pm 3.7$ & 2 & $4.3 \pm 3.5$ & $1.1 \pm 2.5$ & $-0.60 \pm 0.79$ & $9.1 \pm 6.2$ & $0.49 \pm 0.33$ \\
32 & 94.11818 & -21.3609 & $15 \pm 5.3$ & 7.2 & $11 \pm 4.7$ & $4.4 \pm 3.5$ & $-0.42 \pm 0.37$ & $24 \pm 8.5$ & $1.30 \pm 0.46$ \\
33 & 94.12197 & -21.3680 & $5.5 \pm 3.8$ & 2 & $5.5 \pm 3.8$ & $0 \pm 2.1$ & $-1.00 \pm 0.75$ & $8.1 \pm 5.5$ & $0.43 \pm 0.29$ \\
34 & 94.12488 & -21.3790 & $15 \pm 5.5$ & 8.3 & $11 \pm 4.9$ & $4.1 \pm 3.5$ & $-0.45 \pm 0.38$ & $21 \pm 7.9$ & $1.14 \pm 0.42$ \\
\enddata
\tablecomments{(1) Source number, (2) Right Ascension (RA), (3)
  Declination (Dec). (4) Net counts in broad (0.5--8 keV) band. The uncertainty expressed here takes into
  account the fluctuations in the source as well as in the
  background. (5) Broad band source detection significance from {\tt
    wavdetect}.  This computes how unlikely it is for the background
  in the customized psf region to fluctuate to yield the detected
  number of counts.  Note that the psf region is optimized differently
  in {\tt wavdetect} than in the calculation of column (4) and is
  typically larger than in the latter. (6)-(7) Net counts in soft
  (0.5--2 keV) and hard (2--8 keV) bands respectively. Uncertainties in net counts are quoted to
  $1\,\sigma$. (7) Hardness ratio, computed with eq. (\ref{eq:hr}). Uncertainties were obtained by applying error propagation to the uncertainties in the net counts. (8) X-ray luminosity in the 0.5--8 keV band, (9) X-ray flux in the 0.5--8 keV band. $\dagger$ Central Active Galactic Nucleus, $\ddagger$ Extended soft X-ray source.}
\end{deluxetable}
\begin{deluxetable}{lccccccccc}
\tablewidth{0pt}
\tabletypesize{\tiny}
\tablecaption{\label{tab:14914_properties} NGC2207/IC2163: X-Ray Source Properties
for ObsID~14914}
\tablehead{
	\colhead{Source} &
	\colhead{$\alpha_{J2000}$} &
	\colhead{$\delta_{J2000}$}  & 
	\colhead{$0.5-8\,\rm{keV}$} &
	\colhead{Signif} &
	\colhead{$0.5-2\,\rm{keV}$} &
	\colhead{$2-8\,\rm{keV}$} &
	\colhead{HR} &
    	\colhead{$L_{\rm{X}}$} &
	\colhead{$F_{\rm{X}}$} \\
	\colhead{} & 
	\colhead{(deg)} &
	\colhead{(deg)} & 
      	\colhead{(cts)} &
      	\colhead{($\sigma$)} &
	\colhead{(cts)} &
	\colhead{(cts)} &
         \colhead{} &
	\colhead{($10^{38}\,\rm{erg}\,\rm{s}^{-1}$)} &
	\colhead{($10^{-14}\,\rm{erg}\,\rm{cm}^{-2}\,\rm{s}^{-1}$)}\\
	(1)  & (2) & (3) & (4) & (5) & (6) & (7) & (8)  & (9) & (10)
}
\startdata 
1 & 94.05535 & -21.3669 & $7.8 \pm 4.3$ & 3.2 & $7.8 \pm 4.3$ & $0 \pm 2.1$ & $-1.00 \pm 0.54$ & $7 \pm 3.9$ & $0.37 \pm 0.21$ \\
2$\ddagger$ & 94.06608 & -21.3677 & $18 \pm 6.1$ & 8.3 & $12 \pm 5.2$ & $6.3 \pm 4$ & $-0.30 \pm 0.35$ & $16 \pm 5.3$ & $0.85 \pm 0.28$ \\
3 & 94.06618 & -21.3757 & $130 \pm 13$ & 45 & $89 \pm 11$ & $46 \pm 8.2$ & $-0.32 \pm 0.10$ & $120 \pm 12$ & $6.26 \pm 0.62$ \\
4 & 94.06938 & -21.3743 & $38 \pm 7.7$ & 15 & $28 \pm 6.8$ & $9.6 \pm 4.5$ & $-0.49 \pm 0.20$ & $33 \pm 6.7$ & $1.76 \pm 0.36$ \\
5 & 94.07044 & -21.3759 & $16 \pm 5.5$ & 6.4 & $8.3 \pm 4.4$ & $7.4 \pm 4.2$ & $-0.06 \pm 0.38$ & $14 \pm 4.8$ & $0.73 \pm 0.25$ \\
6 & 94.07048 & -21.3527 & $46 \pm 8.4$ & 19 & $33 \pm 7.3$ & $13 \pm 5.1$ & $-0.44 \pm 0.18$ & $40 \pm 7.3$ & $2.14 \pm 0.39$ \\
7 & 94.07074 & -21.3809 & $8.3 \pm 4.4$ & 3.5 & $5.3 \pm 3.8$ & $3.1 \pm 3.3$ & $-0.27 \pm 0.60$ & $7.3 \pm 3.9$ & $0.39 \pm 0.21$ \\
8 & 94.07178 & -21.3806 & $68 \pm 9.9$ & 26 & $42 \pm 8$ & $26 \pm 6.6$ & $-0.23 \pm 0.15$ & $59 \pm 8.6$ & $3.15 \pm 0.46$ \\
9 & 94.07187 & -21.3599 & $14 \pm 5.2$ & 6.1 & $9.9 \pm 4.6$ & $4.4 \pm 3.5$ & $-0.38 \pm 0.39$ & $12 \pm 4.5$ & $0.66 \pm 0.24$ \\
10 & 94.07493 & -21.3678 & $26 \pm 6.7$ & 10 & $17 \pm 5.6$ & $8.7 \pm 4.5$ & $-0.32 \pm 0.28$ & $22 \pm 5.8$ & $1.18 \pm 0.31$ \\
11 & 94.07506 & -21.3647 & $19 \pm 5.8$ & 8.6 & $9.6 \pm 4.5$ & $9.5 \pm 4.5$ & $-0.01 \pm 0.33$ & $17 \pm 5$ & $0.88 \pm 0.27$ \\
12 & 94.07529 & -21.3688 & $9.6 \pm 4.5$ & 4.1 & $6.4 \pm 3.9$ & $3.3 \pm 3.2$ & $-0.32 \pm 0.52$ & $8.3 \pm 3.9$ & $0.44 \pm 0.21$ \\
13 & 94.07817 & -21.3743 & $30 \pm 7$ & 13 & $19 \pm 5.8$ & $11 \pm 4.7$ & $-0.29 \pm 0.24$ & $26 \pm 6$ & $1.38 \pm 0.32$ \\
14 & 94.08011 & -21.3827 & $7.5 \pm 4.1$ & 3.4 & $0 \pm 2$ & $7.6 \pm 4.1$ & $1.00 \pm 0.54$ & $6.8 \pm 3.7$ & $0.36 \pm 0.20$ \\
15 & 94.08054 & -21.3642 & $9.4 \pm 4.6$ & 4 & $8.4 \pm 4.4$ & $0.96 \pm 2.6$ & $-0.80 \pm 0.51$ & $8.1 \pm 4$ & $0.43 \pm 0.21$ \\
16 & 94.08268 & -21.3629 & $8.7 \pm 4.4$ & 3.6 & $4.4 \pm 3.5$ & $4.3 \pm 3.5$ & $-0.02 \pm 0.57$ & $7.9 \pm 4$ & $0.42 \pm 0.21$ \\
17 & 94.08540 & -21.3697 & $23 \pm 6.3$ & 9.8 & $12 \pm 4.8$ & $12 \pm 4.8$ & $0.00 \pm 0.30$ & $21 \pm 5.6$ & $1.10 \pm 0.30$ \\
18 & 94.08613 & -21.3737 & $14 \pm 5.2$ & 5.2 & $13 \pm 5.1$ & $1.1 \pm 2.6$ & $-0.84 \pm 0.35$ & $12 \pm 4.5$ & $0.63 \pm 0.24$ \\
19 & 94.08674 & -21.3808 & $39 \pm 7.7$ & 17 & $26 \pm 6.5$ & $13 \pm 5$ & $-0.33 \pm 0.20$ & $33 \pm 6.6$ & $1.77 \pm 0.35$ \\
20 & 94.08898 & -21.3738 & $9.3 \pm 4.6$ & 3.6 & $6.2 \pm 4$ & $3.2 \pm 3.2$ & $-0.32 \pm 0.54$ & $8 \pm 3.9$ & $0.43 \pm 0.21$ \\
21$\dagger$ & 94.09175 & -21.3727 & $86 \pm 11$ & 30 & $19 \pm 6.4$ & $68 \pm 9.9$ & $0.56 \pm 0.13$ & $74 \pm 9.7$ & $3.94 \pm 0.52$ \\
22 & 94.09426 & -21.3609 & $31 \pm 7.1$ & 13 & $22 \pm 6.2$ & $8.7 \pm 4.4$ & $-0.43 \pm 0.24$ & $26 \pm 6.1$ & $1.41 \pm 0.32$ \\
23 & 94.09786 & -21.3719 & $39 \pm 7.7$ & 15 & $22 \pm 6.1$ & $17 \pm 5.5$ & $-0.12 \pm 0.21$ & $34 \pm 6.6$ & $1.80 \pm 0.35$ \\
24 & 94.09819 & -21.3708 & $22 \pm 6.3$ & 8 & $12 \pm 5$ & $9.8 \pm 4.7$ & $-0.10 \pm 0.31$ & $19 \pm 5.4$ & $0.99 \pm 0.29$ \\
25 & 94.09921 & -21.3710 & $6.7 \pm 4.1$ & 3 & $6.9 \pm 4.1$ & $0 \pm 2$ & $-1.00 \pm 0.59$ & $5.8 \pm 3.6$ & $0.31 \pm 0.19$ \\
26 & 94.10011 & -21.3697 & $7.5 \pm 4.2$ & 3.2 & $2 \pm 3$ & $5.6 \pm 3.8$ & $0.48 \pm 0.64$ & $6.5 \pm 3.7$ & $0.35 \pm 0.19$ \\
27 & 94.10093 & -21.3627 & $18 \pm 5.7$ & 7.3 & $14 \pm 5.1$ & $4.2 \pm 3.5$ & $-0.53 \pm 0.33$ & $15 \pm 4.9$ & $0.82 \pm 0.26$ \\
28 & 94.10493 & -21.3749 & $13 \pm 5.1$ & 5.5 & $5.4 \pm 3.8$ & $7.5 \pm 4.2$ & $0.16 \pm 0.44$ & $11 \pm 4.4$ & $0.59 \pm 0.23$ \\
29$\star$ & 94.10608 & -21.3733 & $10 \pm 4.7$ & 4.4 & $7.2 \pm 4.1$ & $3.3 \pm 3.2$ & $-0.38 \pm 0.49$ & $9 \pm 4$ & $0.48 \pm 0.21$ \\
30 & 94.10819 & -21.3771 & $15 \pm 5.5$ & 7.1 & $11 \pm 4.8$ & $4.4 \pm 3.6$ & $-0.42 \pm 0.38$ & $13 \pm 4.7$ & $0.70 \pm 0.25$ \\
31 & 94.10869 & -21.3750 & $7.4 \pm 4.1$ & 3.2 & $3 \pm 3.2$ & $4.4 \pm 3.5$ & $0.18 \pm 0.64$ & $6.4 \pm 3.6$ & $0.34 \pm 0.19$ \\
32 & 94.10971 & -21.3702 & $12 \pm 5$ & 6.1 & $8.3 \pm 4.3$ & $4.2 \pm 3.5$ & $-0.33 \pm 0.44$ & $11 \pm 4.3$ & $0.58 \pm 0.23$ \\
33 & 94.11052 & -21.3701 & $33 \pm 7.2$ & 13 & $19 \pm 5.8$ & $14 \pm 5.1$ & $-0.16 \pm 0.23$ & $29 \pm 6.3$ & $1.53 \pm 0.33$ \\
34 & 94.11171 & -21.3698 & $33 \pm 7.4$ & 13 & $19 \pm 5.9$ & $15 \pm 5.3$ & $-0.13 \pm 0.24$ & $29 \pm 6.4$ & $1.54 \pm 0.34$ \\
35 & 94.11250 & -21.3632 & $23 \pm 6.2$ & 11 & $15 \pm 5.3$ & $7.5 \pm 4.2$ & $-0.34 \pm 0.29$ & $20 \pm 5.4$ & $1.05 \pm 0.29$ \\
36 & 94.11316 & -21.3785 & $4.9 \pm 3.8$ & 2.3 & $1.6 \pm 2.9$ & $3.3 \pm 3.2$ & $0.34 \pm 0.92$ & $4.4 \pm 3.3$ & $0.23 \pm 0.18$ \\
37 & 94.11416 & -21.3713 & $6.3 \pm 4$ & 2.7 & $5.2 \pm 3.7$ & $1.1 \pm 2.6$ & $-0.65 \pm 0.70$ & $5.5 \pm 3.5$ & $0.30 \pm 0.19$ \\
38 & 94.11693 & -21.3798 & $6.5 \pm 4$ & 2.3 & $5.5 \pm 3.8$ & $0.96 \pm 2.6$ & $-0.71 \pm 0.70$ & $6.5 \pm 4$ & $0.35 \pm 0.21$ \\
39 & 94.11809 & -21.3609 & $34 \pm 7.3$ & 14 & $20 \pm 5.9$ & $14 \pm 5.2$ & $-0.17 \pm 0.23$ & $29 \pm 6.4$ & $1.56 \pm 0.34$ \\
40 & 94.12501 & -21.3791 & $28 \pm 6.9$ & 15 & $25 \pm 6.6$ & $3.2 \pm 3.2$ & $-0.78 \pm 0.21$ & $28 \pm 6.8$ & $1.50 \pm 0.36$ \\
\enddata
\tablecomments{(1) Source number, (2) Right Ascension (RA), (3)
  Declination (Dec). (4) Net counts in broad (0.5--8 keV) band. The uncertainty expressed here takes into
  account the fluctuations in the source as well as in the
  background. (5) Broad band source detection significance from {\tt
    wavdetect}.  This computes how unlikely it is for the background
  in the customized psf region to fluctuate to yield the detected
  number of counts.  Note that the psf region is optimized differently
  in {\tt wavdetect} than in the calculation of column (4) and is
  typically larger than in the latter. (6)-(7) Net counts in soft
  (0.5--2 keV) and hard (2--8 keV) bands respectively. Uncertainties in net counts are quoted to
  $1\,\sigma$. (7) Hardness ratio, computed with
  eq. (\ref{eq:hr}). Uncertainties were obtained by applying error
  propagation to the uncertainties in the net counts. (8) X-ray
  luminosity in the 0.5--8 keV band, (9) X-ray flux in the 0.5--8 keV
  band. $\dagger$ Central Active Galactic Nucleus, $\ddagger$ Extended
  soft X-ray source, $\star$ SN~2003H ($3.1\arcsec$ match).}
\end{deluxetable}
\begin{deluxetable}{lccccccccc}
\tablewidth{0pt}
\tabletypesize{\tiny}
\tablecaption{\label{tab:14799_properties} NGC2207/IC2163: X-Ray Source Properties
for ObsID~14799}
\tablehead{
	\colhead{Source} &
	\colhead{$\alpha_{J2000}$} &
	\colhead{$\delta_{J2000}$}  & 
	\colhead{$0.5-8\,\rm{keV}$} &
	\colhead{Signif} &
	\colhead{$0.5-2\,\rm{keV}$} &
	\colhead{$2-8\,\rm{keV}$} &
	\colhead{HR} &
    	\colhead{$L_{\rm{X}}$} &
	\colhead{$F_{\rm{X}}$} \\
	\colhead{} & 
	\colhead{(deg)} &
	\colhead{(deg)} & 
      	\colhead{(cts)} &
      	\colhead{($\sigma$)} &
	\colhead{(cts)} &
	\colhead{(cts)} &
         \colhead{} &
	\colhead{($10^{38}\,\rm{erg}\,\rm{s}^{-1}$)} &
	\colhead{($10^{-14}\,\rm{erg}\,\rm{cm}^{-2}\,\rm{s}^{-1}$)}\\
	(1)  & (2) & (3) & (4) & (5) & (6) & (7) & (8)  & (9) & (10)
}
\startdata 
1 & 94.05536 & -21.3669 & $6.5 \pm 4$ & 3 & $4.4 \pm 3.5$ & $2.1 \pm 3$ & $-0.36 \pm 0.71$ & $11 \pm 6.9$ & $0.60 \pm 0.37$ \\
2 & 94.06615 & -21.3757 & $31 \pm 7$ & 14 & $22 \pm 6.1$ & $8.8 \pm 4.4$ & $-0.43 \pm 0.23$ & $54 \pm 12$ & $2.87 \pm 0.66$ \\
3$\ddagger$ & 94.06619 & -21.3676 & $8 \pm 4.4$ & 3.7 & $4.7 \pm 3.8$ & $3.3 \pm 3.2$ & $-0.17 \pm 0.62$ & $15 \pm 8.3$ & $0.81 \pm 0.44$ \\
4 & 94.06937 & -21.3742 & $12 \pm 4.9$ & 5.1 & $5.5 \pm 3.7$ & $6.2 \pm 4$ & $0.06 \pm 0.47$ & $20 \pm 8.5$ & $1.08 \pm 0.45$ \\
5 & 94.07055 & -21.3527 & $33 \pm 7.3$ & 15 & $24 \pm 6.4$ & $8.7 \pm 4.4$ & $-0.47 \pm 0.22$ & $59 \pm 13$ & $3.12 \pm 0.69$ \\
6 & 94.07061 & -21.3598 & $5.1 \pm 3.8$ & 2.4 & $4.4 \pm 3.5$ & $0.66 \pm 2.6$ & $-0.74 \pm 0.91$ & $8.8 \pm 6.5$ & $0.47 \pm 0.35$ \\
7 & 94.07176 & -21.3806 & $53 \pm 8.8$ & 22 & $31 \pm 7.1$ & $22 \pm 6.1$ & $-0.18 \pm 0.18$ & $92 \pm 15$ & $4.89 \pm 0.82$ \\
8 & 94.07309 & -21.3786 & $8.2 \pm 4.4$ & 3.9 & $7.4 \pm 4.2$ & $0.82 \pm 2.6$ & $-0.80 \pm 0.57$ & $14 \pm 7.6$ & $0.76 \pm 0.40$ \\
9 & 94.07486 & -21.3679 & $13 \pm 5$ & 5.7 & $9.6 \pm 4.5$ & $3.3 \pm 3.2$ & $-0.49 \pm 0.41$ & $23 \pm 9$ & $1.24 \pm 0.48$ \\
10 & 94.07502 & -21.3647 & $12 \pm 4.8$ & 5.4 & $7.3 \pm 4.1$ & $4.2 \pm 3.5$ & $-0.27 \pm 0.46$ & $20 \pm 8.4$ & $1.06 \pm 0.45$ \\
11 & 94.07530 & -21.3686 & $11 \pm 4.7$ & 4.9 & $7.6 \pm 4.1$ & $3.3 \pm 3.2$ & $-0.40 \pm 0.47$ & $20 \pm 8.5$ & $1.05 \pm 0.45$ \\
12 & 94.07536 & -21.3708 & $12 \pm 4.9$ & 5.9 & $5.4 \pm 3.8$ & $6.6 \pm 4$ & $0.09 \pm 0.46$ & $21 \pm 8.5$ & $1.10 \pm 0.45$ \\
13$\star$ & 94.07653 & -21.3758 & $36 \pm 7.7$ & 18 & $5.4 \pm 3.8$ & $31 \pm 7.2$ & $0.70 \pm 0.19$ & $63 \pm 13$ & $3.34 \pm 0.71$ \\
14 & 94.07819 & -21.3743 & $22 \pm 6.2$ & 11 & $12 \pm 4.8$ & $11 \pm 4.7$ & $-0.04 \pm 0.30$ & $39 \pm 11$ & $2.05 \pm 0.57$ \\
15 & 94.08047 & -21.3642 & $9.9 \pm 4.7$ & 5.3 & $8.7 \pm 4.5$ & $1.1 \pm 2.6$ & $-0.77 \pm 0.48$ & $17 \pm 8.1$ & $0.91 \pm 0.43$ \\
16 & 94.08548 & -21.3697 & $8.8 \pm 4.4$ & 4.4 & $6.7 \pm 4$ & $2.1 \pm 3$ & $-0.52 \pm 0.56$ & $16 \pm 8$ & $0.85 \pm 0.43$ \\
17 & 94.08617 & -21.3776 & $4.2 \pm 3.5$ & 2 & $1.1 \pm 2.5$ & $3.1 \pm 3.2$ & $0.48 \pm 0.98$ & $7.3 \pm 6$ & $0.39 \pm 0.32$ \\
18 & 94.08675 & -21.3809 & $13 \pm 5.2$ & 6.5 & $5.6 \pm 3.8$ & $7.8 \pm 4.3$ & $0.16 \pm 0.43$ & $23 \pm 8.9$ & $1.24 \pm 0.48$ \\
19 & 94.08907 & -21.3738 & $4.2 \pm 3.5$ & 1.9 & $2 \pm 2.9$ & $2.2 \pm 2.9$ & $0.03 \pm 0.98$ & $7.2 \pm 6$ & $0.39 \pm 0.32$ \\
20$\dagger$ & 94.09177 & -21.3728 & $39 \pm 8$ & 16 & $8.9 \pm 4.6$ & $30 \pm 7.1$ & $0.55 \pm 0.20$ & $70 \pm 14$ & $3.74 \pm 0.76$ \\
21 & 94.09791 & -21.3719 & $28 \pm 6.8$ & 11 & $15 \pm 5.4$ & $12 \pm 4.9$ & $-0.12 \pm 0.26$ & $50 \pm 12$ & $2.67 \pm 0.65$ \\
22 & 94.10066 & -21.3865 & $12 \pm 4.9$ & 5.5 & $6.6 \pm 3.9$ & $5.5 \pm 3.7$ & $-0.09 \pm 0.45$ & $21 \pm 8.4$ & $1.11 \pm 0.45$ \\
23 & 94.10100 & -21.3627 & $6.4 \pm 4$ & 2.6 & $4.3 \pm 3.5$ & $2.1 \pm 3$ & $-0.35 \pm 0.72$ & $11 \pm 7$ & $0.60 \pm 0.37$ \\
24 & 94.10381 & -21.3639 & $4 \pm 3.6$ & 1.7 & $0 \pm 2.1$ & $4.5 \pm 3.6$ & $1.00 \pm 0.94$ & $7 \pm 6.3$ & $0.37 \pm 0.34$ \\
25 & 94.10484 & -21.3748 & $8.9 \pm 4.6$ & 5.5 & $6.1 \pm 4$ & $2.9 \pm 3.3$ & $-0.35 \pm 0.57$ & $17 \pm 8.8$ & $0.93 \pm 0.47$ \\
26 & 94.10820 & -21.3771 & $7.7 \pm 4.3$ & 3.4 & $5.5 \pm 3.9$ & $2.1 \pm 3$ & $-0.44 \pm 0.63$ & $13 \pm 7.4$ & $0.71 \pm 0.40$ \\
27 & 94.11050 & -21.3702 & $15 \pm 5.4$ & 6.7 & $9.2 \pm 4.5$ & $6.2 \pm 3.9$ & $-0.19 \pm 0.38$ & $27 \pm 9.4$ & $1.44 \pm 0.50$ \\
28 & 94.11194 & -21.3698 & $10 \pm 4.7$ & 5.3 & $6.2 \pm 3.9$ & $4.2 \pm 3.5$ & $-0.19 \pm 0.50$ & $18 \pm 8.1$ & $0.97 \pm 0.43$ \\
29 & 94.11255 & -21.3631 & $5.4 \pm 3.7$ & 2.4 & $3.3 \pm 3.2$ & $2.2 \pm 2.9$ & $-0.20 \pm 0.80$ & $9.5 \pm 6.5$ & $0.51 \pm 0.35$ \\
30 & 94.11814 & -21.3609 & $8.8 \pm 4.4$ & 3.7 & $5.5 \pm 3.7$ & $3.3 \pm 3.2$ & $-0.25 \pm 0.56$ & $15 \pm 7.7$ & $0.83 \pm 0.41$ \\
31 & 94.12123 & -21.3808 & $3.9 \pm 3.6$ & 2.3 & $4.2 \pm 3.6$ & $0 \pm 2.1$ & $-1.00 \pm 1.01$ & $6.7 \pm 6.2$ & $0.36 \pm 0.33$ \\
32 & 94.12492 & -21.3788 & $15 \pm 5.6$ & 9.5 & $9.1 \pm 4.7$ & $6.3 \pm 4.1$ & $-0.18 \pm 0.40$ & $28 \pm 10$ & $1.52 \pm 0.55$ \\
\enddata
\tablecomments{(1) Source number, (2) Right Ascension (RA), (3)
  Declination (Dec). (4) Net counts in broad (0.5--8 keV) band. The uncertainty expressed here takes into
  account the fluctuations in the source as well as in the
  background. (5) Broad band source detection significance from {\tt
    wavdetect}.  This computes how unlikely it is for the background
  in the customized psf region to fluctuate to yield the detected
  number of counts.  Note that the psf region is optimized differently
  in {\tt wavdetect} than in the calculation of column (4) and is
  typically larger than in the latter. (6)-(7) Net counts in soft
  (0.5--2 keV) and hard (2--8 keV) bands respectively. Uncertainties in net counts are quoted to
  $1\,\sigma$. (7) Hardness ratio, computed with
  eq. (\ref{eq:hr}). Uncertainties were obtained by applying error
  propagation to the uncertainties in the net counts. (8) X-ray
  luminosity in the 0.5--8 keV band, (9) X-ray flux in the 0.5--8 keV
  band. $\dagger$ Central Active Galactic Nucleus, $\ddagger$ Extended
  soft X-ray source, $\star$ SN~2013ai ($0.23\arcsec$ match).}
\end{deluxetable}

\begin{deluxetable}{lccccccccc}
\tablewidth{0pt}
\tabletypesize{\tiny}
\tablecaption{\label{tab:14915_properties} NGC2207/IC2163: X-Ray Source Properties
  for ObsID~14915}
\tablehead{
	\colhead{Source} &
	\colhead{$\alpha_{J2000}$} &
	\colhead{$\delta_{J2000}$}  & 
	\colhead{$0.5-8\,\rm{keV}$} &
	\colhead{Signif} &
	\colhead{$0.5-2\,\rm{keV}$} &
	\colhead{$2-8\,\rm{keV}$} &
	\colhead{HR} &
    	\colhead{$L_{\rm{X}}$} &
	\colhead{$F_{\rm{X}}$} \\
	\colhead{} & 
	\colhead{(deg)} &
	\colhead{(deg)} & 
      	\colhead{(cts)} &
      	\colhead{($\sigma$)} &
	\colhead{(cts)} &
	\colhead{(cts)} &
         \colhead{} &
	\colhead{($10^{38}\,\rm{erg}\,\rm{s}^{-1}$)} &
	\colhead{($10^{-14}\,\rm{erg}\,\rm{cm}^{-2}\,\rm{s}^{-1}$)}\\
	(1)  & (2) & (3) & (4) & (5) & (6) & (7) & (8)  & (9) & (10)
}
\startdata 
1 & 94.05543 & -21.3669 & $5.6 \pm 3.8$ & 2.3 & $4.5 \pm 3.6$ & $1.1 \pm 2.6$ & $-0.60 \pm 0.79$ & $4.9 \pm 3.3$ & $0.26 \pm 0.18$ \\
2$\ddagger$ & 94.06613 & -21.3675 & $20 \pm 6.3$ & 11 & $10 \pm 5$ & $9.8 \pm 4.6$ & $-0.01 \pm 0.34$ & $17 \pm 5.4$ & $0.91 \pm 0.29$ \\
3 & 94.06623 & -21.3757 & $39 \pm 7.9$ & 16 & $27 \pm 6.7$ & $12 \pm 5$ & $-0.38 \pm 0.21$ & $33 \pm 6.7$ & $1.78 \pm 0.36$ \\
4 & 94.06792 & -21.3725 & $12 \pm 5$ & 4.5 & $7.7 \pm 4.2$ & $4.2 \pm 3.6$ & $-0.29 \pm 0.46$ & $10 \pm 4.3$ & $0.54 \pm 0.23$ \\
5 & 94.06939 & -21.3743 & $23 \pm 6.3$ & 8 & $13 \pm 5.1$ & $10 \pm 4.6$ & $-0.11 \pm 0.30$ & $19 \pm 5.4$ & $1.03 \pm 0.29$ \\
6 & 94.06973 & -21.3766 & $9.3 \pm 4.6$ & 4 & $8.4 \pm 4.4$ & $0.95 \pm 2.6$ & $-0.80 \pm 0.51$ & $8 \pm 3.9$ & $0.42 \pm 0.21$ \\
7 & 94.07043 & -21.3757 & $6.2 \pm 4$ & 2.7 & $3 \pm 3.3$ & $3.2 \pm 3.3$ & $0.02 \pm 0.74$ & $5.3 \pm 3.4$ & $0.28 \pm 0.18$ \\
8 & 94.07057 & -21.3527 & $13 \pm 5.2$ & 6 & $3.2 \pm 3.3$ & $10 \pm 4.6$ & $0.52 \pm 0.41$ & $23 \pm 9$ & $1.24 \pm 0.48$ \\
9 & 94.07180 & -21.3807 & $41 \pm 8$ & 17 & $29 \pm 6.9$ & $12 \pm 4.9$ & $-0.43 \pm 0.20$ & $35 \pm 6.9$ & $1.89 \pm 0.37$ \\
10 & 94.07292 & -21.3790 & $5.8 \pm 4$ & 2.6 & $3.7 \pm 3.5$ & $2.1 \pm 2.9$ & $-0.29 \pm 0.79$ & $5 \pm 3.4$ & $0.26 \pm 0.18$ \\
11 & 94.07498 & -21.3678 & $15 \pm 5.3$ & 6.5 & $11 \pm 4.7$ & $4.2 \pm 3.5$ & $-0.44 \pm 0.38$ & $13 \pm 4.5$ & $0.69 \pm 0.24$ \\
12 & 94.07507 & -21.3646 & $23 \pm 6.4$ & 10 & $16 \pm 5.6$ & $6.7 \pm 4.1$ & $-0.41 \pm 0.29$ & $20 \pm 5.5$ & $1.06 \pm 0.29$ \\
13 & 94.07534 & -21.3686 & $14 \pm 5.2$ & 6.1 & $10 \pm 4.7$ & $3.1 \pm 3.2$ & $-0.54 \pm 0.40$ & $12 \pm 4.4$ & $0.62 \pm 0.24$ \\
14 & 94.07819 & -21.3743 & $54 \pm 8.9$ & 23 & $37 \pm 7.5$ & $17 \pm 5.6$ & $-0.37 \pm 0.17$ & $46 \pm 7.6$ & $2.47 \pm 0.41$ \\
15 & 94.08049 & -21.3642 & $14 \pm 5.3$ & 7.2 & $10 \pm 4.7$ & $4.3 \pm 3.5$ & $-0.40 \pm 0.39$ & $12 \pm 4.5$ & $0.66 \pm 0.24$ \\
16 & 94.08225 & -21.3784 & $6.3 \pm 4.1$ & 3.2 & $1.8 \pm 3$ & $4.5 \pm 3.6$ & $0.42 \pm 0.75$ & $5.4 \pm 3.5$ & $0.29 \pm 0.19$ \\
17 & 94.08265 & -21.3628 & $11 \pm 4.8$ & 5.7 & $8.8 \pm 4.5$ & $1.7 \pm 2.9$ & $-0.67 \pm 0.48$ & $9 \pm 4.2$ & $0.48 \pm 0.22$ \\
18 & 94.08532 & -21.3740 & $13 \pm 5.1$ & 5 & $3.2 \pm 3.3$ & $9.7 \pm 4.6$ & $0.50 \pm 0.42$ & $11 \pm 4.4$ & $0.59 \pm 0.24$ \\
19 & 94.08548 & -21.3697 & $12 \pm 5$ & 5.3 & $8.9 \pm 4.5$ & $3.2 \pm 3.3$ & $-0.46 \pm 0.45$ & $10 \pm 4.3$ & $0.55 \pm 0.23$ \\
20 & 94.08675 & -21.3808 & $25 \pm 6.4$ & 11 & $21 \pm 5.9$ & $4.1 \pm 3.5$ & $-0.67 \pm 0.25$ & $21 \pm 5.5$ & $1.13 \pm 0.29$ \\
21 & 94.08708 & -21.3711 & $8.6 \pm 4.5$ & 4.2 & $5.5 \pm 3.8$ & $3.1 \pm 3.3$ & $-0.28 \pm 0.59$ & $7.4 \pm 3.8$ & $0.39 \pm 0.20$ \\
22$\dagger$ & 94.09180 & -21.3727 & $160 \pm 15$ & 51 & $13 \pm 5.6$ & $150 \pm 14$ & $0.83 \pm 0.07$ & $140 \pm 13$ & $7.31 \pm 0.68$ \\
23 & 94.09431 & -21.3608 & $8.7 \pm 4.4$ & 4.7 & $6.4 \pm 4$ & $2.2 \pm 3$ & $-0.49 \pm 0.56$ & $7.5 \pm 3.9$ & $0.40 \pm 0.21$ \\
24 & 94.09820 & -21.3709 & $22 \pm 6.2$ & 9.1 & $14 \pm 5.1$ & $8.1 \pm 4.3$ & $-0.25 \pm 0.30$ & $19 \pm 5.3$ & $1.00 \pm 0.28$ \\
25 & 94.10011 & -21.3696 & $7.6 \pm 4.2$ & 3.1 & $3.1 \pm 3.3$ & $4.5 \pm 3.6$ & $0.18 \pm 0.65$ & $6.6 \pm 3.7$ & $0.35 \pm 0.20$ \\
26 & 94.10064 & -21.3818 & $7.1 \pm 4.1$ & 3.4 & $2.8 \pm 3.2$ & $4.2 \pm 3.5$ & $0.19 \pm 0.67$ & $6.5 \pm 3.8$ & $0.35 \pm 0.20$ \\
27 & 94.10071 & -21.3865 & $22 \pm 6.2$ & 9.7 & $20 \pm 6$ & $2.2 \pm 3$ & $-0.80 \pm 0.25$ & $19 \pm 5.4$ & $1.02 \pm 0.29$ \\
28 & 94.10097 & -21.3627 & $10 \pm 4.7$ & 4.4 & $9.1 \pm 4.5$ & $1.1 \pm 2.5$ & $-0.79 \pm 0.45$ & $8.9 \pm 4.1$ & $0.47 \pm 0.22$ \\
29 & 94.10385 & -21.3640 & $15 \pm 5.2$ & 6.4 & $9.6 \pm 4.5$ & $5.3 \pm 3.7$ & $-0.29 \pm 0.39$ & $13 \pm 4.6$ & $0.69 \pm 0.24$ \\
30 & 94.10485 & -21.3748 & $25 \pm 6.5$ & 10 & $14 \pm 5.2$ & $11 \pm 4.8$ & $-0.12 \pm 0.28$ & $21 \pm 5.6$ & $1.14 \pm 0.30$ \\
31 & 94.10568 & -21.3770 & $5.4 \pm 3.7$ & 2.2 & $5.4 \pm 3.7$ & $0 \pm 2$ & $-1.00 \pm 0.75$ & $4.9 \pm 3.3$ & $0.26 \pm 0.18$ \\
32 & 94.10635 & -21.3703 & $7.3 \pm 4.1$ & 3.5 & $0.8 \pm 2.5$ & $6.5 \pm 3.9$ & $0.78 \pm 0.63$ & $6.3 \pm 3.5$ & $0.34 \pm 0.19$ \\
33 & 94.10823 & -21.3770 & $6.2 \pm 4.1$ & 2.9 & $4.1 \pm 3.6$ & $2.1 \pm 3$ & $-0.32 \pm 0.75$ & $5.6 \pm 3.6$ & $0.30 \pm 0.19$ \\
34 & 94.11048 & -21.3701 & $70 \pm 9.9$ & 29 & $45 \pm 8.2$ & $25 \pm 6.4$ & $-0.29 \pm 0.15$ & $60 \pm 8.6$ & $3.20 \pm 0.46$ \\
35 & 94.11177 & -21.3698 & $35 \pm 7.6$ & 14 & $25 \pm 6.6$ & $10 \pm 4.6$ & $-0.43 \pm 0.22$ & $30 \pm 6.5$ & $1.60 \pm 0.35$ \\
36 & 94.11363 & -21.3738 & $4.2 \pm 3.6$ & 2.1 & $3.1 \pm 3.3$ & $1.1 \pm 2.6$ & $-0.47 \pm 1.00$ & $3.8 \pm 3.2$ & $0.20 \pm 0.17$ \\
37 & 94.11431 & -21.3713 & $9.1 \pm 4.5$ & 4 & $7.3 \pm 4.1$ & $1.8 \pm 2.9$ & $-0.61 \pm 0.55$ & $7.8 \pm 3.9$ & $0.42 \pm 0.21$ \\
38 & 94.11816 & -21.3609 & $23 \pm 6.2$ & 9.5 & $18 \pm 5.6$ & $5.2 \pm 3.8$ & $-0.54 \pm 0.28$ & $20 \pm 5.4$ & $1.06 \pm 0.29$ \\
39 & 94.12033 & -21.3768 & $7 \pm 4.2$ & 3.2 & $6.2 \pm 4$ & $0.83 \pm 2.6$ & $-0.76 \pm 0.66$ & $6.1 \pm 3.6$ & $0.32 \pm 0.19$ \\
40 & 94.12192 & -21.3680 & $19 \pm 5.8$ & 8.7 & $18 \pm 5.7$ & $0.95 \pm 2.5$ & $-0.90 \pm 0.25$ & $17 \pm 5$ & $0.88 \pm 0.27$ \\
41 & 94.12480 & -21.3786 & $6.5 \pm 4$ & 2.5 & $4.3 \pm 3.5$ & $2.2 \pm 2.9$ & $-0.32 \pm 0.70$ & $5.7 \pm 3.5$ & $0.30 \pm 0.19$ \\
\enddata
\tablecomments{(1) Source number, (2) Right Ascension (RA), (3)
  Declination (Dec). (4) Net counts in broad (0.5--8 keV) band. The uncertainty expressed here takes into
  account the fluctuations in the source as well as in the
  background. (5) Broad band source detection significance from {\tt
    wavdetect}.  This computes how unlikely it is for the background
  in the customized psf region to fluctuate to yield the detected
  number of counts.  Note that the psf region is optimized differently
  in {\tt wavdetect} than in the calculation of column (4) and is
  typically larger than in the latter. (6)-(7) Net counts in soft
  (0.5--2 keV) and hard (2--8 keV) bands respectively. Uncertainties in net counts are quoted to
  $1\,\sigma$. (7) Hardness ratio, computed with eq. (\ref{eq:hr}). Uncertainties were obtained by applying error propagation to the uncertainties in the net counts. (8) X-ray luminosity in the 0.5--8 keV band, (9) X-ray flux in the 0.5--8 keV band. $\dagger$ Central Active Galactic Nucleus, $\ddagger$ Extended soft X-ray source.}
\end{deluxetable}

\begin{deluxetable}{lccccccccc}
\tablewidth{0pt}
\tabletypesize{\tiny}
\tablecaption{\label{tab:merged_properties} NGC2207/IC2163: X-Ray Source Properties
for the combined image}
\tablehead{
	\colhead{Source} &
	\colhead{$\alpha_{J2000}$} &
	\colhead{$\delta_{J2000}$}  & 
	\colhead{$0.5-8\,\rm{keV}$} &
	\colhead{Signif} &
	\colhead{$0.5-2\,\rm{keV}$} &
	\colhead{$2-8\,\rm{keV}$} &
	\colhead{HR} &
    	\colhead{$L_{\rm{X}}$} &
	\colhead{$F_{\rm{X}}$} \\
	\colhead{} & 
	\colhead{(deg)} &
	\colhead{(deg)} & 
      	\colhead{(cts)} &
      	\colhead{($\sigma$)} &
	\colhead{(cts)} &
	\colhead{(cts)} &
         \colhead{} &
	\colhead{($10^{38}\,\rm{erg}\,\rm{s}^{-1}$)} &
	\colhead{($10^{-14}\,\rm{erg}\,\rm{cm}^{-2}\,\rm{s}^{-1}$)}\\
	(1)  & (2) & (3) & (4) & (5) & (6) & (7) & (8)  & (9) & (10)
}
\startdata 
1 & 94.05537 & -21.3669 & $22 \pm 6.3$ & 8.8 & $18 \pm 5.7$ & $4 \pm 3.6$ & $-0.64 \pm 0.29$ & $6 \pm 1.7$ & $0.32 \pm 0.09$ \\
2 & 94.06618 & -21.3757 & $300 \pm 19$ & 82 & $200 \pm 16$ & $95 \pm 11$ & $-0.36 \pm 0.06$ & $81 \pm 5.3$ & $4.32 \pm 0.28$ \\
3$\ddagger$ & 94.06618 & -21.3675 & $67 \pm 10$ & 19 & $43 \pm 8.8$ & $24 \pm 6.4$ & $-0.29 \pm 0.15$ & $19 \pm 2.9$ & $0.99 \pm 0.15$ \\
4 & 94.06674 & -21.3581 & $6.9 \pm 4.4$ & 3.3 & $6 \pm 4$ & $0.93 \pm 3$ & $-0.73 \pm 0.76$ & $2 \pm 1.2$ & $0.10 \pm 0.07$ \\
5 & 94.06788 & -21.3726 & $17 \pm 6.1$ & 5.4 & $11 \pm 5.2$ & $5.6 \pm 4.1$ & $-0.33 \pm 0.39$ & $4.6 \pm 1.7$ & $0.25 \pm 0.09$ \\
6$\ast$ & 94.06856 & -21.3692 & $11 \pm 5.4$ & 3.4 & $5.3 \pm 4.6$ & $5.2 \pm 3.9$ & $-0.01 \pm 0.57$ & $2.9 \pm 1.5$ & $0.15 \pm 0.08$ \\
7 & 94.06937 & -21.3743 & $95 \pm 12$ & 26 & $67 \pm 10$ & $28 \pm 7$ & $-0.41 \pm 0.12$ & $26 \pm 3.2$ & $1.40 \pm 0.17$ \\
8 & 94.06969 & -21.3766 & $14 \pm 5.5$ & 5 & $13 \pm 5.2$ & $1.3 \pm 3.1$ & $-0.81 \pm 0.41$ & $3.8 \pm 1.5$ & $0.20 \pm 0.08$ \\
9 & 94.07010 & -21.3726 & $18 \pm 6.2$ & 6.2 & $8.2 \pm 4.7$ & $10 \pm 4.8$ & $0.11 \pm 0.37$ & $5.1 \pm 1.7$ & $0.27 \pm 0.09$ \\
10 & 94.07048 & -21.3758 & $47 \pm 8.7$ & 14 & $26 \pm 6.7$ & $22 \pm 6.3$ & $-0.09 \pm 0.19$ & $13 \pm 2.4$ & $0.70 \pm 0.13$ \\
11 & 94.07050 & -21.3527 & $120 \pm 13$ & 45 & $80 \pm 11$ & $40 \pm 8.1$ & $-0.33 \pm 0.11$ & $41 \pm 4.3$ & $2.19 \pm 0.23$ \\
12 & 94.07052 & -21.3598 & $13 \pm 5.3$ & 5 & $8.5 \pm 4.5$ & $4.1 \pm 3.9$ & $-0.35 \pm 0.47$ & $3.5 \pm 1.5$ & $0.18 \pm 0.08$ \\
13 & 94.07070 & -21.3809 & $21 \pm 6.5$ & 7.3 & $10 \pm 4.9$ & $11 \pm 5.1$ & $0.02 \pm 0.34$ & $5.8 \pm 1.8$ & $0.31 \pm 0.09$ \\
14 & 94.07177 & -21.3807 & $210 \pm 17$ & 54 & $140 \pm 14$ & $74 \pm 10$ & $-0.30 \pm 0.08$ & $58 \pm 4.6$ & $3.09 \pm 0.24$ \\
15 & 94.07189 & -21.3599 & $41 \pm 8.1$ & 16 & $24 \pm 6.5$ & $17 \pm 5.7$ & $-0.17 \pm 0.21$ & $11 \pm 2.2$ & $0.61 \pm 0.12$ \\
16 & 94.07299 & -21.3724 & $7.9 \pm 4.6$ & 2.7 & $5.3 \pm 4$ & $2.6 \pm 3.3$ & $-0.35 \pm 0.65$ & $2.2 \pm 1.3$ & $0.12 \pm 0.07$ \\
17 & 94.07301 & -21.3787 & $23 \pm 6.8$ & 8.1 & $20 \pm 6.3$ & $3.3 \pm 3.6$ & $-0.72 \pm 0.28$ & $6.4 \pm 1.9$ & $0.34 \pm 0.10$ \\
18 & 94.07493 & -21.3678 & $88 \pm 11$ & 27 & $65 \pm 9.8$ & $24 \pm 6.5$ & $-0.47 \pm 0.12$ & $24 \pm 3.1$ & $1.31 \pm 0.17$ \\
19 & 94.07506 & -21.3647 & $75 \pm 10$ & 23 & $45 \pm 8.4$ & $30 \pm 7$ & $-0.19 \pm 0.14$ & $21 \pm 2.9$ & $1.11 \pm 0.15$ \\
20 & 94.07530 & -21.3687 & $41 \pm 8.1$ & 14 & $31 \pm 7.2$ & $10 \pm 4.8$ & $-0.49 \pm 0.19$ & $12 \pm 2.3$ & $0.61 \pm 0.12$ \\
21 & 94.07536 & -21.3708 & $20 \pm 6.1$ & 7.2 & $12 \pm 5.1$ & $7.9 \pm 4.4$ & $-0.21 \pm 0.33$ & $5.6 \pm 1.7$ & $0.30 \pm 0.09$ \\
22$\diamond$ & 94.07649 & -21.3758 & $41 \pm 8.1$ & 13 & $6.4 \pm 4.2$ & $35 \pm 7.4$ & $0.69 \pm 0.18$ & $11 \pm 2.2$ & $0.60 \pm 0.12$ \\
23 & 94.07707 & -21.3632 & $6.5 \pm 4.2$ & 2.8 & $3.9 \pm 3.5$ & $2.6 \pm 3.3$ & $-0.19 \pm 0.75$ & $1.8 \pm 1.2$ & $0.10 \pm 0.06$ \\
24 & 94.07818 & -21.3743 & $130 \pm 13$ & 42 & $85 \pm 11$ & $49 \pm 8.6$ & $-0.27 \pm 0.10$ & $37 \pm 3.7$ & $1.95 \pm 0.20$ \\
25 & 94.08016 & -21.3827 & $15 \pm 5.5$ & 6.4 & $0 \pm 2.1$ & $15 \pm 5.5$ & $1.00 \pm 0.27$ & $4.2 \pm 1.5$ & $0.22 \pm 0.08$ \\
26 & 94.08054 & -21.3642 & $49 \pm 8.8$ & 17 & $40 \pm 8.1$ & $8.7 \pm 4.5$ & $-0.64 \pm 0.16$ & $13 \pm 2.4$ & $0.71 \pm 0.13$ \\
27 & 94.08208 & -21.3620 & $14 \pm 5.5$ & 5.8 & $9.3 \pm 4.6$ & $4.8 \pm 3.8$ & $-0.32 \pm 0.42$ & $4 \pm 1.5$ & $0.21 \pm 0.08$ \\
28 & 94.08222 & -21.3784 & $15 \pm 5.6$ & 6.3 & $2.3 \pm 3.4$ & $13 \pm 5.2$ & $0.70 \pm 0.38$ & $4.3 \pm 1.6$ & $0.23 \pm 0.08$ \\
29 & 94.08267 & -21.3629 & $22 \pm 6.4$ & 9 & $16 \pm 5.7$ & $6.2 \pm 4$ & $-0.45 \pm 0.29$ & $6.2 \pm 1.8$ & $0.33 \pm 0.10$ \\
30 & 94.08410 & -21.3736 & $11 \pm 4.9$ & 3.5 & $5.9 \pm 4$ & $4.7 \pm 3.8$ & $-0.12 \pm 0.52$ & $2.9 \pm 1.4$ & $0.16 \pm 0.07$ \\
31 & 94.08431 & -21.3851 & $20 \pm 6.2$ & 8.4 & $13 \pm 5.2$ & $7.3 \pm 4.3$ & $-0.28 \pm 0.33$ & $5.6 \pm 1.7$ & $0.30 \pm 0.09$ \\
32 & 94.08528 & -21.3741 & $24 \pm 6.7$ & 7.4 & $9.7 \pm 4.8$ & $15 \pm 5.4$ & $0.21 \pm 0.29$ & $6.7 \pm 1.8$ & $0.36 \pm 0.10$ \\
33 & 94.08544 & -21.3697 & $71 \pm 10$ & 23 & $46 \pm 8.4$ & $25 \pm 6.5$ & $-0.31 \pm 0.14$ & $20 \pm 2.8$ & $1.05 \pm 0.15$ \\
34 & 94.08568 & -21.3719 & $11 \pm 5.1$ & 4 & $5.2 \pm 4$ & $5.8 \pm 4$ & $0.05 \pm 0.52$ & $3 \pm 1.4$ & $0.16 \pm 0.07$ \\
35 & 94.08611 & -21.3737 & $24 \pm 6.5$ & 7.1 & $23 \pm 6.4$ & $0.63 \pm 2.6$ & $-0.95 \pm 0.21$ & $6.5 \pm 1.8$ & $0.34 \pm 0.09$ \\
36 & 94.08616 & -21.3776 & $9.6 \pm 4.8$ & 3.4 & $4.7 \pm 3.8$ & $4.9 \pm 3.8$ & $0.03 \pm 0.56$ & $2.6 \pm 1.3$ & $0.14 \pm 0.07$ \\
37 & 94.08675 & -21.3808 & $77 \pm 10$ & 28 & $52 \pm 8.7$ & $25 \pm 6.5$ & $-0.34 \pm 0.14$ & $21 \pm 2.8$ & $1.13 \pm 0.15$ \\
38 & 94.08706 & -21.3712 & $21 \pm 6.3$ & 7.6 & $15 \pm 5.6$ & $6 \pm 4$ & $-0.43 \pm 0.31$ & $5.8 \pm 1.8$ & $0.31 \pm 0.09$ \\
39 & 94.08906 & -21.3738 & $18 \pm 6$ & 6.1 & $11 \pm 4.9$ & $7.5 \pm 4.4$ & $-0.19 \pm 0.35$ & $5 \pm 1.6$ & $0.27 \pm 0.09$ \\
40 & 94.09023 & -21.3654 & $8.9 \pm 4.8$ & 3.7 & $2.2 \pm 3.3$ & $6.7 \pm 4.2$ & $0.51 \pm 0.60$ & $2.5 \pm 1.3$ & $0.13 \pm 0.07$ \\
41$\dagger$ & 94.09177 & -21.3727 & $350 \pm 22$ & 74 & $46 \pm 9.4$ & $300 \pm 20$ & $0.74 \pm 0.05$ & $97 \pm 6$ & $5.15 \pm 0.32$ \\
42 & 94.09429 & -21.3608 & $70 \pm 10$ & 24 & $50 \pm 8.8$ & $20 \pm 6.1$ & $-0.44 \pm 0.14$ & $19 \pm 2.8$ & $1.03 \pm 0.15$ \\
43 & 94.09447 & -21.3775 & $15 \pm 5.5$ & 5.7 & $10 \pm 4.7$ & $4.5 \pm 3.7$ & $-0.39 \pm 0.41$ & $4 \pm 1.5$ & $0.21 \pm 0.08$ \\
44 & 94.09474 & -21.3863 & $10 \pm 4.8$ & 4.5 & $5.9 \pm 4$ & $4.1 \pm 3.5$ & $-0.18 \pm 0.53$ & $2.8 \pm 1.3$ & $0.15 \pm 0.07$ \\
45 & 94.09764 & -21.3541 & $9.6 \pm 4.6$ & 4.1 & $3.3 \pm 3.3$ & $6.3 \pm 4$ & $0.30 \pm 0.53$ & $2.7 \pm 1.3$ & $0.14 \pm 0.07$ \\
46 & 94.09788 & -21.3719 & $69 \pm 10$ & 20 & $37 \pm 7.7$ & $32 \pm 7.1$ & $-0.07 \pm 0.15$ & $19 \pm 2.8$ & $1.02 \pm 0.15$ \\
47 & 94.09820 & -21.3708 & $51 \pm 9$ & 15 & $30 \pm 7.3$ & $21 \pm 6.1$ & $-0.19 \pm 0.18$ & $14 \pm 2.5$ & $0.76 \pm 0.13$ \\
48 & 94.09843 & -21.3665 & $15 \pm 5.5$ & 5.4 & $13 \pm 5.2$ & $1.5 \pm 2.9$ & $-0.80 \pm 0.37$ & $4.1 \pm 1.5$ & $0.22 \pm 0.08$ \\
49 & 94.10012 & -21.3697 & $17 \pm 5.8$ & 5.9 & $5.7 \pm 4.1$ & $11 \pm 4.8$ & $0.31 \pm 0.38$ & $4.7 \pm 1.6$ & $0.25 \pm 0.09$ \\
50 & 94.10064 & -21.3819 & $8.3 \pm 4.6$ & 3.6 & $3.2 \pm 3.5$ & $5.1 \pm 3.7$ & $0.23 \pm 0.63$ & $2.3 \pm 1.3$ & $0.12 \pm 0.07$ \\
51 & 94.10071 & -21.3865 & $41 \pm 8$ & 16 & $33 \pm 7.3$ & $7.3 \pm 4.2$ & $-0.64 \pm 0.18$ & $12 \pm 2.3$ & $0.62 \pm 0.12$ \\
52 & 94.10097 & -21.3627 & $44 \pm 8.3$ & 15 & $32 \pm 7.3$ & $12 \pm 4.8$ & $-0.47 \pm 0.18$ & $12 \pm 2.3$ & $0.65 \pm 0.12$ \\
53 & 94.10377 & -21.3641 & $26 \pm 6.6$ & 10 & $12 \pm 5$ & $14 \pm 5.1$ & $0.08 \pm 0.28$ & $7.1 \pm 1.8$ & $0.38 \pm 0.10$ \\
54 & 94.10483 & -21.3749 & $56 \pm 9.4$ & 19 & $34 \pm 7.6$ & $22 \pm 6.4$ & $-0.23 \pm 0.17$ & $16 \pm 2.6$ & $0.84 \pm 0.14$ \\
55 & 94.10548 & -21.3720 & $7.3 \pm 4.3$ & 3 & $5.4 \pm 3.9$ & $1.9 \pm 2.9$ & $-0.48 \pm 0.65$ & $2.1 \pm 1.2$ & $0.11 \pm 0.06$ \\
56$\star$ & 94.10608 & -21.3733 & $24 \pm 6.4$ & 7.7 & $18 \pm 5.8$ & $5.3 \pm 3.7$ & $-0.55 \pm 0.27$ & $6.6 \pm 1.8$ & $0.35 \pm 0.10$ \\
57 & 94.10637 & -21.3704 & $20 \pm 5.9$ & 7 & $2.6 \pm 3.2$ & $17 \pm 5.5$ & $0.73 \pm 0.30$ & $5.5 \pm 1.7$ & $0.29 \pm 0.09$ \\
58 & 94.10726 & -21.3758 & $9 \pm 4.7$ & 3.6 & $4.5 \pm 3.7$ & $4.5 \pm 3.7$ & $-0.00 \pm 0.59$ & $2.6 \pm 1.4$ & $0.14 \pm 0.07$ \\
59 & 94.10817 & -21.3771 & $43 \pm 8.1$ & 15 & $27 \pm 6.7$ & $16 \pm 5.4$ & $-0.27 \pm 0.20$ & $12 \pm 2.3$ & $0.66 \pm 0.12$ \\
60 & 94.10868 & -21.3750 & $13 \pm 5.1$ & 4.9 & $7 \pm 4.1$ & $6.1 \pm 3.9$ & $-0.07 \pm 0.43$ & $3.9 \pm 1.5$ & $0.21 \pm 0.08$ \\
61 & 94.10971 & -21.3701 & $16 \pm 5.8$ & 6.9 & $12 \pm 5.2$ & $3.6 \pm 3.7$ & $-0.55 \pm 0.38$ & $4.5 \pm 1.6$ & $0.24 \pm 0.09$ \\
62 & 94.11047 & -21.3701 & $130 \pm 13$ & 40 & $83 \pm 11$ & $46 \pm 8.3$ & $-0.28 \pm 0.10$ & $36 \pm 3.7$ & $1.93 \pm 0.20$ \\
63 & 94.11082 & -21.3769 & $9.1 \pm 4.7$ & 4 & $7.4 \pm 4.3$ & $1.7 \pm 2.9$ & $-0.62 \pm 0.55$ & $2.7 \pm 1.4$ & $0.14 \pm 0.07$ \\
64 & 94.11177 & -21.3698 & $89 \pm 11$ & 27 & $60 \pm 9.4$ & $29 \pm 6.8$ & $-0.35 \pm 0.12$ & $26 \pm 3.2$ & $1.36 \pm 0.17$ \\
65 & 94.11218 & -21.3762 & $6.1 \pm 4.1$ & 2.4 & $0 \pm 2.1$ & $6.4 \pm 4.1$ & $1.00 \pm 0.66$ & $1.9 \pm 1.2$ & $0.10 \pm 0.07$ \\
66 & 94.11249 & -21.3632 & $31 \pm 7.2$ & 12 & $21 \pm 6.2$ & $9.8 \pm 4.7$ & $-0.37 \pm 0.24$ & $8.7 \pm 2$ & $0.46 \pm 0.11$ \\
67 & 94.11423 & -21.3713 & $22 \pm 6.3$ & 8 & $18 \pm 5.8$ & $4 \pm 3.6$ & $-0.64 \pm 0.28$ & $6.5 \pm 1.8$ & $0.35 \pm 0.10$ \\
68 & 94.11599 & -21.3825 & $9.5 \pm 4.8$ & 4 & $7.9 \pm 4.4$ & $1.6 \pm 3$ & $-0.65 \pm 0.54$ & $2.7 \pm 1.4$ & $0.14 \pm 0.07$ \\
69 & 94.11693 & -21.3759 & $8 \pm 4.5$ & 3.1 & $6 \pm 4.2$ & $2 \pm 2.9$ & $-0.49 \pm 0.60$ & $2.3 \pm 1.3$ & $0.12 \pm 0.07$ \\
70 & 94.11814 & -21.3609 & $79 \pm 11$ & 29 & $53 \pm 8.8$ & $26 \pm 6.6$ & $-0.35 \pm 0.13$ & $23 \pm 3$ & $1.21 \pm 0.16$ \\
71 & 94.12022 & -21.3768 & $9.8 \pm 4.7$ & 3.9 & $8 \pm 4.4$ & $1.8 \pm 2.9$ & $-0.63 \pm 0.51$ & $3.4 \pm 1.6$ & $0.18 \pm 0.09$ \\
72 & 94.12192 & -21.3680 & $30 \pm 7$ & 13 & $29 \pm 6.8$ & $0.95 \pm 2.6$ & $-0.94 \pm 0.17$ & $10 \pm 2.3$ & $0.54 \pm 0.12$ \\
73 & 94.12476 & -21.3786 & $39 \pm 7.9$ & 14 & $25 \pm 6.6$ & $14 \pm 5.2$ & $-0.28 \pm 0.21$ & $12 \pm 2.4$ & $0.64 \pm 0.13$ \\
74 & 94.12505 & -21.3792 & $46 \pm 8.5$ & 16 & $41 \pm 8$ & $5.2 \pm 3.8$ & $-0.77 \pm 0.15$ & $13 \pm 2.5$ & $0.71 \pm 0.13$ \\
\enddata
\tablecomments{(1) Source number, (2) Right Ascension (RA), (3)
  Declination (Dec). (4) Net counts in broad (0.5--8 keV) band. The uncertainty expressed here takes into
  account the fluctuations in the source as well as in the
  background. (5) Broad band source detection significance from {\tt
    wavdetect}.  This computes how unlikely it is for the background
  in the customized psf region to fluctuate to yield the detected
  number of counts.  Note that the psf region is optimized differently
  in {\tt wavdetect} than in the calculation of column (4) and is
  typically larger than in the latter. (6)-(7) Net counts in soft
  (0.5--2 keV) and hard (2--8 keV) bands respectively. Uncertainties in net counts are quoted to
  $1\,\sigma$. (7) Hardness ratio, computed with eq. (\ref{eq:hr}). Uncertainties were obtained by applying error propagation to the uncertainties in the net counts. (8) X-ray luminosity in the 0.5--8 keV band, (9) X-ray flux in the 0.5--8 keV band. $\dagger$ Central Active Galactic Nucleus, $\ddagger$ Extended soft X-ray source, $\star$ SN~2003H ($3.1\arcsec$ match), $\diamond$ SN~2013ai ($0.12\arcsec$ match), $\ast$ SN~1999ec ($4.2\arcsec$ match).}
\end{deluxetable}

\end{appendix}

\end{document}